\documentclass[aps,prc,twocolumn,showpacs,preprintnumbers,nofootinbib,float,superscriptaddress,longbibliography]{revtex4-2}
\usepackage{graphicx,fancybox}
\usepackage{amsmath,amssymb}
\usepackage[colorlinks=true, pdfstartview=FitV, linkcolor=red, citecolor=blue, urlcolor=blue]{hyperref}
\usepackage{color}
\usepackage{soul}
\usepackage{array}
\usepackage{url}
\usepackage[utf8]{inputenc}

\usepackage{lineno}
\newcommand*\patchAmsMathEnvironmentForLineno[1]{%
  \expandafter\let\csname old#1\expandafter\endcsname\csname #1\endcsname
  \expandafter\let\csname oldend#1\expandafter\endcsname\csname end#1\endcsname
  \renewenvironment{#1}%
     {\linenomath\csname old#1\endcsname}%
     {\csname oldend#1\endcsname\endlinenomath}}%
\newcommand*\patchBothAmsMathEnvironmentsForLineno[1]{%
  \patchAmsMathEnvironmentForLineno{#1}%
  \patchAmsMathEnvironmentForLineno{#1*}}%
\AtBeginDocument{%
\patchBothAmsMathEnvironmentsForLineno{equation}%
\patchBothAmsMathEnvironmentsForLineno{align}%
\patchBothAmsMathEnvironmentsForLineno{flalign}%
\patchBothAmsMathEnvironmentsForLineno{alignat}%
\patchBothAmsMathEnvironmentsForLineno{gather}%
\patchBothAmsMathEnvironmentsForLineno{multline}%
}
\usepackage[normalem]{ulem}

\begin{document}

\title{Inclusive jet and hadron suppression in a multistage approach}


\author{A.~Kumar}
\affiliation{Department of Physics, McGill University, Montr\'{e}al QC H3A\,2T8, Canada.}
\affiliation{Department of Physics and Astronomy, Wayne State University, Detroit MI 48201.}

\author{Y.~Tachibana}
\affiliation{Akita International University, Yuwa, Akita-city 010-1292, Japan.}
\affiliation{Department of Physics and Astronomy, Wayne State University, Detroit MI 48201.}

\author{C.~Sirimanna}
\affiliation{Department of Physics and Astronomy, Wayne State University, Detroit MI 48201.}

\author{G.~Vujanovic}
\affiliation{Department of Physics and Astronomy, Wayne State University, Detroit MI 48201.}
\affiliation{Department of Physics, University of Regina, Regina, SK S4S 0A2, Canada}

\author{S.~Cao}
\affiliation{Institute of Frontier and Interdisciplinary Science, Shandong University, Qingdao, Shandong 266237, China}
\affiliation{Department of Physics and Astronomy, Wayne State University, Detroit MI 48201.}

\author{A.~Majumder}
\affiliation{Department of Physics and Astronomy, Wayne State University, Detroit MI 48201.}

\author{Y.~Chen}
\affiliation{Laboratory for Nuclear Science, Massachusetts Institute of Technology, Cambridge MA 02139.}
\affiliation{Department of Physics, Massachusetts Institute of Technology, Cambridge MA 02139.}

\author{L.~Du}
\affiliation{Department of Physics, McGill University, Montr\'{e}al QC H3A\,2T8, Canada.}

\author{R.~Ehlers}
\affiliation{Department of Physics and Astronomy, University of Tennessee, Knoxville TN 37996.}
\affiliation{Physics Division, Oak Ridge National Laboratory, Oak Ridge TN 37830.}

\author{D.~Everett}
\affiliation{Department of Physics, The Ohio State University, Columbus OH 43210.}

\author{W.~Fan}
\affiliation{Department of Physics, Duke University, Durham NC 27708.}

\author{Y.~He}
\affiliation{Guangdong Provincial Key Laboratory of Nuclear Science, Institute of Quantum Matter, South China Normal University, Guangzhou 510006, China.}
\affiliation{Guangdong-Hong Kong Joint Laboratory of Quantum Matter, Southern Nuclear Science Computing Center, South China Normal University, Guangzhou 510006, China.}

\author{J.~Mulligan}
\affiliation{Department of Physics, University of California, Berkeley CA 94270.}
\affiliation{Nuclear Science Division, Lawrence Berkeley National Laboratory, Berkeley CA 94270.}

\author{C.~Park}
\affiliation{Department of Physics and Astronomy, Wayne State University, Detroit MI 48201.}


\author{A.~Angerami}
\affiliation{Lawrence Livermore National Laboratory, Livermore CA 94550.}

\author{R.~Arora}
\affiliation{Research Computing Group, University Technology Solutions, The University of Texas at San Antonio, San Antonio TX 78249.}

\author{S.~A.~Bass}
\affiliation{Department of Physics, Duke University, Durham NC 27708.}

\author{T.~Dai}
\affiliation{Department of Physics, Duke University, Durham NC 27708.}

\author{H.~Elfner}
\affiliation{GSI Helmholtzzentrum f\"{u}r Schwerionenforschung, 64291 Darmstadt, Germany.}
\affiliation{Institute for Theoretical Physics, Goethe University, 60438 Frankfurt am Main, Germany.}
\affiliation{Frankfurt Institute for Advanced Studies, 60438 Frankfurt am Main, Germany.}

\author{R.~J.~Fries}
\affiliation{Cyclotron Institute, Texas A\&M University, College Station TX 77843.}
\affiliation{Department of Physics and Astronomy, Texas A\&M University, College Station TX 77843.}

\author{C.~Gale}
\affiliation{Department of Physics, McGill University, Montr\'{e}al QC H3A\,2T8, Canada.}

\author{F.~Garza}
\affiliation{Cyclotron Institute, Texas A\&M University, College Station TX 77843.}
\affiliation{Department of Physics and Astronomy, Texas A\&M University, College Station TX 77843.}

\author{M.~Heffernan}
\affiliation{Department of Physics, McGill University, Montr\'{e}al QC H3A\,2T8, Canada.}

\author{U.~Heinz}
\affiliation{Department of Physics, The Ohio State University, Columbus OH 43210.}

\author{B.~V.~Jacak}
\affiliation{Department of Physics, University of California, Berkeley CA 94270.}
\affiliation{Nuclear Science Division, Lawrence Berkeley National Laboratory, Berkeley CA 94270.}

\author{P.~M.~Jacobs}
\affiliation{Department of Physics, University of California, Berkeley CA 94270.}
\affiliation{Nuclear Science Division, Lawrence Berkeley National Laboratory, Berkeley CA 94270.}

\author{S.~Jeon}
\affiliation{Department of Physics, McGill University, Montr\'{e}al QC H3A\,2T8, Canada.}

\author{K.~Kauder}
\affiliation{Department of Physics, Brookhaven National Laboratory, Upton NY 11973.}

\author{L.~Kasper}
\affiliation{Department of Physics and Astronomy, Vanderbilt University, Nashville TN 37235.}

\author{W.~Ke}
\affiliation{Los Alamos National Laboratory, Theoretical Division, Los Alamos, NM 87545.}

\author{M.~Kelsey}
\affiliation{Department of Physics and Astronomy, Wayne State University, Detroit MI 48201.}

\author{B.~Kim}
\affiliation{Cyclotron Institute, Texas A\&M University, College Station TX 77843.}
\affiliation{Department of Physics and Astronomy, Texas A\&M University, College Station TX 77843.}

\author{M.~Kordell~II}
\affiliation{Cyclotron Institute, Texas A\&M University, College Station TX 77843.}
\affiliation{Department of Physics and Astronomy, Texas A\&M University, College Station TX 77843.}

\author{J.~Latessa}
\affiliation{Department of Computer Science, Wayne State University, Detroit MI 48202.}

\author{Y.-J.~Lee}
\affiliation{Laboratory for Nuclear Science, Massachusetts Institute of Technology, Cambridge MA 02139.}
\affiliation{Department of Physics, Massachusetts Institute of Technology, Cambridge MA 02139.}

\author{D.~Liyanage}
\affiliation{Department of Physics, The Ohio State University, Columbus OH 43210.}

\author{A.~Lopez}
\affiliation{Instituto  de  F\`{i}sica,  Universidade  de  S\~{a}o  Paulo,  C.P.  66318,  05315-970  S\~{a}o  Paulo,  SP,  Brazil. }

\author{M.~Luzum}
\affiliation{Instituto  de  F\`{i}sica,  Universidade  de  S\~{a}o  Paulo,  C.P.  66318,  05315-970  S\~{a}o  Paulo,  SP,  Brazil. }

\author{S.~Mak}
\affiliation{Department of Statistical Science, Duke University, Durham NC 27708.}

\author{A.~Mankolli}
\affiliation{Department of Physics and Astronomy, Vanderbilt University, Nashville TN 37235.}

\author{C.~Martin}
\affiliation{Department of Physics and Astronomy, University of Tennessee, Knoxville TN 37996.}

\author{H.~Mehryar}
\affiliation{Department of Computer Science, Wayne State University, Detroit MI 48202.}

\author{T.~Mengel}
\affiliation{Department of Physics and Astronomy, University of Tennessee, Knoxville TN 37996.}

\author{C.~Nattrass}
\affiliation{Department of Physics and Astronomy, University of Tennessee, Knoxville TN 37996.}

\author{D.~Oliinychenko}
\affiliation{Nuclear Science Division, Lawrence Berkeley National Laboratory, Berkeley CA 94270.}

\author{J.-F. Paquet}
\affiliation{Department of Physics, Duke University, Durham NC 27708.}

\author{J.~H.~Putschke}
\affiliation{Department of Physics and Astronomy, Wayne State University, Detroit MI 48201.}

\author{G.~Roland}
\affiliation{Laboratory for Nuclear Science, Massachusetts Institute of Technology, Cambridge MA 02139.}
\affiliation{Department of Physics, Massachusetts Institute of Technology, Cambridge MA 02139.}

\author{B.~Schenke}
\affiliation{Physics Department, Brookhaven National Laboratory, Upton NY 11973.}

\author{L.~Schwiebert}
\affiliation{Department of Computer Science, Wayne State University, Detroit MI 48202.}

\author{A.~Sengupta}
\affiliation{Cyclotron Institute, Texas A\&M University, College Station TX 77843.}
\affiliation{Department of Physics and Astronomy, Texas A\&M University, College Station TX 77843.}

\author{C.~Shen}
\affiliation{Department of Physics and Astronomy, Wayne State University, Detroit MI 48201.}
\affiliation{RIKEN BNL Research Center, Brookhaven National Laboratory, Upton NY 11973.}

\author{A.~Silva}
\affiliation{Department of Physics and Astronomy, University of Tennessee, Knoxville TN 37996.}

\author{D.~Soeder}
\affiliation{Department of Physics, Duke University, Durham NC 27708.}

\author{R.~A.~Soltz}
\affiliation{Department of Physics and Astronomy, Wayne State University, Detroit MI 48201.}
\affiliation{Lawrence Livermore National Laboratory, Livermore CA 94550.}

\author{J.~Staudenmaier}
\affiliation{GSI Helmholtzzentrum f\"{u}r Schwerionenforschung, 64291 Darmstadt, Germany.}

\author{M.~Strickland}
\affiliation{Department of Physics, Kent State University, Kent, OH 44242.}

\author{J.~Velkovska}
\affiliation{Department of Physics and Astronomy, Vanderbilt University, Nashville TN 37235.}

\author{X.-N.~Wang}
\affiliation{Key Laboratory of Quark and Lepton Physics (MOE) and Institute of Particle Physics, Central China Normal University, Wuhan 430079, China.}
\affiliation{Department of Physics, University of California, Berkeley CA 94270.}
\affiliation{Nuclear Science Division, Lawrence Berkeley National Laboratory, Berkeley CA 94270.}

\author{R.~L.~Wolpert}
\affiliation{Department of Statistical Science, Duke University, Durham NC 27708.}

\collaboration{The JETSCAPE Collaboration}

\begin{abstract}
We present a new study of jet interactions in the quark-gluon plasma created in high-energy heavy-ion collisions, using a multistage event generator within the \textsc{jetscape} framework. We focus on medium-induced modifications in the rate of inclusive jets and high transverse momentum (high-$p_{\mathrm{T}}$) hadrons. Scattering-induced jet energy loss is calculated in two stages: A high virtuality stage based on the \textsc{matter} model, in which scattering of highly virtual partons modifies the vacuum radiation pattern, and a second stage at lower jet virtuality based on the \textsc{lbt} model, in which leading partons gain and lose virtuality by scattering and radiation. 
Coherence effects that reduce the medium-induced emission rate in the \textsc{matter} phase are also included. The \textsc{trento} model is used for initial conditions, and the (2+1)dimensional \textsc{vishnu} model is used for viscous hydrodynamic evolution. Jet interactions with the medium are modeled via 2-to-2 scattering with Debye screened potentials, in which the recoiling partons are tracked, hadronized, and included in the jet clustering. Holes left in the medium are also tracked and subtracted to conserve transverse momentum. 
Calculations of the nuclear modification factor ($R_{\mathrm{AA}}$) for inclusive jets and high-$p_{\mathrm{T}}$ hadrons are compared to experimental measurements at the BNL Relativistic Heavy Ion Collider (RHIC) and the CERN Large Hadron Collider (LHC). Within this framework, we find that with one extra parameter which codifies the transition between stages of jet modification---along with the typical parameters such as the coupling in the medium, the start and stop criteria etc.---we can describe these data at all energies for central and semicentral collisions without a rescaling of the jet transport coefficient $\hat{q}$.
\end{abstract}

\maketitle


\section{Introduction}
\label{Section:Intro}
Jet modification in high-energy heavy-ion collisions~\cite{Majumder:2010qh,Cao:2020wlm} is currently one of the leading mechanisms to study the properties of the quark-gluon plasma (QGP)~\cite{Shuryak:1978ij,Shuryak:1980tp} created at the Relativistic Heavy-Ion Collider (RHIC) in Brookhaven National Laboratory (BNL) and the Large Hadron Collider (LHC) in CERN. 
Due to the much larger momentum scales associated with partons in a jet, these partons typically exchange momenta with the medium that are much larger than the thermal momentum scale. 
As a result, they probe the medium at much shorter distance scales than the thermal scale~\cite{Akiba:2015jwa,Mehtar-Tani:2011hma,Armesto:2011ir, Casalderrey-Solana:2011ule,Kumar:2019uvu}. 
The primary observable for studying jet energy loss is the nuclear modification factor, $R_{\mathrm{AA}}$, defined as the ratio of the yield in heavy-ion collisions (typically in bins of transverse momentum $p_{\mathrm{T}}$) to the corresponding yield in proton-proton collisions, scaled by the number of binary nucleon-nucleon collisions for a specified class of heavy-ion events. 

Early experimental jet-modification results at RHIC were restricted to single-hadron spectra ~\cite{PHENIX:2001hpc,PHENIX:2003djd,PHENIX:2003qdj,STAR:2002ggv,STAR:2003fka}, di-hadron correlations~\cite{PHENIX:2005zfm,STAR:2002svs}, and $\gamma$-hadron correlation~\cite{STAR:2016jdz, PHENIX:2020alr}. 
Theoretical approaches at the time were likewise restricted to the calculation of energy loss of the leading parton in a jet~\cite{Gyulassy:1993hr}.
Already at that time there existed several different approaches that described the nuclear modification of the single-hadron spectrum within error bars ~\cite{Gyulassy:2000er, Salgado:2003gb, Arnold:2002ja, Wang:2001ifa}. The differences in formalism between the approaches applied at RHIC manifested in the widely different values of the jet transport coefficient $\hat{q}$, that was extracted by these different approaches when compared to the same data~\cite{Majumder:2007iu,Bass:2008rv}. 
The jet transport coefficient $\hat{q}$ is the mean squared momentum exchanged between a jet parton and the medium, per unit length traversed by the jet parton, in a direction transverse to the momentum of the jet parton:
\begin{align}
    \hat{q} = \frac{1}{N_\mathrm{events}} \sum_{i=1}^{N_\mathrm{events}}\frac{ (k_\perp^i)^2 }{L_i} \approxeq \int d^2 k_\perp k_\perp^2 \frac{d^2 \Gamma^\mathrm{el}}{ dk_\perp^2 }.
    \label{eq:qhat-definition}
\end{align}

The equation above sums over heavy-ion events where jet partons encounter varying momentum exchanges with the medium. The meaning of the expression above is that in event $i$ we consider the propagation of a jet parton a distance $L_i$, shorter than its lifetime $\tau_i$ ($L_i<\tau_i$). It may encounter several scatterings in this length yielding a net transverse momentum $k_\perp^i$ (different in each event). 
Under the assumption of short Debye screening lengths, multiple scatterings can be factorized into uncorrelated single scatterings. One can then reduce the length $L_i$, such that the hard parton will at most engender one single scattering. In this limit, $\hat{q}$ becomes a local property of the QGP.

In the equation above, $\Gamma^\mathrm{el}$ is the scattering rate between a jet parton and constituents from the QGP. In the limit of single scattering, it will include 2-to-2 matrix elements that describe the scattering off a single constituent in the medium. In principle, this rate, which encompasses the energy-momentum exchange between the jet and the medium, is not known \emph{a priori}. 
It may be nonperturbative~\cite{Liu:2006he, CasalderreySolana:2007qw, Kumar:2020wvb} or perturbative~\cite{Arnold:2008vd} in nature, or a combination of both~\cite{Kumar:2019uvu}: nonperturbative for softer exchanges and perturbative for harder exchanges. 
One of the central goals of the study of jet quenching is the determination of this rate or distribution, and by extension, to determine the dynamical behavior of the medium constituents, off which the hard partons scatter. Observables that strongly depend on the soft component of the jet are sensitive to physics beyond the scattering rate, e.g., the energy-momentum deposition and thermalization in the medium~\cite{Neufeld:2009ep, Qin:2009uh, Chesler:2007an}. However, these effects should be separable by comparing a sufficiently comprehensive simulator with a wide range of data.

Subsequent measurements of hadron production at the LHC, extending the transverse momentum ($p_{\mathrm{T}}$) reach by an order of magnitude~\cite{ALICE:2012aqc,CMS:2012aa}, revealed a reduction in the nuclear modification at the LHC, even accounting for the change in the shape of the hard spectrum, suggesting a weakening of the interaction strength between the medium and the hard parton, typically indicated by the ratio $\hat{q}/T^3$ (where $T$ is the ambient temperature). 
This effective reduction in the suppression at LHC, compared to RHIC, was established by the JET Collaboration via a comparison with the nuclear modification factor for high-$p_{\mathrm{T}}$ single-hadron spectra for the most-central (0-5\%) events at RHIC and LHC~\cite{JET:2013cls}. 
A wide range of approaches~\cite{Majumder:2011uk,Qin:2009bk,Chen:2011vt,Schenke:2009gb,Xu:2014ica} to jet modification were constrained to calculate energy loss while propagating through an identical fluid medium, constructed using a realistic equation of state and by comparison with bulk observables. All these approaches were compared to the nuclear modification data, and all comparisons required a reduction in the overall normalization of $\hat{q}$ at LHC compared to RHIC. 

Hard sector measurements at the LHC and RHIC have since been extended to cover a variety of observables related to jets over a range of collision energies and centralities, presenting an opportunity to further constrain and refine the theoretical approach to modeling jet quenching in heavy-ion collisions. 
To systematically compare theoretical models with this growing assortment of observables requires a comprehensive and extendable simulation framework.
The JETSCAPE Collaboration has developed such a framework for a multistage event generator to study and interpret bulk medium, jet-quenching, and heavy-flavor measurements in heavy-ion collisions~\cite{Putschke:2019yrg}. 
The \textsc{jetscape} framework has been benchmarked against $p+p$  collisions~\cite{JETSCAPE:2019udz} and used for Bayesian parameter estimation of the bulk properties of the QGP~\cite{JETSCAPE:2020mzn,JETSCAPE:2020shq,JETSCAPE:2022cob}. An earlier, simplified multistage generator was used to carry out a Bayesian evaluation of the jet transport coefficient $\hat{q}$~\cite{JETSCAPE:2021ehl} [comparing to central and semicentral $R_{\mathrm{AA}}$ at RHIC ($\pi^0$ at $\sqrt{s_\mathrm{NN}}=200$~GeV) and the LHC ($h^\pm$ at $\sqrt{s_\mathrm{NN}}=2.76$~TeV and 5.02~TeV)]. 

In this paper, we present results from a new calculation for nuclear modification factors for inclusive jets and charged particles, calculated for a range of centralities and collision energies using the publicly available \textsc{jetscape} 3 series~\cite{jetscape}. 
This version incorporates modifications of a hard thermal-loop (HTL) $\hat{q}$ for fixed coupling, running coupling, and with a virtuality dependent factor that effectively modulates $\hat{q}$ 
to account for a reduced medium-induced emission in the high virtuality phase, due to coherence effects. For the description of the medium response in this study, the energy and momentum exchanged between jet partons and the medium are tracked using a recoil and hole scheme in both the high-virtuality and the transport stages.

Results are presented in the form of a sensitivity study in which we vary the parameters governing these new features, along with parameters for formation time, hadronization temperatures, and switching virtuality. 
These results will be compared to nuclear modification factors for inclusive jets and hadrons over a range of collision centralities for $\sqrt{s_\mathrm{NN}}=2.76$ and 5.02~TeV, 
measured by the ALICE, ATLAS, and CMS experiments, and for $\sqrt{s_\mathrm{NN}}=200$~GeV measured by the PHENIX and STAR experiments. 
This effort will demonstrate that our multistage framework, with an in-medium coupling strength and a transition scale between the stages, 
set by comparison to data at one energy and centrality along with parameters typical of energy loss in a fluid medium, such as the energy-loss start and end time, is sufficient to describe $R_{\mathrm{AA}}$ data for inclusive jets and hadrons, simultaneously at all centralities and energies. 
This work will set the stage for a future Bayesian parameter estimation over a wider range of parameters.

The remainder of the paper is organized as follows:  In Sec.~\ref{Section:Framework}, we will describe the various components of the simulation framework: the evolution of the bulk medium (Sec.~\ref{Subsection:SoftSectorAA}) that provides a substrate for the propagation of jets, carried out by a combination of the high-virtuality stage (\textsc{matter}, described in Sec.~\ref{Subsection:MATTERmodule}), and a lower-virtuality stage (\textsc{lbt}, described in Sec.~\ref{Subsection:LBTmodule}). 
Jet partons and partons scattered by jets are then hadronized using a variant of the Lund \textsc{pythia} model described in Sec.~\ref{Subsection:HadronizationModule}. The parametrized interaction of the jet partons with the medium is described in Sec.~\ref{Section:Jetmodel}. 
This involves both a description of the theory behind the jet transport coefficient $\hat{q}$ in Sec.~\ref{Subsection:Q2-dependentQhat}, and the parametric dependencies of coherence effects, as well as a description of the recoil formalism used in Sec.~\ref{Subsection:medium-response}. A multistage simulator will engender multiple parameters, these are recapitulated and discussed in Sec.~\ref{Section:sim} along with various technical details of the simulation. Results of the simulation are presented in Sec.~\ref{Section:Results}. A summary of our findings is presented in Sec.~\ref{Section:Summary}, followed by a discussion of alternate parameter choices, different background subtraction mechanisms, and other considerations in the appendices. 

\section{Overview of \textsc{jetscape} framework}
\label{Section:Framework}
To explore the multiscale dynamics of jets within the framework of \textsc{jetscape}, we embed the space-time information of the bulk medium in the parton shower and set up an effective parton evolution within this background.
Thus, the fluid dynamical simulation is run first, the space-time profile of the energy-momentum tensor, along with local fluid velocities and temperature, are stored and then recalled in a second run of the framework, which simulates hard parton formation, energy loss, and fragmentation.

The high-virtuality parton evolution is handled by the \textsc{matter} event generator (Sec.~\ref{Subsection:MATTERmodule}) and the low-virtuality parton evolution is handled by \textsc{lbt} event generator (Sec.~\ref{Subsection:LBTmodule}). The transfer of a parton from one energy loss model to another is performed on a parton-by-parton basis. 
In this first attempt to simultaneously describe the nuclear modification factor for inclusive jets and leading hadrons we use a single switching virtuality $Q_{\rm sw}$. 
Partons with a virtuality $Q>Q_{\rm sw}$ are handled by the \textsc{matter} generator, while those below $Q_{\rm sw}$ are handled by the \textsc{lbt} generator. Partons that escape the medium are transferred back to the \textsc{matter} generator to continue showering in the vacuum. For $p+p$ simulations, the entire parton vacuum evolution is carried out in the \textsc{matter} generator by switching off the medium effect.

A very straightforward means to understand the cause of the transition from high virtuality to a lower virtuality gen- erator is given in Ref.~\cite{Cao:2021rpv}, where one can see the medium modified contribution to the radiation spectrum steadily grow and surpass the vacuum contribution as the virtuality is reduced. As a parton propagates in a medium it undergoes multiple scattering, which iteratively adds virtuality up to the medium generated scale
\begin{equation}
    Q^2_{\rm Med} = \int_0^\tau d \xi \hat{q}(\xi) \approx \hat{q} \tau, 
\end{equation}
where $\tau$ is the lifetime of the parton and $\hat{q}(\xi)$ is the local value of the transport coefficient.

Besides the energy $E$ of the parton, emissions (or splits) depend on the virtuality of the parton $Q^2$: they have a transverse momentum $l_\perp^2 \approxeq y(1-y) Q^2$ (where $y$ is the forward momentum fraction of one of the daughter partons in a split), and occur in a time, 
\begin{equation}
    \tau \approxeq \frac{2E}{Q^2}.
\end{equation}
If $Q$ is large, $\tau$ is small, and the accumulated virtuality from scattering is small compared to the actual virtuality of the parton $Q_{\rm Med} \ll Q $. In this regime, scattering from the medium is at most a perturbative correction to the process of vacuum emission. This portion is simulated with the \textsc{matter} generator.

As the virtuality $Q$ continues to drop with each successive emission, the formation time of splits increases and the medium generated virtuality becomes comparable to the virtuality of the parton. The time of onset of this stage can be estimated by setting $Q=Q_{\rm Med}$: 
\begin{equation}
    \tau_{\rm  sw} \approxeq \frac{2E}{Q^2_{\rm Med}} \approxeq \sqrt{  \frac{2E}{\hat{q}} }.
\end{equation}
By this point, the medium modified kernel far exceeds the vacuum emission kernel~\cite{Cao:2021rpv}. Thus, generators such as \textsc{martini} and \textsc{lbt} which simulate this phase typically ignore the vacuum contribution. Since the $T$ and $E$ dependent transition scale 
($Q_{\rm Med}$), between the high and low virtuality regimes, is only known approximately, we replace it with a parameter $Q_{\rm sw}$, which is then tuned in comparison with data. 

As the shower proceeds through the dynamical medium, the ambient temperature will eventually begin to drop and this causes $\hat{q}$ and the medium generated scale to also drop. If this drop is sudden, e.g., in the case of the jet passing the parton-hadron transition surface, the jet parton may once again be in a regime where its virtuality is much larger than that generated by multiple scatterings in the medium. To mimic this effect, partons that cross the phase transition surface with $Q^2 > 1$~GeV$^2$ are fed back to the \textsc{matter} generator.

In the following subsections, we describe the basic features of the simulation of the bulk medium. 
This is followed by a discussion of the salient features of the \textsc{matter} and \textsc{lbt} event generators. 
Both of these generators have been extensively discussed in the literature. Hence, the descriptions that follow will be terse. The last subsection (\ref{Subsection:HadronizationModule}) concludes with a discussion of the hard hadronization module, which fragments the jet partons, recoil partons, and hole partons into regular and hole hadrons respectively.

The simulations described in this paper have been carried out using the publicly available \textsc{jetscape} 3.0~\cite{jetscape} version of the event generator framework. 
Unlike prior versions, the current framework can simultaneously account for light and heavy flavor energy loss. It also contains modules that can nonperturbatively deal with the energy and momentum deposited from a jet to the medium. There are three separate modules for the hadronization of the hard sector and one for the soft sector. 
In the main body of this paper, we use the simple recoil description of the medium response (described in the next section), with other approaches discussed in the appendices. 

\subsection{Dynamics of the soft sector in A+A collisions}
\label{Subsection:SoftSectorAA}

The QGP medium evolution is modeled by relativistic viscous hydrodynamics. We assume longitudinal boost invariance for heavy-ion collisions at the top RHIC and LHC energies. Event-by-event simulations are set up using the \textsc{trento} initial conditions \cite{Moreland:2014oya} followed by a (2+1)- dimensional [(2+1)D] free-streaming preequilibrium evolution up to a proper time of $\tau_{\rm hydro}$ (=1.2~fm/$c$ at LHC, and 0.5~fm/$c$ at top RHIC energy~\cite{Liu:2015nwa}).

After matching the system's energy-momentum tensor between the preequilibrium and fluid phase, the QGP medium evolution is described by the \textsc{vishnu} (2+1)D hydrodynamics~\cite{Song:2007ux, Shen:2014vra}. 
As the system evolves to dilute densities, individual fluid cells are converted to hadrons via the Cooper-Frye prescription~\cite{Cooper:1974mv, Bernhard:2018hnz}. This conversion is performed on an isothermal hypersurface at $T_{\rm sw} = 151$~MeV~\cite{Huovinen:2012is}. The produced hadrons are transferred to a hadronic transport model \textsc{urqmd} for scatterings and decay~\cite{Bass:1998ca, Bleicher:1999xi}. We point out that while the default \textsc{jetscape} settings use the \textsc{music} generator~\cite{Schenke:2010nt} for the fluid dynamical simulation and the \textsc{smash} generator~\cite{Petersen:2018jag} for the hadronic cascade, the framework is devised to work with a variety of interchangeable generators (detailed comparisons between \textsc{vishnu} and \textsc{music} and between \textsc{urqmd} and \textsc{smash} are presented in Appendix H of Ref.~\cite{JETSCAPE:2020mzn}).

\begin{table*}[!ht]
\caption{Transport coefficients in Eqs. (\ref{eq:bulk_relax}) and (\ref{eq:shear_relax}).} 
\centering
\renewcommand{\arraystretch}{2}
\begin{tabular}{l l l l l l l l l l}
\hline \hline
Bulk & & $\tau_\Pi=\zeta[15(\epsilon+P)(\frac{1}{3}-c^2_{\mathrm{s}})^2]^{-1}$ & $\delta_{\Pi\Pi}=\frac{2}{3}\tau_\Pi$ & & $\lambda_{\Pi\pi}=\frac{8}{5}(\frac{1}{3}-c^2_{\mathrm{s}})\tau_\Pi$ & & \\
Shear & & $\tau_\pi=5\eta[\epsilon+P]^{-1}$  & $\delta_{\pi\pi}=\frac{4}{3}\tau_\pi$ &  & $\lambda_{\pi\Pi}=\frac{6}{5}\tau_\pi$ & & $\tau_{\pi\pi}=\frac{10}{7}\tau_\pi$ & & $\phi_7=\frac{18}{175} \frac{\tau_\pi}{\eta}$\\
\hline \hline
\end{tabular}
\label{table:bulk_shear_relax} 
\end{table*}

The causal relativistic hydrodynamical equation of motion is given by the second-order Israel-Stewart theory \cite{Israel1976310, Israel:1979wp}. In addition to conservation of energy and momentum, second-order hydrodynamical equations also include relaxation-type equations for six independent viscous degrees of freedom, namely five in the shear stress tensor $\pi^{\mu\nu}$ with the remaining being the bulk viscous pressure $\Pi$. Energy-momentum conservation is expressed as:
\begin{align}
\partial_\mu T^{\mu\nu}=0,
\label{eq:energy-momentum_conserv}
\end{align}
with the energy-momentum tensor
\begin{align}
T^{\mu\nu} = \epsilon u^\mu u^\nu - (P+\Pi)\Delta^{\mu\nu}+\pi^{\mu\nu},
\end{align}  
where $\epsilon$ is the energy density, $u^\mu$ is the flow four velocity, $P$ is the thermodynamic pressure related to $\epsilon$ by the lattice QCD equation of state $P(\epsilon)$ at vanishing net baryon density~\cite{Bazavov:2014pvz,Bernhard:2018hnz}. We define the spatial projection tensor as $\Delta^{\mu\nu}=g^{\mu\nu}-u^\mu u^\nu$, where $g^{\mu\nu}={\rm diag}(1,-1,-1,-1)$ is the metric tensor.
The dissipative degrees of freedom are evolved according to
\begin{align}
\tau_\Pi \dot{\Pi}+\Pi &= -\zeta\theta - \delta_{\Pi\Pi}\Pi\theta + \lambda_{\Pi\pi}\pi^{\alpha\beta}\sigma_{\alpha\beta}, \label{eq:bulk_relax}\\
\tau_\pi \dot{\pi}^{\langle\mu\nu\rangle}+\pi^{\mu\nu} &= 2\eta\sigma^{\mu\nu}-\delta_{\pi\pi}\pi^{\mu\nu}\theta + \lambda_{\pi\Pi}\Pi\sigma^{\mu\nu}\nonumber\\&\quad  - \tau_{\pi\pi} \pi^{\langle\mu}_\alpha\sigma^{\nu\rangle\alpha} + \phi_7 \pi^{\langle\mu}_\alpha\pi^{\nu\rangle\alpha},
\label{eq:shear_relax}
\end{align}  
where 
$\dot{\Pi}\equiv u^\alpha\partial_\alpha\Pi$, $\dot{\pi}^{\langle\mu\nu\rangle}\equiv\Delta^{\mu\nu}_{\alpha\beta}u^\lambda\partial_\lambda\pi^{\alpha\beta}$, $\Delta_{\alpha\beta}^{\mu\nu}\equiv\left(\Delta_{\alpha}^{\mu}\Delta_{\beta}^{\nu}+\Delta_{\beta}^{\mu}\Delta_{\alpha}^{\nu}\right)/2-\left(\Delta_{\alpha\beta}\Delta^{\mu\nu}\right)/3$, 
$\theta\equiv\partial_\alpha u^\alpha$, and  $\sigma^{\mu\nu}\equiv\partial^{\langle\mu} u^{\nu\rangle}$ 
with $A^{\langle\mu\nu\rangle}\equiv\Delta^{\mu\nu}_{\alpha\beta}A^{\alpha\beta}$. 
The second-order transport coefficients present in Eqs.~(\ref{eq:bulk_relax}) and (\ref{eq:shear_relax}) were computed in Refs.~\cite{Denicol:2012cn, Denicol:2014vaa} assuming a single-component gas of constituent particles in the limit $m/T \ll 1$, where $m$ is their mass and $T$ is the temperature. Table \ref{table:bulk_shear_relax} summarizes these transport coefficients, where $c^2_\mathrm{s} = \partial P /\partial \epsilon$ is the speed of sound squared. 
The temperature dependent specific shear viscosity $\eta/s(T)$ --- where $s$ is the entropy density --- and the specific bulk viscosity $\zeta/s(T)$ are taken from a recent Bayesian model-to-data comparison~\cite{Bernhard:2019bmu}.

The time evolution of hydrodynamic fields, such as local energy density, temperature, and flow velocity, are stored on disk event-by-event for the second stage for jet showering. During the preequilibrium stage $\tau < \tau_{\rm hydro}$, the Landau matching procedure \cite{Liu:2015nwa} is performed at every time step to compute the local energy density and flow velocity from the system's energy-momentum tensor. Then the local temperature is estimated by the ideal massless quark-gluon gas equation of state with $N_f = 3$ and $N_c = 3$.

Individual jet partons start to interact with the QCD medium at a longitudinal proper time $\tau_0 = 0.6$~fm/$c$. When $\tau_0$ is smaller than the proper time of the fluid dynamical simulation $\tau_{\mathrm{hydro}}$, 
which is the case for the LHC simulation in this work, the energy loss calculation is performed using the local temperature and flow velocity obtained via the Landau matching in the preequilibrium phase. 
We will vary this jet energy loss starting time $\tau_0$ to quantify its effects on the $R_{\mathrm{AA}}$ observables in Fig.~\ref{fig:Type3-q-hat-Effect_of_tau0} below.

As the jet develops its shower inside a dynamically evolving QCD medium, 
$\hat{q}$ is sampled according to the local
temperature information for the jet partons boosted to the
local rest frame of the fluid cell. 
We stop the jet-medium interactions at the energy loss termination temperature $T_{\mathrm{c}} = 160$~MeV, below which the partons propagate only with vacuum emissions in \textsc{matter} followed by fragmentation. In the \textsc{jetscape} two-stage approach, wwe note that neither the jet partons nor the fragmented hadrons interact with hadrons
from the soft sector in the hadronic phase. To quantify the
uncertainty from this approximation on $R_{\mathrm{AA}}$ observables, we will vary the value of the energy loss termination temperature between 150~MeV and 170~MeV (see Fig.~\ref{fig:Type3-q-hat-Effect_of_Tc} below). We remind the reader that the transition temperature at which the fluid simulation undergoes Cooper-Frye particlization is $T=151$~MeV. We do not take into account the parton energy loss in the phase space between $T = 150-151$~MeV after the entire fireball is frozen out.

\subsection{MATTER event generator}
\label{Subsection:MATTERmodule}
The Modular All Twist Transverse-scattering Elastic-drag and Radiation (\textsc{matter}) is a higher-twist formalism-based event generator that simulates the parton modification both in a vacuum and within a medium. In this instance, a parton in \textsc{matter} will engender an arbitrary number of emissions, where stimulated emissions are calculated in the one-rescattering or twist-4 approximation.
It is primarily applicable to the high-virtuality, the high-energy epoch of the parton shower, where the virtuality of the parton $Q^2 \gg \sqrt{\hat{q}E}$. In this phase, the medium-modified radiative processes are \emph{not} dominant, and the successive emissions from the parton are ordered in virtuality.

In \textsc{matter}, 
the distribution of the medium-modified radiated gluon from a single scattering with the medium is given as 
\begin{align}
\frac{dN^a_g}{dydQ^{2}} &= \frac{\alpha_\mathrm{s}}{2\pi} \frac{\tilde{P}_a(y,Q^2)}{Q^2}, 
\label{eq:HT-single-emission-single-scattering} 
\end{align} 
where $\alpha_\mathrm{s}$ is the strong coupling constant and
\begin{align}
\!\!\Tilde{P}_a(y,Q^2)&=P^{\mathrm{vac}}_a(y) \!\left[ 1 +\!\! \int^{\xi^{+}_{o} + \tau^{+}}_{\xi^{+}_{o}}  \!\!\! d\xi^{+}\,K_a(\xi^{+},\xi^{+}_{o},  y, p^{+},Q^{2} )   \right]\!.
\label{eq:HT-splitting function} 
\end{align}
Here $a=(g,q,\bar{q})$ is the parent parton species, 
$P^{\mathrm{vac}}_a(y)$ is the standard Altarelli-Parisi vacuum splitting function, 
$y$ is the momentum fraction carried away by the emitted daughter parton, 
$p^{+}=(p^{0}+p^{3})/\sqrt{2}$ is the light-cone momentum for the parent parton traveling along $z$-direction, 
and $\tau^{+}=2p^{+}/Q^{2}$ is the formation time of the radiated gluon. The parent parton started at $\xi^+_o$ and completes the split at $\xi^{+}$ which lies between $\xi^{+}_{o}$ and $\xi^{+}_{o} + \tau^{+}$.
The quantity $K_a(\xi^{+},\xi^{+}_{o},  y, p^{+},Q^{2} )$ is single-emission-single-scattering kernel given as~\cite{Abir:2014sxa,Abir:2015hta,Cao:2017crw,Kumar:2019uvu}
\begin{align}
 K_a(\xi^{+},\xi^{+}_{o},  y, p^{+},Q^{2} ) & = \frac{C^a_{1} \hat{q}_a}{y(1-y) Q^{2} (1+\chi_a)^{2}}  \nonumber\\
 &\quad \times f \left \{ 2 - 2\cos \left( \frac{\xi^{+} - \xi^{+}_{o} }{\tau^{+}} \right) \right \},
\label{matter_kernel}
\end{align}
where, 
\begin{align}
C^a_{1}&=\left\{\left[1-\frac{y}{2}\left(\delta_{a,q}+\delta_{a,\bar{q}}\right)\right] - \chi_a \left[1 - \left(1-\frac{y}{2}\right) \chi_a\right] \right\}.
\end{align}
In above equation, $\delta_{a,q}$ and $\delta_{a,\bar{q}}$ are Kronecker delta functions, while $\chi_a=(\delta_{a,q}+\delta_{a,\bar{q}})y^2m_a^2/(y(1-y)Q^2 -y^2m_a^{2})$ with $m_a$ being the mass of the parent parton $a$. The jet transport coefficient $\hat{q}_a$ measures the average squared transverse momentum broadening per unit length of the medium. 
If the value of $\hat{q}$ is zero, 
the distribution of the emitted gluon in Eq.~(\ref{eq:HT-single-emission-single-scattering}) reduces to a vacuumlike distribution. 
The factor $f$ in Eq. (\ref{matter_kernel}) accounts for the transverse size of the jet parton and is discussed in detail in the next section.

The virtuality-ordered shower is generated based on the Sudakov formalism where we solve the in-medium Dokshitzer-Gribov-Lipatov-Altarelli-Parisi ({DGLAP}) equation using Monte Carlo (MC) method. The shower is initiated by a single hard parton produced at a space-time location $r$ carrying the forward light-cone momentum $p^{+}=(p^{0}+\hat{n}\cdot\vec{p})/\sqrt{2}$, where $p\equiv (p^{0},\vec{p})$ is 4-vector momentum of the parton and $\hat{n}=\vec{p}/|\vec{p}|$ denotes the direction of the jet.
Then, given a maximum allowed virtuality $t_\mathrm{max}$ and minimum virtuality $t_\mathrm{min}$, one determines the virtuality of the parent parton $a$ by sampling the Sudakov form factor given as
\begin{align}
\!S_a(t_\mathrm{max},t)
&=
\exp \!\!\left[ -\!\int^{t_\mathrm{max}}_{t}  \!\!\!dt' C_F \frac{\alpha_\mathrm{s}(t')}{2\pi t'} \int\limits^{y_\mathrm{ max}}_{y_\mathrm{min}} \!\!dy \tilde{P}_a(y,t') \right]\!\!,
\label{eq:SudakovFormFactor}
\end{align}
where the Sudakov form factor represents the probability for a parton to transition from virtuality $t_\mathrm{max}$ to $t$ via ``unresolvable" emissions. Here the path length integration in the in-medium splitting function of Eq.~(\ref{eq:HT-splitting function}) is performed along the direction of the jet $\hat{n}$. 
The virtuality of the parent parton is determined by generating a random number $R$ from a uniform distribution between 0 and 1. If $S_a(t_\mathrm{max},t_\mathrm{min}) > R $, then the parton is assigned $t=t_\mathrm{min}$ and propagates freely to next energy loss routine. However, if $S_a(t_\mathrm{max},t_\mathrm{min}) < R $, then the virtuality $t$ is obtained by solving $S_a(t_\mathrm{max},t)=R$, and the splitting may happen. With the determined virtuality $t$, the splitting function $\tilde{P}_a(y,t)$ is sampled to determine the momentum fraction $y$ shared by the two daughter. This sets the daughters' momenta to be $(1-y)p^{+}$ (daughter 1) and $yp^{+}$ (daughter 2). Now, the daughter parton's virtuality is determined using the Sudakov factor with $(1-y)^2 t$ as maximum virtuality for first daughter and $y^2t$ for the second daughter. Once the actual virtuality $t_{1}$ and $t_{2}$ of the daughters are known, their transverse momentum with respect to the parent is calculated from 
\begin{align}
l^2_{\perp} = y(1-y)t - yt_{1} -(1-y)t_{2}.
\end{align}
Next, the $l^{-}$ component is determined using energy-momentum relation: $l^2_1=t_{1}$ and $l^2_{2}=t_{2}$. Finally, the location ($\vec{r}+\hat{n}\xi$) of medium-induced splitting is determined by sampling a Gaussian distribution given as
\begin{align}
f(\xi^+) = \frac{2}{\tau^{+}_\mathrm{f}\pi} \exp \left[ - \left(\frac{\xi^{+}}{\tau^{+}_\mathrm{f}\sqrt{\pi}} \right)^2  \right],
\end{align}
where $\tau^{+}_\mathrm{f}$ is the mean life time given as $\tau^{+}_\mathrm{f}=2p^{+}/t$.

The above procedure is repeated iteratively for each shower initiating parton until the virtuality reaches or goes below a switching scale $Q_\mathrm{sw}$. At this scale, the parton is transferred to the \textsc{lbt} event generator, discussed in the next subsection. 
The minimum virtuality $t_\mathrm{min}$ 
in the Sudakov sampling is always fixed to 
$1~\mathrm{GeV}^2$. If the parton exits the medium and the lower virtuality generator (in this case, \textsc{lbt}), it will return to \textsc{matter} and continue vacuumlike showering until the virtuality reaches or below $1~\mathrm{GeV}^2$.

\subsection{LBT event generator}
\label{Subsection:LBTmodule}
The linear Boltzmann transport (\textsc{lbt}) model is a parton transport generator that is used to simulate the propagation and interaction of both the jet shower and recoil partons with elastic and inelastic collisions in the QGP medium~\cite{He:2015pra,Cao:2016gvr}. It is primarily applicable to the low-virtuality, high-energy epoch of the parton shower. In this phase, multiple scattering-induced emission is the dominant mechanism of parton energy loss. Vacuumlike emission is ignored in this stage.

The phase space distribution $f_{a}(x_{a},p_{a})$ of the parton with momentum $p^{\mu}_{a}(E_{a},\vec{p}_{a})$ is  determined by solving the linear Boltzmann equation:
\begin{align}
p_{a}.\partial f_{a}(x_{a},p_{a}) = E_{a} (\mathcal{C}^\mathrm{el}_{a} + \mathcal{C}^\mathrm{inel}_{a} ),
\end{align}
where $\mathcal{C}^\mathrm{el}_{a}$ and $\mathcal{C}^\mathrm{inel}_{a}$ are the collision integrals for elastic and inelastic scatterings. 
In this formalism, the total scattering probability is expressed as $P^\mathrm{tot}_{a}=P^\mathrm{el}_{a} + P^\mathrm{inel}_{a} - P^\mathrm{el}_{a}\cdot P^\mathrm{inel}_{a}$, where $P^\mathrm{el}_{a}$ and $P^\mathrm{inel}_{a}$ are elastic and inelastic scattering probability, respectively. These probabilities are sampled using Monte Carlo techniques to determine the type of scattering channel. The probability for a parton to undergo elastic scattering ($2\rightarrow 2$) in the given time step $\Delta t$ is given by $P^\mathrm{el}_{a}=\Gamma^\mathrm{el}_{a} \Delta t$, where the elastic rate is given as
\begin{align}
\begin{split}
\Gamma^\mathrm{el}_{a} =& \sum \limits_{b,c,d} \frac{g_{b}}{2E_{a}} \int \prod \limits_{i=b,c,d} d[p_{i}] f_{b}(\vec{p}_{b}) S_{2}(s,t,u)  \\
& \times (2\pi)^4 \delta^{(4)} (p_{a}+p_{b}-p_{c}-p_{d}) |\mathcal{M}_{ab\rightarrow cd}|^2,
\end{split}
\label{eq:ElasticScatteringRate}
\end{align}
where $g_{b}$ represents spin-color degeneracy, $f_{b}$ is the thermal distribution of parton $b$ in the plasma, $d[p_{i}]=d^{3}p_{i}/[2E_{i} (2\pi)^{3}]$, and $|\mathcal{M}_{ab\rightarrow cd}|^2$ is the amplitude square of the process $a+b\rightarrow c+d$. In the interaction kernel, $S_{2}(s,t,u) =\theta(s \ge 2 m^{2}_\mathrm{D}) \theta (-s+m^{2}_\mathrm{D} \le t \le - m^{2}_\mathrm{D})$ is imposed to regularize the divergence in the matrix element $|\mathcal{M}_{ab\rightarrow cd}|^2$ arising from small angle, $u,t \rightarrow 0$. The Debye screening mass is given as
\begin{align}
    m^2_\mathrm{D} = \frac{4\pi\alpha_\mathrm{s}T^2}{3} \left(N_{c}+\frac{N_{f}}{2}\right),
    \label{eq:DebyMassEquation}
\end{align}
where $N_{f}$ is the active quark flavors in the QGP.

Currently, the \textsc{lbt} model is set up to simulate inelastic channels via single scattering ($2\rightarrow 2+n$) causing a multiple gluon emission ($n$) processes, where the medium-induced gluons are independent. 
At each time step $\Delta t$, the number of gluon emissions is sampled using the Poisson distribution,
\begin{align}
P(n) = \frac{\langle N^{a}_{g} \rangle^{n}}{n!} e^{- \langle N^{a}_{g} \rangle},
\end{align}
where the mean number of gluon $\langle N^{a}_{g} \rangle = \Gamma^\mathrm{inel}_{a}\Delta t$. Thus, the probability for total inelastic scattering process to occur is given as $P^\mathrm{inel}_{a}= 1-P(0) = 1 - e^{-<N^{a}_{g}>}$.
The inelastic rate for medium-induced gluon radiation is given by
\begin{align}
\Gamma^\mathrm{inel}_{a} 
= \frac{1}{1+\delta^{a}_{g}} \int dy dl^{2}_{\perp} \frac{dN^{a}_{g}}{dydl^{2}_{\perp}dt},
\label{eq:InelasticRate}
\end{align}
where $\delta^{a}_{g}$ is the correction term for double counting of the process $g\rightarrow gg$.

The medium-induced gluon spectrum in Eq.~(\ref{eq:InelasticRate}) is derived using the higher-twist energy loss formalism and given as
\begin{align}
\frac{dN^{a}_{g}}{dydl^{2}_{\perp} dt} =  \frac{2\alpha_\mathrm{s}(l^2_{\perp}) P^\mathrm{vac}_{a}(y) l^{4}_{\perp} }{ \pi (l^2_{\perp} + y^2 m^2_{a})^4} \hat{q}_a \sin^{2} \left( \frac{t-t_{i}}{2\tau_{\mathrm{f}}}\right),
\label{eq:HT-mediumInducedRadiation}
\end{align}
where $y$ is the momentum fraction, 
$l_{\perp}$ is the transverse momentum of the radiated gluon, 
$t_{i}$ is the initial time of the parent parton, 
and $\tau_{\mathrm{f}}$ is the formation time of the radiated gluon.

Based on the probabilities $P^\mathrm{tot}_{a}$, $P^\mathrm{el}_{a}$, and $P^\mathrm{inel}_{a}$, 
we first determine whether scattering occurs and whether the scattering is elastic or inelastic. 
Once these are known, the differential spectra given in Eq.~(\ref{eq:ElasticScatteringRate}) and Eq.~(\ref{eq:HT-mediumInducedRadiation}) are sampled to determine the energies and momenta of the outgoing partons. The \textsc{lbt} model has one free parameter, the jet-medium coupling constant $\alpha_\mathrm{s}$ that controls both elastic and inelastic parton energy loss.

In the low virtuality transport stage of a heavy-ion collision, one expects multiple scattering per emission. The \textsc{lbt} generator, however, only includes one re-scattering (or two scatterings) per emission. 
In contrast to this, the \textsc{martini} generator~\cite{Schenke:2009gb} includes an arbitrary number of scatterings per emission. In Ref.~\cite{JETSCAPE:2017eso} we studied multistage energy loss in a static medium using a combination of \textsc{matter} and \textsc{lbt}, as well as \textsc{matter} and \textsc{martini}. In these studies it became clear that for static media with lengths that lie in the range $2\lesssim L \lesssim 8$~fm, there is a negligible difference in the final results if \textsc{martini} is replaced by \textsc{lbt}. Further studies in fluid dynamical media were reported in Ref.~\cite{Park:2019sdn}, where a combination of \textsc{matter}+\textsc{lbt} was compared with \textsc{matter}+\textsc{martini} for jets and leading hadrons at 2.76~TeV for two different centralities. The differences between these two implementations were less than 5\%. Due to the order of magnitude longer compute time required by the \textsc{martini} generator, we carry out this first study of the nuclear modification of jet and leading hadrons in \textsc{jetscape} using the \textsc{matter}+\textsc{lbt} generator.

\subsection{String hadronization}
\label{Subsection:HadronizationModule}
\textsc{jetscape} 3.0 has three different hadronization modules: \textsc{colored hadronization} \cite{Putschke:2019yrg,JETSCAPE:2019udz}, \textsc{colorless hadronization} \cite{Putschke:2019yrg,JETSCAPE:2019udz}, and \textsc{hybrid hadronization} \cite{Kordella:2020nwi}. Both \textsc{colored} and \textsc{colorless hadronization} use the default Lund string fragmentation from \textsc{pythia} 8. The \textsc{hybrid hadronization} model is a combination of Lund string fragmentation and recombination. Since \textsc{colorless hadronization} is the only hadronization used in this study, we provide a brief explanation of \textsc{colorless hadronization} in this subsection.

Even though \textsc{colorless hadronization} uses string hadronization through \textsc{pythia}, it removes all color information prior to the hadronization process. All the partons generated from the collection of shower-initiating partons---the radiated partons, the recoils, etc.---are collected in one list. The module then reconstructs one to several strings depending on the number of quarks and antiquarks in the combined set of showers in that event. 
Depending on whether the total number of quarks or antiquarks are even or odd, extra quarks or antiquarks are added at beam rapidities to make the net quark number of all the showers be zero. 
Gluons are assigned to a string with a quark and an antiquark at either end. Once all partons have been assigned to strings, color tags are generated for all partons, such that each string remains a color singlet. These are then hadronized.

The collection of all hole partons, which are particles introduced for the description of the medium response explained in the next section, is then combined and treated similarly to generate hole hadrons, which can then be subtracted from jet cones within which they appear. 
As will be discussed further in Appendix~\ref{parton_jet}, forming strings out of a large number of partons, especially where there are a lot of soft partons, may run into issues with \textsc{pythia} string-breaking algorithms. 
In cases where two partons have a $| \delta \vec{p} | \lesssim 4 \Lambda_{\mathrm{QCD}}$, the $p_z$ component of the parton at larger rapidity is increased until the bound is reached.

\section{Jet transport coefficient and medium response}
\label{Section:Jetmodel}
In the previous section, we divided the history of a jet shower into the production, the initial propagation of high virtuality partons (in \textsc{matter}), the evolution of lower virtuality partons (in \textsc{lbt}), and their fragmentation into hadrons. In both cases of \textsc{matter} and \textsc{lbt}, 
the scattering in the medium identifies and correlates medium partons which were nudged forward in the direction of the jet. Along with the partons generated by showering, the entire collection of jet-correlated partons now includes the incoming partons from the medium, referred to as \emph{holes} and their scattered versions called \emph{recoils}.

Given the differences in virtuality, there is some difference between stimulated emission in the \textsc{matter} phase versus the \textsc{lbt} phase. Due to the small transverse size of the emission antenna in \textsc{matter}, the effect of scattering in the medium is diminished. 
This leads to a reduction of the effective value of the transport coefficient $\hat{q}$ and the distribution of recoils and holes. These details, along with the method of subtraction of the holes, are discussed in the following subsections.

\subsection{Jet transport coefficient and coherence effects at high virtuality}
\label{Subsection:Q2-dependentQhat}
The dominant mechanism of jet energy loss in the plasma is due to bremsstrahlung radiation, triggered by soft gluon exchanges with the medium. 
The jet transport coefficient $\hat{q}$ defined in Eq.~\eqref{eq:qhat-definition} effectively characterizes the momentum broadening of a single parton due to the in-medium scattering, leading to induced emission.

In the last decade, several attempts have been made to compute $\hat{q}$ using first principles and model-to-data comparisons.
In the limit of high temperature and weak-coupling approximation, the hard-thermal-loop (HTL) based calculation yields a $\hat{q}$ given by~\cite{Arnold:2008vd}:
\begin{equation}
    \hat{q} \approxeq m_{\mathrm{D}}^2 C_a \alpha_{\mathrm{s}} T \left[ \ln \left( \frac{ET}{m_{\mathrm{D}}^2}  \right) + C \right].
    \label{eq:HTL-qhat-formula}
\end{equation}
In the equation above, $C_a$ is the representation specific Casimir, the constant $C$ arises from different choices of the upper limit of the $k_\perp$ integral in Eq.~\eqref{eq:qhat-definition}, which leads to the logarithmic term.

The weak coupling calculation has been extended to next-to-leading order (NLO) by the author of Ref. \cite{Caron-Huot:2008zna}, where a large additive contribution to the above equation was found. Quantitative determinations of $\hat{q}$ based on lattice gauge theory have also been  formulated~\cite{Majumder:2012sh,Laine:2013apa,Panero:2013pla,Benzke:2012sz}. Recently, $\hat{q}$ has been computed for (2+1)-flavor QCD plasma on 4D lattices \cite{Kumar:2020wvb}. These results are similar to and somewhat lower than those extracted from phenomenology. 
The calculations of transport coefficient $\hat{q}$ based on $\mathcal{N}$=4 Super-Yang-Mills theory have also been carried out, however, these yield an order of magnitude higher results compared to phenomenology-based extractions~\cite{Liu:2006ug,Lin:2006au,Avramis:2006ip,Armesto:2006zv}.

A first systematic extraction of $\hat{q}$ based on phenomenology was carried out by the JET Collaboration~\cite{JET:2013cls}.
Extractions were based on a comparison of jet quenching model calculations to the experimental measurement of only the hadron $R_\mathrm{AA}$, in only the most-central collisions at RHIC and LHC energies. These were performed independently, for five different parton energy loss approaches: GLV-CUJET \cite{Buzzatti:2011vt, Gyulassy:2000er}, HT-M \cite{Majumder:2011uk}, HT-BW \cite{Chen:2011vt}, \textsc{martini} \cite{Schenke:2009gb}, and McGill-AMY \cite{Qin:2007rn}. These jet energy loss calculations were run on identical (2+1)D viscous hydrodynamical media~\cite{Song:2007fn,Song:2007ux,Qiu:2011hf,Qiu:2012uy}.
The striking result of this work was that the interaction strength $\hat{q}/T^{3}$ for the QGP at RHIC energy appeared to be higher compared to that at LHC energy. 
In other words, one has to re-adjust the fit parameter in $\hat{q}$ when comparing with data from LHC collision energies versus RHIC energies.

Even within the work of the JET Collaboration, it was clear that different energy loss models had made slightly different assumptions regarding the temperature and parton energy dependence of $\hat{q}$. While some used a variant of Eq.~\eqref{eq:HTL-qhat-formula}, other models assumed a scaling with some density profiles of the medium, e.g., entropy density. 
As a result of the JET effort, it becomes necessary to generalize the functional dependence of $\hat{q}$ on $T$ and $E$ and use a more data-driven approach to isolate this dependence. This was carried out recently by our collaboration in Ref.~\cite{JETSCAPE:2021ehl}. In that effort, the formula for $\hat{q}$ was generalized 
to allow for an \emph{additive} dependence on logarithms of energy along with the generic HTL form [Eq.~\eqref{eq:HTL-qhat-formula}]. Comparisons were carried out with the hadron $R_{\mathrm{AA}}$ at two centralities over three different collision energies ($\sqrt{s_{\rm NN}} = 0.2$~TeV, 2.76~TeV and 5~TeV). The outcome of the effort in Ref.~\cite{JETSCAPE:2021ehl} was that an additive dependence of $\hat{q}$ on logarithms of energy and temperature did allow for a simultaneous description of the hadron $R_{\mathrm{AA}}$ at RHIC and LHC, without the need for refitting. However, there was no noticeable improvement in the fit from a multistage versus a single-stage model.

Up to this point, almost all attempts to extract the transport
coefficient $\hat{q}$ have at most assumed dependence on $E$ and $T$, which are the only possibilities for an on-shell hard parton
propagating through the plasma. This may not be the case
for a highly virtual hard parton. Recently, several authors
have argued that medium-induced radiation should depend on
the resolution scale of the medium~\cite{Caucal:2018dla,Mehtar-Tani:2010ebp,Mehtar-Tani:2011hma,Casalderrey-Solana:2011ule}. 
Early in the history of a jet propagating in a medium, partons are very virtual, and splits engender large transverse-momentum scales. 
As a result, the transverse sizes of the QCD antennae are
very small, compared to the resolution scale in the medium, $Q^2_{\mathrm{med}} \approx \hat{q} \tau$ ($\tau$ is the formation length of the parton). The inability of partons emanating from the medium to resolve such small antennae are often referred to as ``coherence effects" in jet propagation.

While the authors of Refs.~\cite{Caucal:2018dla,Mehtar-Tani:2010ebp,Mehtar-Tani:2011hma,Casalderrey-Solana:2011ule} advocated the use of a sudden approximation---neglecting any medium-induced emission in the high-virtuality phase---the authors of Ref.~\cite{Kumar:2019uvu} derived a more gradual reduction of medium-induced emission as a function of the virtuality of the parton. 
In Ref.~\cite{Kumar:2019uvu}, the reduction in medium-induced emission is cast as a reduction in the effective value of $\hat{q}$ as a function of the parton virtuality $Q^2$. The latter formalism will be used in the current effort as it is a better approximation to the reduced energy loss at high virtuality. Also, with minor modifications, as outlined below, we will be able to study the onset of coherence effects at high virtuality. The reduction in the effective $\hat{q}$ will only take place in the \textsc{matter} phase of the simulation. The \textsc{lbt} phase will include the \emph{full} $\hat{q}$ with running coupling, described below. As part of our analysis, we will vary the transition scale between \textsc{matter} and \textsc{lbt} as well. 

For a high energy on-shell parton ($E\gg m_\mathrm{D}$), the hard-thermal-loop (HTL) calculation yields the following form of transport coefficient $\hat{q}$, given as~\cite{He:2015pra},
\begin{align}
\hat{q}_{\mathrm{HTL}} =C_{a}\frac{42 \zeta(3)}{\pi}  \alpha^{2}_\mathrm{s} T^{3} \ln\left[ \frac{2ET}{6\pi T^{2} \alpha_\mathrm{s}^\mathrm{fix}} \right]. \label{eq:HTL-qhat-formula-C-2}
\end{align}
In the above calculation, Bose and Fermi distributions for plasma constituents are invoked. Comparing with Eq.~\eqref{eq:HTL-qhat-formula}, we have set $C=\ln(2) $ and  $N_c=N_f = 3$ in Eq.~\eqref{eq:DebyMassEquation} for the Debye mass, and $\zeta(3)=1.202$ is Ap\'{e}ry's constant. In the above formula, only the $\alpha_{\mathrm{s}}$ within the logarithm is designated as $\alpha_{\mathrm{s}}^{\rm fix}$, while such a qualification is suppressed for the factors of $\alpha_{\mathrm{s}}$ outside the logarithm. We will present results with these factors of $\alpha_{\mathrm{s}}$ either  \emph{fixed} at the Debye scale or running with the scale of the exchanged $k_\perp$ in an improved perturbation theory calculation of Eq.~\eqref{eq:qhat-definition}.

Incorporating the reduction in the medium induced emission described in Ref.~\cite{Kumar:2019uvu}, we propose a virtuality ($Q^2$) dependent modulation factor $f(Q^2)$ as the $f$ in Eq. (\ref{matter_kernel}). 
One could incorporate the factor $f$ within $\hat{q}$, yielding a virtuality dependent $\hat{q} (T,E,Q^2) = \hat{q} (T,E) \cdot f(Q^2) $.  
This $f$ factor effectively decreases $\hat{q}$ as virtuality increases, in the high virtuality stage, 
and reduces to the HTL result ($f=1$) 
in the low virtuality stage of the parton shower. 
In this effort, we shall explore the following three formulations 
and carry out a sensitivity study of the in-medium coupling constant ($\alpha^\mathrm{fix}_\mathrm{s}$) and switching scale ($Q_\mathrm{sw}$) parameters in the simultaneous description of inclusive jet and charged-particle $R_{\mathrm{AA}}$:
\begin{enumerate}
\item  Type 1: HTL $\hat{q}$ with fixed coupling ($f=1$ applies for any $Q^2$),
\begin{align}
\hat{q} \cdot f &=\hat{q}^\mathrm{fix}_\mathrm{HTL} \!=\!C_{a}\frac{50.484}{\pi} \alpha^\mathrm{fix}_\mathrm{s} \alpha^\mathrm{fix}_\mathrm{s} T^{3} \ln\left[ \frac{2ET}{m^2_\mathrm{D}} \right], 
\label{eq:type1-q-hatform}
\end{align}
where $m^2_\mathrm{D}=6\pi T^{2} \alpha_\mathrm{s}^\mathrm{fix}$ is the Debye mass for $N_{f}=3$ flavors.
\item  Type 2: HTL $\hat{q}$ with running coupling ($f=1$ applies for any $Q^2$),
\begin{align}
\hat{q} \cdot f &=\hat{q}^\mathrm{run}_\mathrm{HTL} \!=\!C_{a}\frac{50.484}{\pi} \alpha^\mathrm{run}_\mathrm{s}(Q_{\rm max}^2) \alpha^\mathrm{fix}_\mathrm{s} T^{3} \ln\left[\! \frac{2ET}{m^2_\mathrm{D}}\! \right],\!
\label{eq:type2-q-hatform}
\end{align}
where  $Q_{\rm max}^2=2ET$
\item  Type 3: HTL $\hat{q}$ with a virtuality ($Q^2$) dependence factor 
\begin{align}
\hat{q} \cdot f &= \hat{q}^\mathrm{run}_\mathrm{HTL}f(Q^2) \label{eq:q2-dep-qhat} \\
f(Q^2) & = \left\{ \begin{array}{cc} \frac{1+10\ln^{2}(Q^2_\mathrm{sw}) + 100\ln^{4}(Q^2_\mathrm{sw})}{1+10\ln^{2}(Q^2) + 100\ln^{4}(Q^2)} & Q^2 > Q_{\rm sw}^2 \\ 1 & Q^2 \le Q_{\rm sw}^2 \end{array} \right.,
\label{eq:qhatSuppressionFactor}
\end{align}
\end{enumerate}
where $E$ is the energy of the hard parton, $T$ is the local temperature of the medium, and $Q^2$ is the running virtuality of the hard parton. The form for $f(Q^2)$ is a simplified form of the formula derived in Ref.~\cite{Kumar:2019uvu}. We point out here that the $\hat{q}$ is the same between type-2 and type-3. The extra factor of $f(Q^2)$ in type-3 diminishes the interaction between the jet and the medium due to coherence effects~\cite{Armesto:2011ir,Mehtar-Tani:2011hma,Caucal:2018dla,Kumar:2019uvu}. The $\hat{q}$ in this case is the same as in type-2.

\begin{figure}[htbp]
\centering
\includegraphics[width=0.45\textwidth]{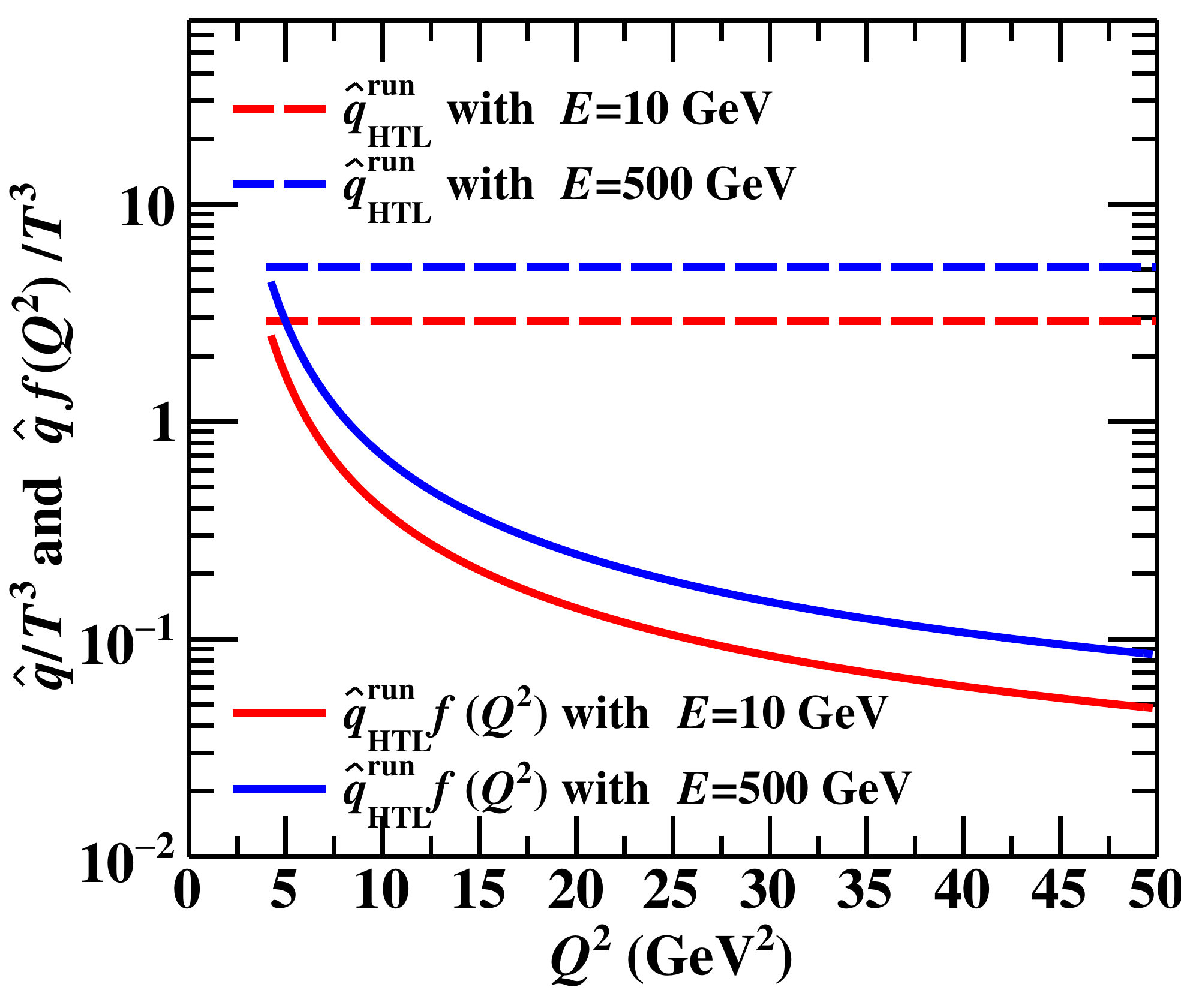}
\caption{Plot of $\hat{q}/T^3$ of Type 2 and $\hat{q}f(Q^2)/T^3$ (multiplied by the coherence factor) of Type 3 for a light quark as a function of virtuality scale $Q^2$. Here, $\hat{q}$ is evaluated by the traditional HTL based formulation at two different energy of the quark traversing a QGP medium at temperature $T=400$~MeV, $\alpha^\mathrm{fix}_\mathrm{s}=0.3$, and the switching virtuality is set to $Q_\mathrm{sw}=2$~GeV. }
\label{fig:Q2_dependent-qhatFunctionPlot}
\end{figure}
To visualize the new functional dependence ($Q^2$) proposed above, we plot $\hat{q}/T^3$ for Type 2 and $\hat{q}f(Q^2)/T^3$ for Type 3 in Fig.~\ref{fig:Q2_dependent-qhatFunctionPlot}. 
In the case of Type 3 (solid line), the virtuality dependence in the \textsc{matter} phase is shown, exhibiting a reduction as $Q^2$ increases.
In addition to this, $\hat{q}f(Q^2)$ of Type 3 reduces to traditional HTL $\hat{q}$ in the \textsc{lbt} phase where the parton's virtuality becomes $Q^2 \le Q^2_\mathrm{sw}$. In Sec.~\ref{Section:Results}, we shall show that the experimental data for the charged-hadron $R_\mathrm{AA}$ strongly favors the one with $Q^2$ dependence.

\subsection{Medium response in a weakly-coupled approach by recoils}
\label{Subsection:medium-response}
Jets exchange energy and momentum with the soft medium, during which they excite medium constituents. Some of these excited partons are clustered with the jet and modify the structure of reconstructed jets. In the current effort, the medium responses are incorporated using perturbation theory (nonperturbative methods are further described in Appendix~\ref{subsect:approximate-medium-response}), with nonperturbative effects incorporated solely during hadronization. 
In this medium response process, some portion of jet energy and momentum are transported through further successive interactions among the medium constituents far beyond the jet cone.
Their contribution appears as jet-correlated particles in the final state and affects the jet medium correlations (these will be described in detail in a future effort).

In this study, the medium response is described as the propagation of recoil partons and their successive interactions in the \textsc{jetscape} framework~\cite{Li:2010ts,Zapp:2012ak,Zapp:2013vla,Cao:2017hhk,Luo:2018pto,Park:2018acg,Tachibana:2020mtb}. 
In both \textsc{matter} and \textsc{lbt}, the energy-momentum transfer between jets and the medium is executed via 2-to-2 partonic scatterings. 
For each scattering, a medium parton is sampled from a thermal bath of  three-flavor ideal QGP gas. The sampled thermal parton after being scattered by the jet parton, referred to as a recoil parton, is assumed to be on shell. Since it is not virtual, its subsequent in-medium evolution will be carried out by \textsc{lbt}, assuming its weak coupling with the QGP. Note, in \textsc{matter} energy loss phase, the elastic scattering probability is also reduced by $f(Q^2)$ to be consistent with the modified kernel for transverse momentum broadening.

The jet shower partons, including these recoil partons, are hadronized together via the Colorless string hadronization routine. On the other hand, the recoil parton leaves an energy-momentum deficit (\emph{hole}) in the medium. We also keep track of the \emph{hole} partons and subtract their contribution to ensure energy-momentum conservation. 
The hole partons are assumed to freestream in the medium and are hadronized separately from other regular jet-shower partons. 
The subtraction of the \emph{hole} contribution for 
the final reconstructed jet momentum is performed as 
\begin{align}
\label{eq:neg_sub}
p^\mu_{\mathrm{jet}}
&=p^\mu_{\mathrm{shower}}
-\sum_{\substack{i\in \mathrm{holes}\\ \Delta r_i < R}} p^\mu_i.
\end{align}
Here $p^\mu_{\mathrm{shower}}$ is the four-momentum of the jet 
reconstructed from all particles from the hadronization of 
jet showering partons including the recoil contribution by the anti-$k_\mathrm{T}$ algorithm~\cite{Cacciari:2008gp} 
implemented in the \textsc{fastjet} package~\cite{Cacciari:2005hq, Cacciari:2011ma} 
with a jet cone size $R$. 
In the second term on the right hand side, 
the sum of four momenta $p^\mu_i$ is taken over 
particles hadronized from hole partons inside the jet cone $\Delta r_i < R$, 
where $\Delta r_i = [(\eta_i - \eta_{\mathrm{shower}})^2 + (\phi_i - \phi_{\mathrm{shower}})^2]^{1/2}$ is the radial distance from the jet center.

This recoil prescription gives a reasonable description of the medium response as long as jet shower partons have sufficiently large energy and are far from the thermalization, where their mean free paths are long enough to apply the kinetic theory. 
This recoil approximation breaks down when the showering partons' energy approaches the typical scale for the thermalized medium constituents~\cite{Tachibana:2019hrn,Cao:2020wlm,Luo:2021iay}. 
To extend this region of applicability, one needs to incorporate the hydrodynamic description for the soft modes of jets as presented in Refs.~\cite{Tachibana:2017syd,Okai:2017ofp,Chen:2017zte,Chang:2019sae,Tachibana:2020mtb,JETSCAPE:2020uew,Chen:2020tbl,Casalderrey-Solana:2020rsj,Zhao:2021vmu,Yang:2021qtl,Kanakubo:2021qcw,Yang:2022nei,Pablos:2022piv}. 
In this study, 
we do not include this hydrodynamic description for the medium response to jets, which exacts a huge computational cost for the systematic studies of jets presented here. Although they are essential for a more precise description of jet-correlated particle distribution, the recoil prescription is still a good approximation for the estimation of jet transverse momentum with typical jet cone sizes $R \approx 0.4$~\cite{Tachibana:2020mtb}. 
We leave systematic jet-quenching studies including more detailed modeling of medium response for future work (however, see Appendix~\ref{subsect:approximate-medium-response} for a discussion of alternative approaches). 

\section{Simulation with multistage energy loss approach in JETSCAPE}
\label{Section:sim}
The \textsc{jetscape} framework is a general-purpose numerical framework to simulate the complete evolution of heavy-ion collisions. Currently, it provides several alternative implementations of physics models to simulate both the QGP evolution and the jet-medium interactions at different epochs of the parton shower. In this paper, we carry out the calculation for both $p+p$ and A+A collisions using the publicly available \textsc{jetscape} 3.0~\cite{jetscape}. In this section, we discuss the choice of modules and parameters that will be explored in the current effort.

The $p+p$ baseline is simulated using the \textsc{jetscape} PP19 tune~\cite{JETSCAPE:2019udz}\footnote{The \textsc{jetscape} PP19 tune is publicly available. The parameters of the \textsc{jetscape} PP19 tune are in {\texttt{jetscape\_user\_PP19.xml}} \cite{jetscape}. A systematic study with various comparisons to experimental data is presented in Ref.~\cite{JETSCAPE:2019udz}.}. 
This tune generates the hard scattering in a $p+p$ collision using the \textsc{pythia} module where initial state radiation (ISR) and multiparton interaction (MPI) switches are enabled, but the final state radiation (FSR) is turned off. Then, the produced partons from \textsc{pythia} hard scattering are transferred to the \textsc{matter} energy loss module for final-state radiation. 
As \textsc{matter} embeds the space-time structure of the bulk medium in the parton shower and is capable of performing both the vacuum ($\hat{q}=0$) and in-medium ($\hat{q}\ne 0$) energy loss, it is the desired choice for final state radiation module to ensure the consistency between $p+p$ and A+A collisions.

\begin{table*}[htpb]
\caption{Default parameter set for simulations of the jet evolution (\textsc{matter}+\textsc{lbt}) in A+A collisions.}
\begin{center}
\renewcommand{\arraystretch}{1.2}
\begin{tabular}{lcc} 
\hline \hline 
Parameter& 
Description & 
Default value\\
\hline
$\hat{q}\cdot f$ 
&
Functional form of transport coefficient $\hat{q}$ 
multiplied by a modulation factor $f$ 
&
Type 3, $\hat{q}^\mathrm{run}_\mathrm{HTL} f(Q^2)$ in Eq.~(\ref{eq:q2-dep-qhat}) \\

$\alpha_\mathrm{s}^\mathrm{fix}$& 
Coupling constant for the jet-medium interaction at the Debye mass scale $m_\mathrm{D}$& 
0.3 (Type 3), 0.25 (Type 1 and 2)\\
$Q_\mathrm{sw}$& 
Virtuality to switch parton energy loss from \textsc{matter} phase to \textsc{lbt} phase& 
2~GeV\\
$\tau_0$&
Starting longitudinal proper time for in-medium jet energy loss & 0.6~fm/$c$
\\
$T_\mathrm{c}$& 
Temperature below which the jet-medium interaction is turned off&
160~MeV\\
\hline \hline
\end{tabular}
\end{center}
\label{tab:aa_parameter_set}
\end{table*}

The soft products in A+A collisions are generated using fluctuating \textsc{trento} initial conditions \cite{Moreland:2014oya}, evolved hydrodynamically using the (2+1)D \textsc{vishnu} code package \cite{Shen:2014vra} (the underlying physics setup is discussed in Sec.~\ref{Subsection:SoftSectorAA}). Events at LHC were simulated using the maximum {\it a posteriori} (MAP) parameters obtained in Ref.~\cite{Bernhard:2019bmu} using Bayesian calibration, while hand-tuned parameters were used for top RHIC energy. To obtain events within different centrality classes, we made those centrality tables for each beam energy first and then simulated these events by inputting entropy ranges corresponding to these centrality bins for \textsc{trento}. To make the centrality tables, we first ran $10^5$ minimum-bias \textsc{trento}, events for each energy using the MAP parameters and then sorted them by entropy for binning.

Once the bulk simulations are complete, the space-time
distribution of the energy-momentum tensor is saved. This
allows simulations of in-medium jet evolution for several hard
scattering events on a single bulk event.\footnote{The \textsc{jetscape} framework allows users to either save and reread bulk profiles or generate them ``on the fly", we prefer the former as it allows for repeat calculations of the hard sector from the same bulk medium.}
In the next step, the initial hard scatterings are generated using \textsc{pythia}. The \textsc{trento} initial-state module generates the initial energy density profile for the following free-streaming evolution. It also provides the binary collision profile to sample the transverse
positions for the hard scatterings. To incorporate multiscale
dynamics of the parton energy loss, the jet evolution is carried
out as follows. 
First, all the partons from \textsc{pythia} hard scatterings are transferred to the \textsc{matter} module, which assigns an initial virtuality limited by the maximum $Q^2_\mathrm{init} \equiv p_{\mathrm{T}}^2/4$, based on the parton's initial transverse momentum 
and $t_{\mathrm{max}}$ in Eq.~(\ref{eq:SudakovFormFactor}) for the first Sudakov sampling is set to $Q^2_\mathrm{init}$~\cite{JETSCAPE:2019udz}. 
In both $p+p$ and A+A collisions, the partons generated during initial state \textsc{pythia} hard scattering with $p_\mathrm{T} < 2$~GeV/c are discarded~\cite{JETSCAPE:2019udz}. This treatment is because the \textsc{matter} module can not create a DGLAP based parton shower if the maximum initial virtuality of the parton $Q^2_\mathrm{init} < 1$~GeV$^2$. 
Second, the \textsc{matter} module generates virtuality-ordered showers for individual partons from \textsc{pythia}. It includes both vacuumlike and medium-induced radiation.
As parton's virtuality drops below a switching scale $Q_\mathrm{sw}$, it is switched to propagate with the \textsc{lbt} module. Once the partons fly outside the medium where the local temperature is below the energy loss termination temperature $T_{\mathrm{c}}$, they are transferred back to \textsc{matter} for vacuumlike radiation if the partons have virtuality larger than the minimum virtuality $Q^2_{0}=1$~GeV$^2$. After all the partons are outside the QGP medium and have virtuality $Q^2 \le Q^2_{0}$, they are hadronized by the Colorless string fragmentation (see Sec.~\ref{Subsection:HadronizationModule} for details).

The parameter set for this multistage jet evolution model in A+A collisions is summarized in Tab.~\ref{tab:aa_parameter_set}. 
The exact functional form of the jet transport coefficient $\hat{q}$ of the first parameter is unknown from theory. 
Given prior efforts in Refs.~\cite{JET:2013cls, JETSCAPE:2021ehl}, it is clear that $\hat{q}$ depends on more than the ambient temperature $T$. The choice of an additive dependence on $T$ and parton energy $E$ in Ref.~\cite{JETSCAPE:2021ehl} showed only modest improvement on describing the charged hadron $R_{\mathrm{AA}}$ measurements. 
Based on the work of Ref.~\cite{Kumar:2019uvu}, we have invoked a multiplicative dependence on the virtuality of a parton $Q$ via the modulation factor $f(Q^2)$. 
In Appendix~\ref{type1_param}, we will show conclusively that the multiplicative dependence 
on the parton virtuality $Q$ is essential for a simultaneous description of the modification of the inclusive jet and charged particle spectra.

The next parameter is the in-medium coupling $\alpha_{\mathrm{s}}^\mathrm{fix}$ at the Debye scale. It also controls the overall normalization of $\hat{q}$ and the strength of recoil scattering of jet partons in the medium. The value of this nonperturbative parameter is unknown and needs to be calibrated by the $R_{\mathrm{AA}}$ data.
The switching scale $Q_\mathrm{sw}$ controls the relative phase spaces between the \textsc{matter} and \textsc{lbt} parton shower. The value of $Q_\mathrm{sw}$ should be close to the medium scale $Q_\mathrm{sw} \approx \hat{q} \tau$ with $\tau$ being the parton formation time. 
In this work, we choose a constant switching scale $Q_\mathrm{sw}$ between the \textsc{matter} and the \textsc{lbt} modules. 
In principle, $Q_\mathrm{sw}$ is a dynamical quantity depending on 
local $\hat{q}$, energy and virtuality of the parton. 
Thus, the constant value of $Q_\mathrm{sw}$ introduced in our current work 
should be interpreted as the typical transition scale averaged over those quantities of all partons. 
We leave studying the effects of a dynamical switching scale $Q_\mathrm{sw} = C \hat{q} \tau$ \cite{JETSCAPE:2017eso} on observables for future work. 
The last two parameters $\tau_0$ and $T_{\mathrm{c}}$ specify the start time and stop temperature conditions for jet-medium interactions. We choose $\tau_0$ to be smaller than the starting time of hydrodynamics $\tau_\mathrm{hydro}$ to take into account parton energy loss in the preequilibrium stage. The $R_{\mathrm{AA}}$'s sensitivity to the parameter $\tau_0$ will be investigate in Fig.~\ref{fig:Type3-q-hat-Effect_of_tau0}.
We stop the jet-medium interactions below the energy loss termination temperature $T_{\mathrm{c}}$. This approximation neglects energy loss in the dilute hadronic phase. The uncertainty from this approximation will be quantified by varying the termination temperature $T_{\mathrm{c}}$ in Fig.~\ref{fig:Type3-q-hat-Effect_of_Tc}.

\section{Results}
\label{Section:Results}
In this work, we cover three collision energies: $\sqrt{s_\mathrm{NN}}=5.02$~TeV, 2.76~TeV, and 200~GeV, and show comparisons with selected experimental data available from ALICE, ATLAS, CMS, PHENIX, and STAR Collaborations. 
Table~\ref{tab:exp_data_list} shows the list of references for the data used in the comparisons. 
We will first tune the model parameters in Table~\ref{tab:aa_parameter_set} to achieve a simultaneous description of the inclusive jet and charged-particle $R_{\mathrm{AA}}$ in 0-10\% Pb+Pb collisions at $\sqrt{s_\mathrm{NN}}=5.02$~TeV. With the optimized model parameters, we will present \textsc{jetscape} postdictions for these observables in semi-peripheral centralities of Pb+Pb collisions at $\sqrt{s_\mathrm{NN}}=5.02$~TeV and in central Pb+Pb and Au+Au collisions at 2.76~TeV and 200~GeV, respectively. We will also present predictions for inclusive jets at 200~GeV. 

Once the $R_{\mathrm{AA}}$ for the most central collisions at 2.76~TeV has been obtained (for inclusive jets or leading hadrons) and compares favorably with experimental data, the variation with centrality is easily obtained. This is discussed in Refs.~\cite{JETSCAPE:2021ehl,Park:2019sdn}, and will not be repeated here. Also semicentral $R_{\mathrm{AA}}$ for jets at $\sqrt{s_\mathrm{NN}}=200$~GeV are not currently available and as a result we do not present centrality dependence of the $R_{\mathrm{AA}}$ here. Nuclear modification data for neutral pions in semicentral events is available. However, once the $R_{\mathrm{AA}}$ for most central collisions compares favorably with the experimental data, the centrality dependence is easily obtained, almost independent of the choice of multistage model used. Such comparisons were already presented in Ref.~\cite{JETSCAPE:2021ehl}.

\begin{table*}[htpb]
\caption{Reference list for the experimental data compared with the simulation results.}
\begin{center}
\renewcommand{\arraystretch}{1.2}
\begin{tabular}{lcccc} 
\hline \hline 
Observable & 
Projectiles & 
Collision energy & 
References\\
\hline
Inclusive jet spectrum&$p+p$&
$5.02$~TeV&
ATLAS, PLB 790, 108 (2019)~\cite{ATLAS:2018gwx}\\
&&&
ALICE, PRC 101 034911 (2020)~\cite{ALICE:2019qyj}\\
&&$2.76$~TeV&
CMS, PRC 96, 015202 (2017)~\cite{CMS:2016uxf}\\
&&$200$~GeV&
STAR, PoS DIS2015, 203 (2015)~\cite{Li:2015gna}\\
Inclusive charged particle spectrum&$p+p$&
$5.02$~TeV&
CMS, JHEP 1704, 039 (2017)~\cite{CMS:2016xef}\\
&&
$2.76$~TeV&
CMS, EPJ C72, 1945 (2012)~\cite{CMS:2012aa}\\
Charged pion spectrum&$p+p$&
$200$~GeV&
PHENIX, PRD 76, 051106 (2007)~\cite{PHENIX:2007kqm}\\
Inclusive jet $R_{\mathrm{AA}}$
&Pb+Pb&$5.02$~TeV&
ATLAS, PLB 790, 108 (2019)~\cite{ATLAS:2018gwx}\\
&&&
CMS, JHEP 05, 284 (2021)~\cite{CMS:2021vui}\\
&&&
ALICE, PRC 101, 034911 (2020)~\cite{ALICE:2019qyj}\\
&&$2.76$~TeV&
CMS, PRC 96, 015202 (2017)~\cite{CMS:2016uxf}
\\
Inclusive charged particle $R_{\mathrm{AA}}$
&Pb+Pb&$5.02$~TeV&
CMS, JHEP 1704, 039 (2017)~\cite{CMS:2016xef}
\\
&&$2.76$~TeV&
CMS, EPJ C72, 1945 (2012)~\cite{CMS:2012aa}
\\
Charged jet $R_{\mathrm{AA}}$&
Au+Au&
$200$~GeV&
STAR, PRC 102, 054913 (2020)~\cite{STAR:2020xiv}
\\
Pion $R_{\mathrm{AA}}$&
Au+Au&
$200$~GeV&
PHENIX, PRC 87, 034911 (2013)~\cite{PHENIX:2012jha}
\\

\hline \hline
\end{tabular}
\end{center}
\label{tab:exp_data_list}
\end{table*}

\subsection{The $p+p$ baseline}
To quantify the medium effects in inclusive jet and charged-particle spectra as the nuclear modification factor $R_{\mathrm{AA}}$, calculations for $p+p$ collisions are necessary to obtain the baseline for A+A collisions. A systematic study of inclusive jet, intra jet, and charged-particle observables has been extensively carried out using the \textsc{jetscape} PP19 tune and presented in Ref.~\cite{JETSCAPE:2019udz}.
Thus, we shall only present a selection of plots for inclusive jets and charged-particle yield for $p+p$ collision energies $\sqrt{s}=$5.02~TeV, 2.76~TeV, and 200~GeV.
The presented results are further restricted to those based on the \textsc{jetscape} Colorless hadronization, as it is the hadronization module employed in the A+A sector. The uncertainty in the final observable spectra from the two different prescriptions, colored and colorless hadronization, are roughly of the order of $10\%$, and we refer readers to Ref.~\cite{JETSCAPE:2019udz} for more details.

\begin{figure}[htbp]
\centering
\includegraphics[width=0.5\textwidth]{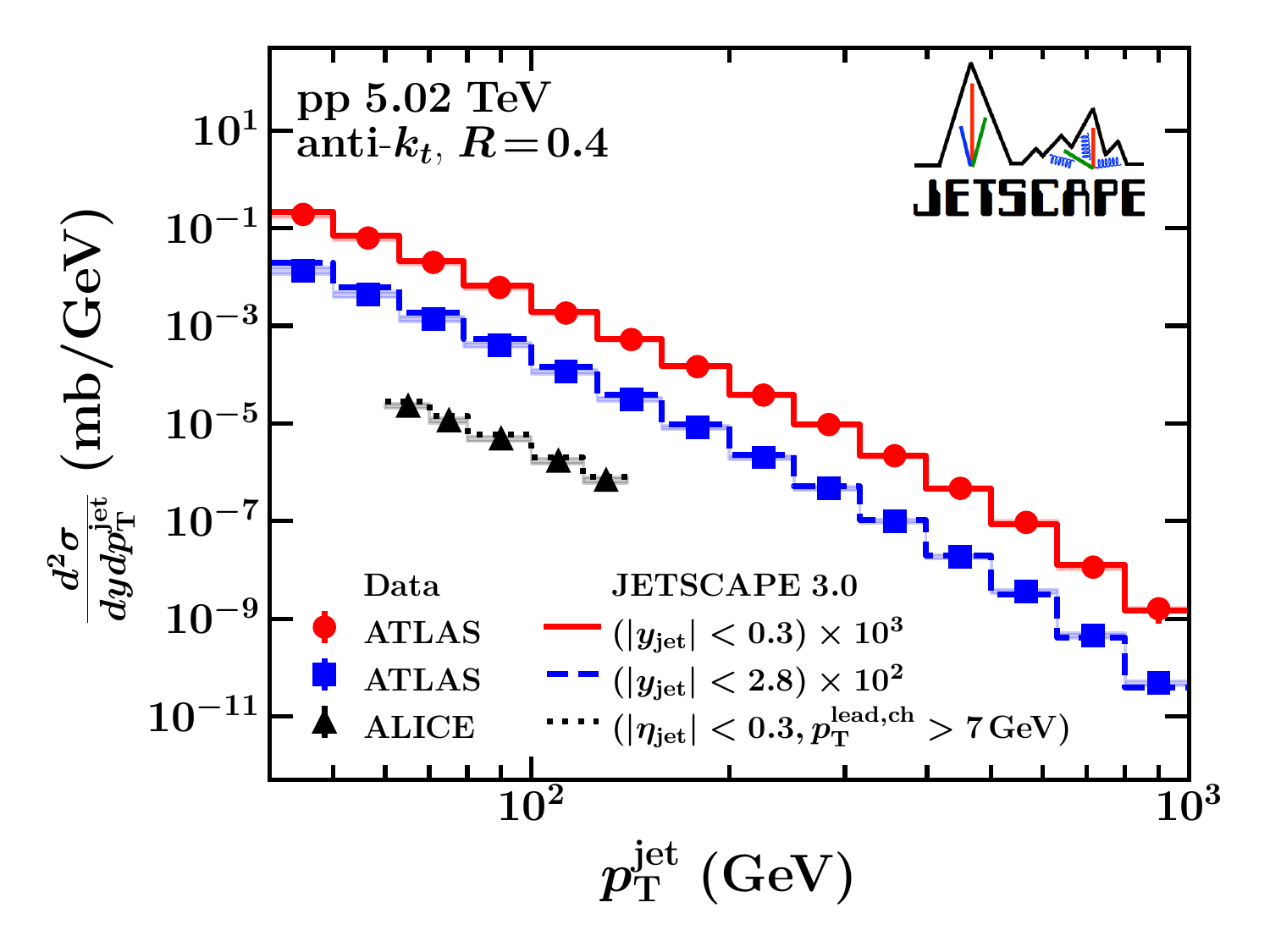}
\caption{Differential cross section of inclusive jets with cone size $R=0.4$ at midrapidity in $p+p$ collisions, at $\sqrt{s}=5.02$~TeV, calculated with \textsc{jetscape}.  
The result for jet rapidity $|y_{\mathrm{jet}}| < 0.3$ (solid red line; scaled up by $10^3$) is compared to ATLAS data~\cite{ATLAS:2018gwx} (red circles). 
The result for $|y_{\mathrm{jet}}| < 2.8$ (dashed blue line; scaled up by $10^2$) is compared to ATLAS data~\cite{ATLAS:2018gwx} (blue squares). 
The result for jet pseudorapidity $|\eta_{\mathrm{jet}}| < 0.3$ with a leading track requirement $p^{\mathrm{lead,\,ch}}_{\mathrm{T}} > 7\,\mathrm{GeV}$ (dotted black line) is compared to ALICE data~\cite{ALICE:2019qyj} (black triangles). 
The shaded boxes indicate the systematic uncertainties of the experimental data. }
\label{fig:spectrum_djcs_R0p4_5020GeV}
\end{figure}
\begin{figure}[htbp]
\centering
\includegraphics[width=0.47\textwidth]{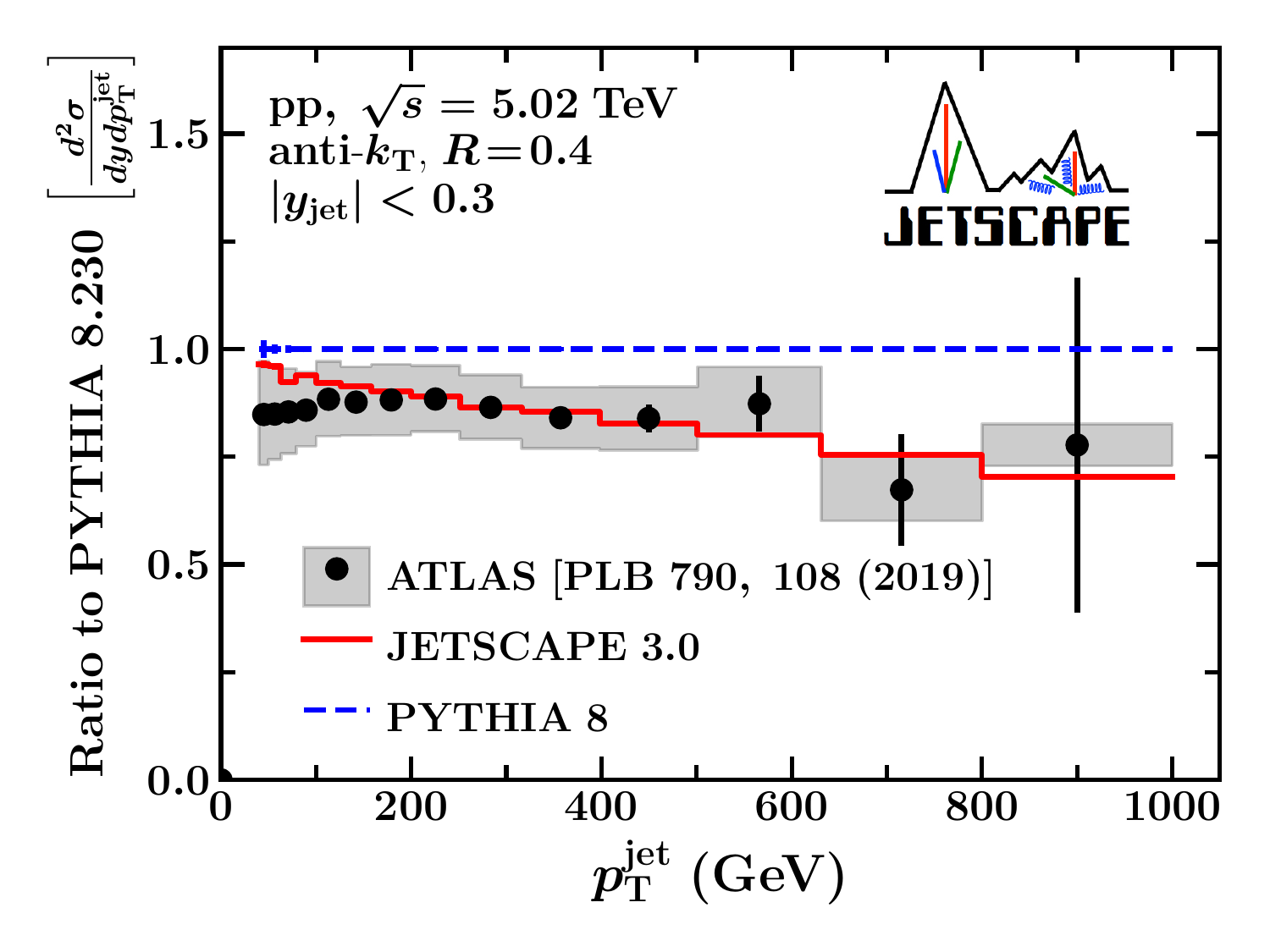}
\includegraphics[width=0.47\textwidth]{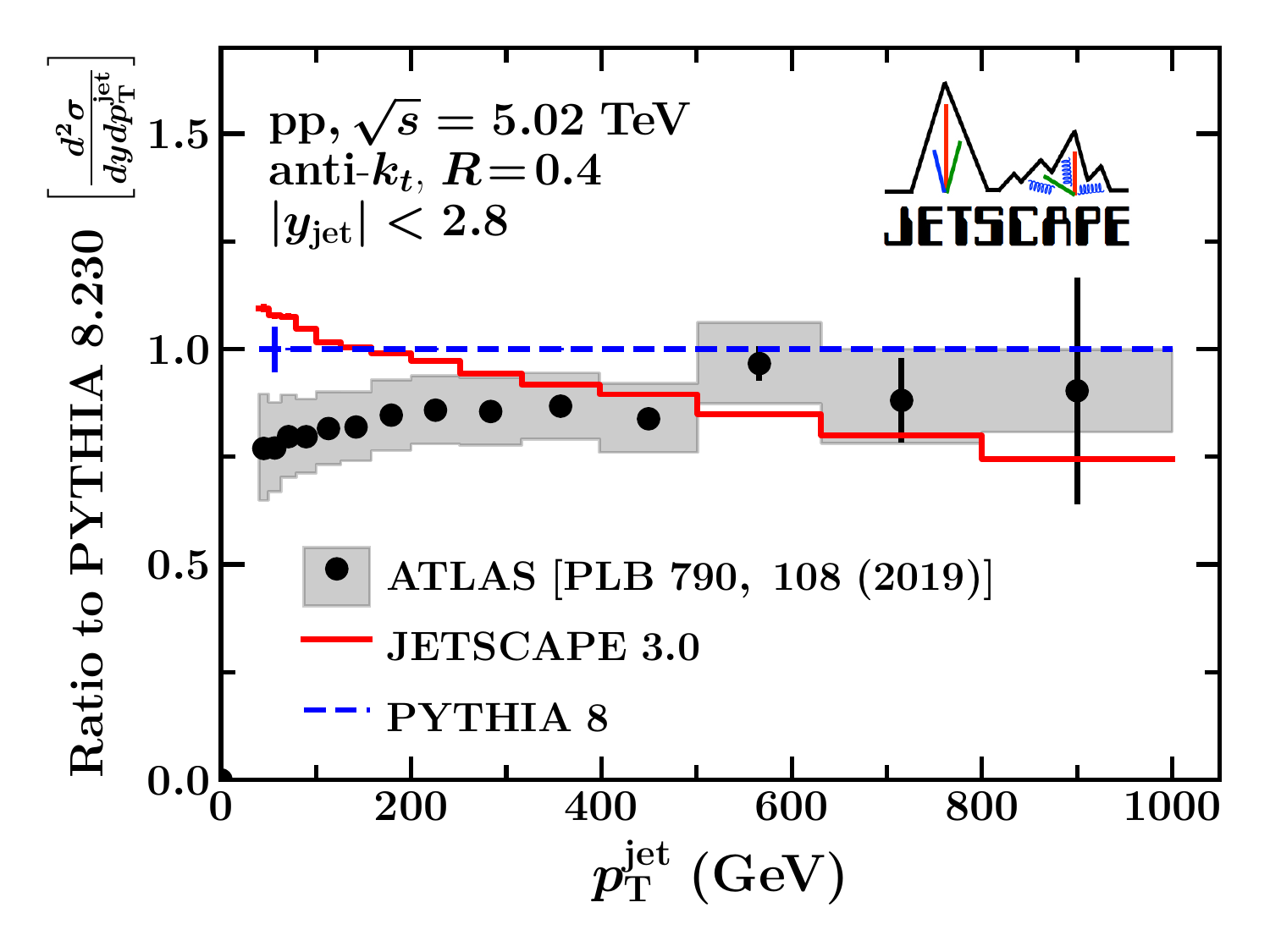}
\includegraphics[width=0.47\textwidth]{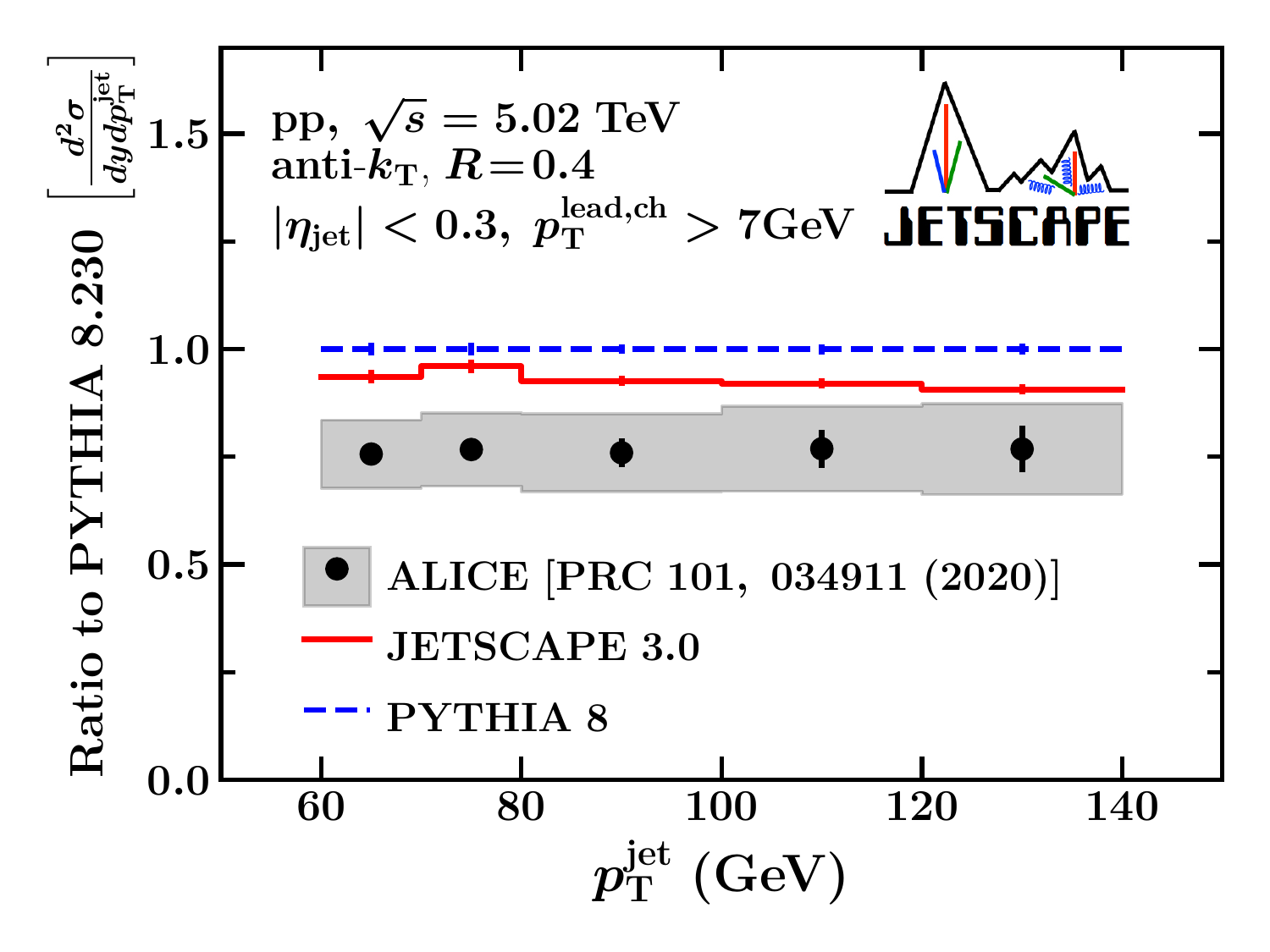}
\caption{Ratio of differential cross section for inclusive jets with cone size $R=0.4$ at midrapidity in $p+p$ collisions at $\sqrt{s}=5.02$~TeV. The ratio is taken w.r.t. the default \textsc{pythia}~8 MC. 
The solid red lines and dashed blue lines show the results from \textsc{jetscape} and \textsc{pythia}~8, respectively. 
Statistical errors (black error bars) and systematic uncertainties (grey bands) are plotted with the experimental data. 
Top panel: 
Results for $y_{\mathrm{jet}} < 0.3$, compared to ATLAS data~\cite{ATLAS:2018gwx}. 
Middle panel: 
Results for $y_{\mathrm{jet}} < 2.8$, compared to ATLAS data~\cite{ATLAS:2018gwx}. 
Bottom panel: 
Results for $\eta_{\mathrm{jet}} < 0.3$ with $p^{\mathrm{lead,\,ch}}_{\mathrm{T}} > 7\,\mathrm{GeV}$, compared to ALICE data~\cite{ALICE:2019qyj}. 
}
\label{fig:ratio_plot_DJCS_5020GeV_ATLAS_ALICE}
\end{figure}

Figure~\ref{fig:spectrum_djcs_R0p4_5020GeV} shows our $p+p$ simulation results for inclusive jet spectra around midrapidity at $\sqrt{s}=5.02$~TeV, compared to experimental data from ATLAS~\cite{ATLAS:2018gwx} and ALICE~\cite{ALICE:2019qyj}. 
The jets are reconstructed with jet cone size $R=0.4$ using the anti-$k_{\mathrm{T}}$ algorithm in \textsc{fastjet}. 
The comparison with the results from \textsc{pythia} 8 with default parameters are shown in Fig.~\ref{fig:ratio_plot_DJCS_5020GeV_ATLAS_ALICE}.

Results from \textsc{jetscape} PP19 describe the experimental data very well for jets with large transverse momentum or small rapidity. 
Our results for $|y_{\mathrm{jet}}|<0.3$ are compatible within uncertainties with data from ATLAS throughout the entire $p_{\mathrm{T}}^{\mathrm{jet}}$ range (up to $1$~TeV). 
The results for the wider rapidity range $|y_{\mathrm{jet}}|<2.8$ agree with ATLAS data in the region with $p_{\mathrm{T}}^{\mathrm{jet}}\gtrsim 300$~GeV but deviate at low-$p_{\mathrm{T}}^{\mathrm{jet}}$. 
In comparison to the ALICE measurements for $|\eta_{\mathrm{jet}}|<0.3$, \textsc{jetscape} PP19 overestimates the data and tends to be similar to \textsc{pythia} 8. 
\begin{figure}[htbp]
\centering
\includegraphics[width=0.48\textwidth]{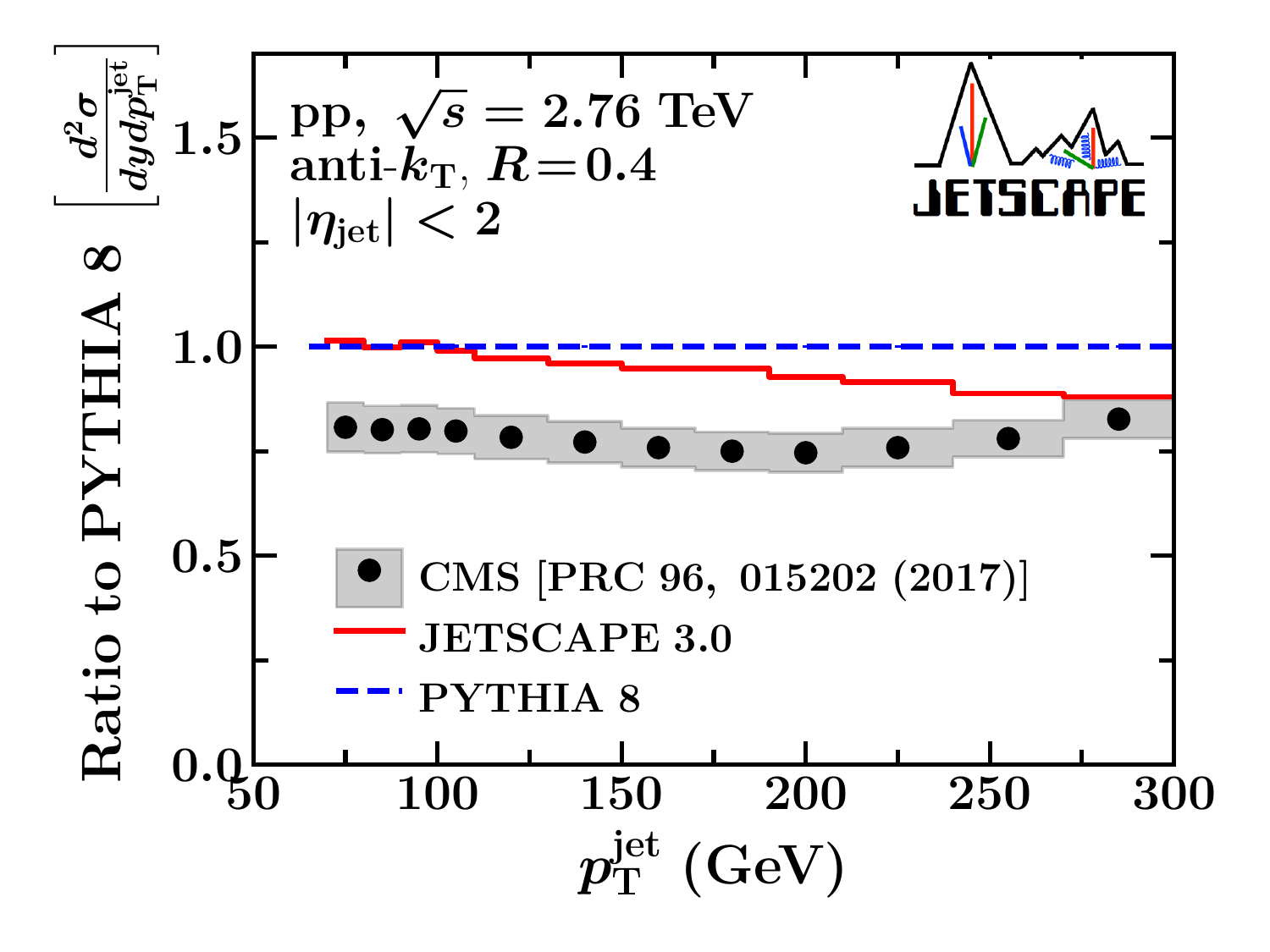}
\includegraphics[width=0.47\textwidth]{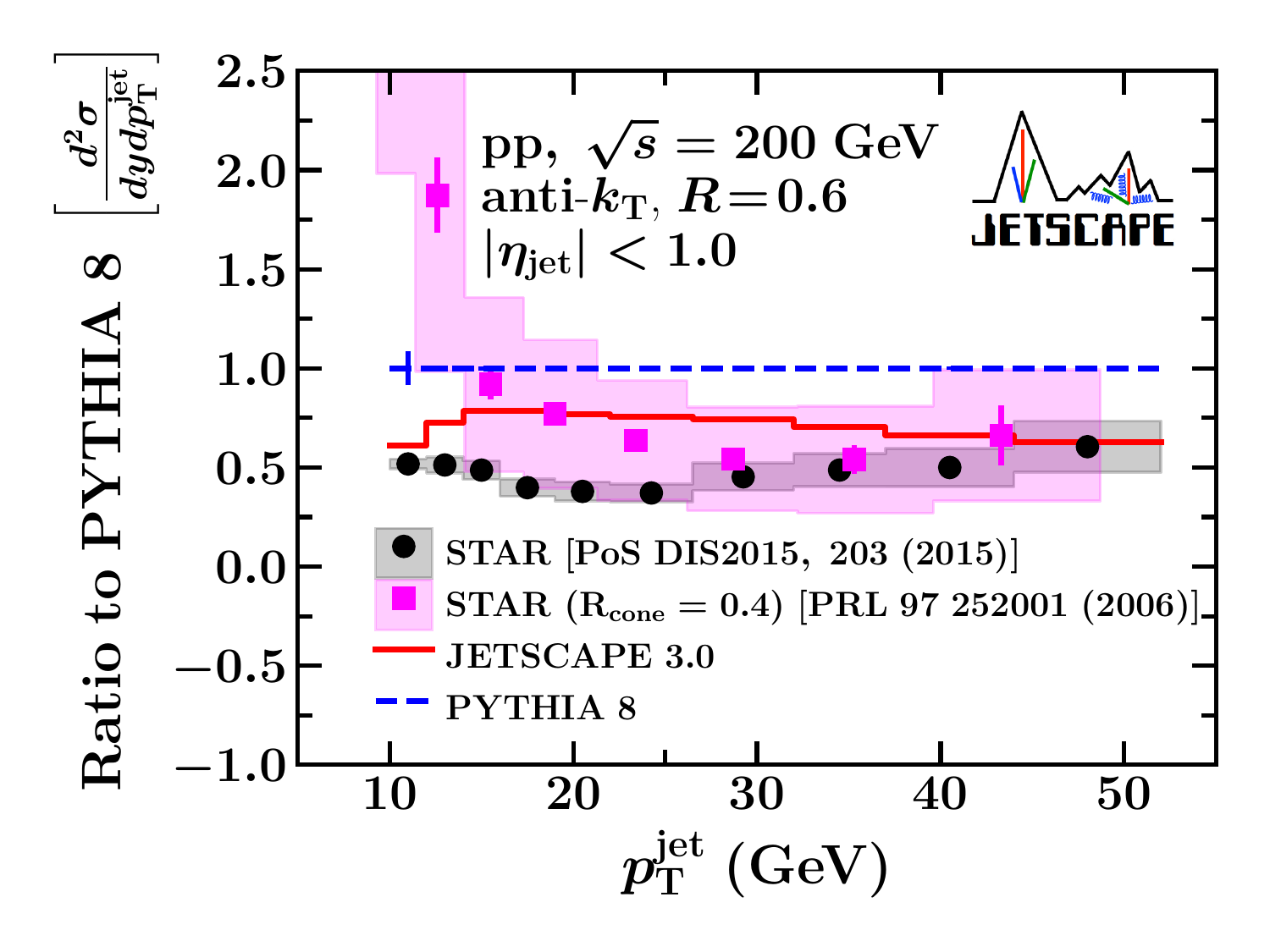}
\caption{
Same as Fig.~\ref{fig:ratio_plot_DJCS_5020GeV_ATLAS_ALICE} for different collision energies. 
Top panel: Jets with $|\eta_{\mathrm{jet}}|<2.0$ at $\sqrt{s}=2.76$~TeV, compared to CMS data~\cite{CMS:2016uxf}. 
Bottom panel: Jets with $R=0.6$ and $|\eta_{\mathrm{jet}}|<1.0$ at $\sqrt{s}=200$~GeV, compared with the STAR data~\cite{Li:2015gna}. Additionally, we show STAR measurements for inclusive jets based on Midpoint-cone algorithm with $R=0.4$~ \cite{STAR:2006opb}. }
\label{fig:ratio_plot_DJCS_2760GeV_CMS}
\end{figure}
We also show the ratios of differential cross section for inclusive jet at midrapidity in $p+p$ collisions at $\sqrt{s}=2.76$~TeV and $200$~GeV in Fig.~\ref{fig:ratio_plot_DJCS_2760GeV_CMS}. 
Both \textsc{jetscape} and \textsc{pythia}~8 tend to overestimate the measured jet cross-section at those collision energies.

\begin{figure}[htbp]
\centering
\includegraphics[width=0.45\textwidth]{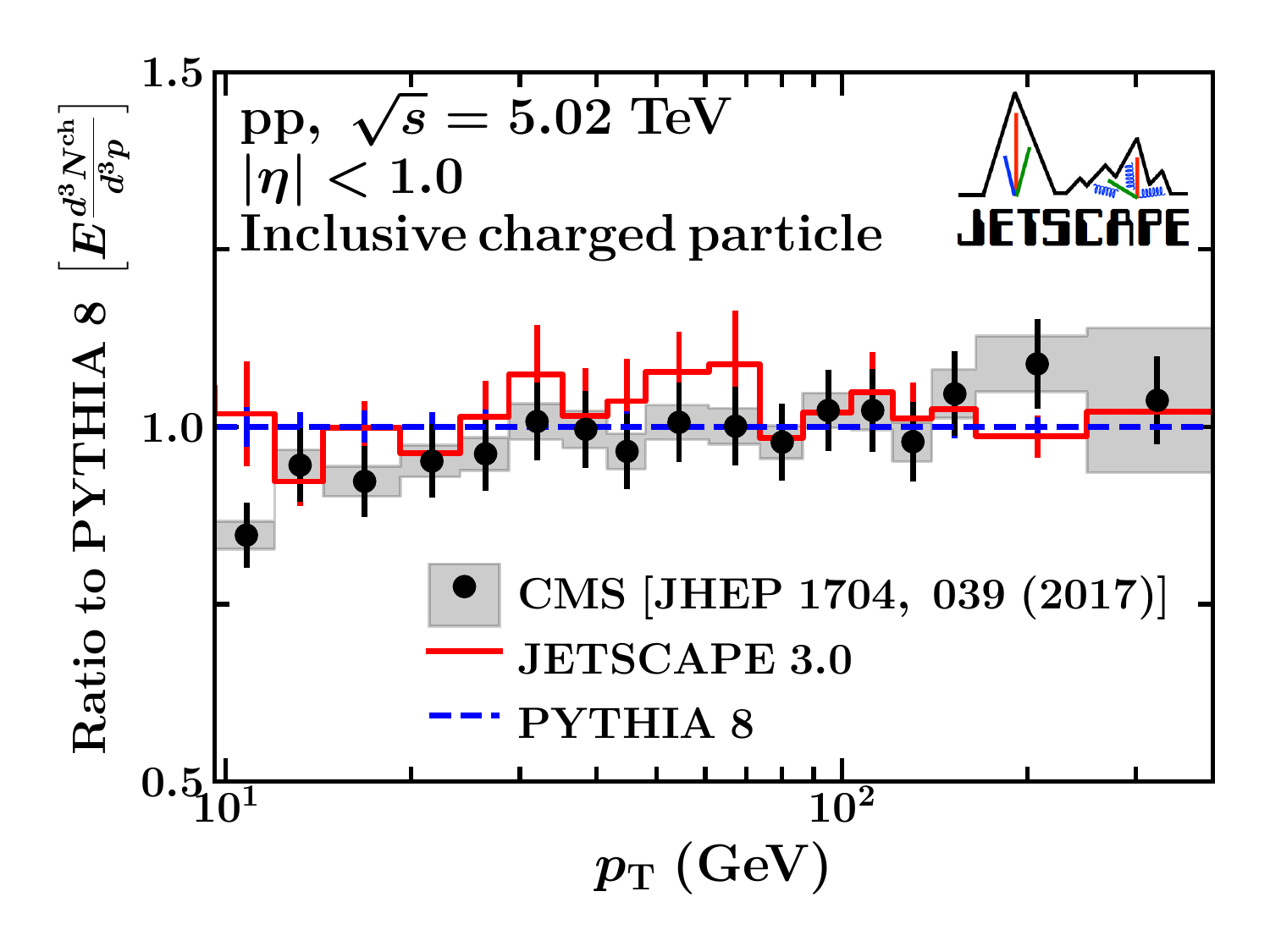}
\includegraphics[width=0.46\textwidth]{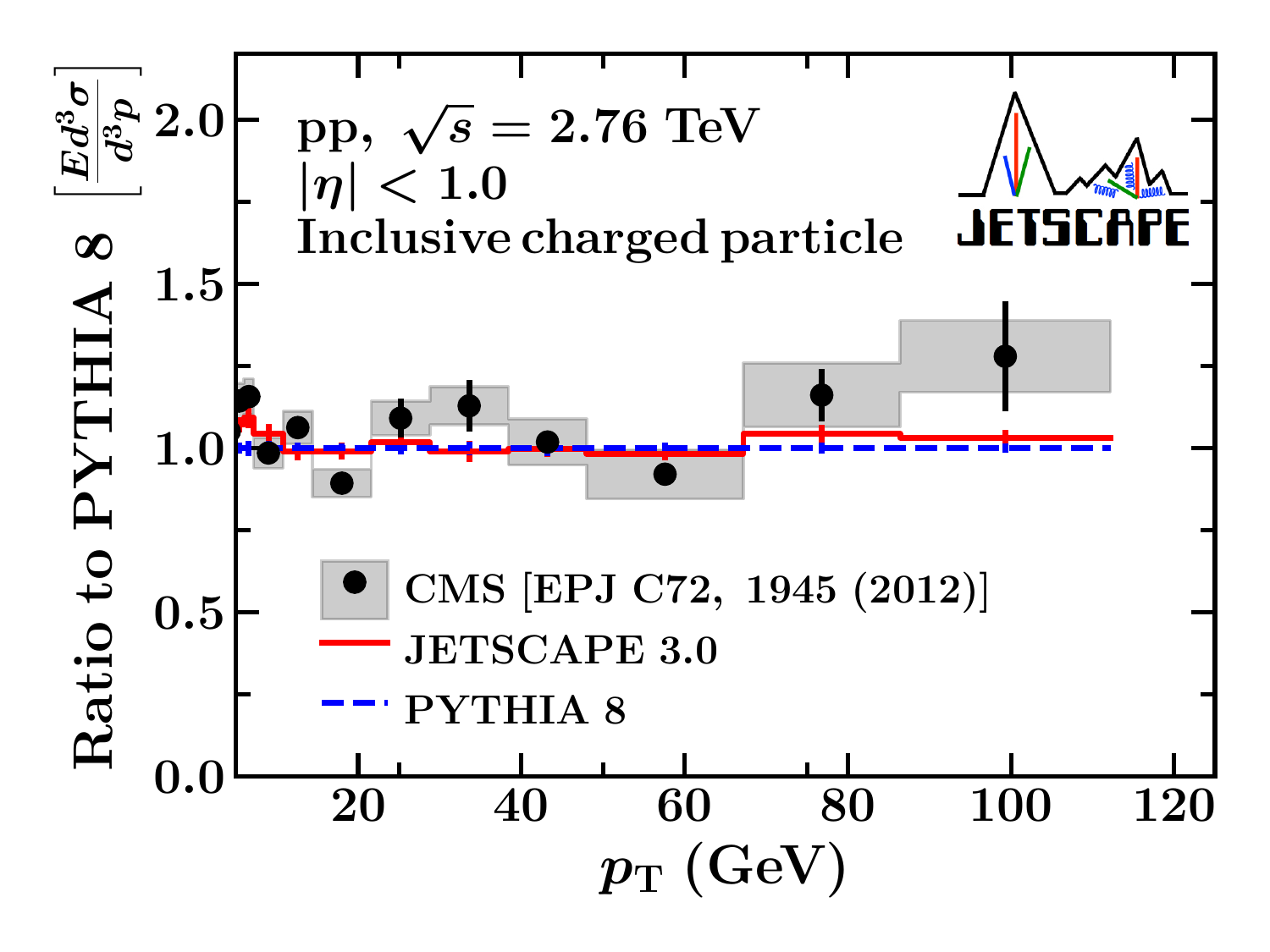}
\includegraphics[width=0.49\textwidth]{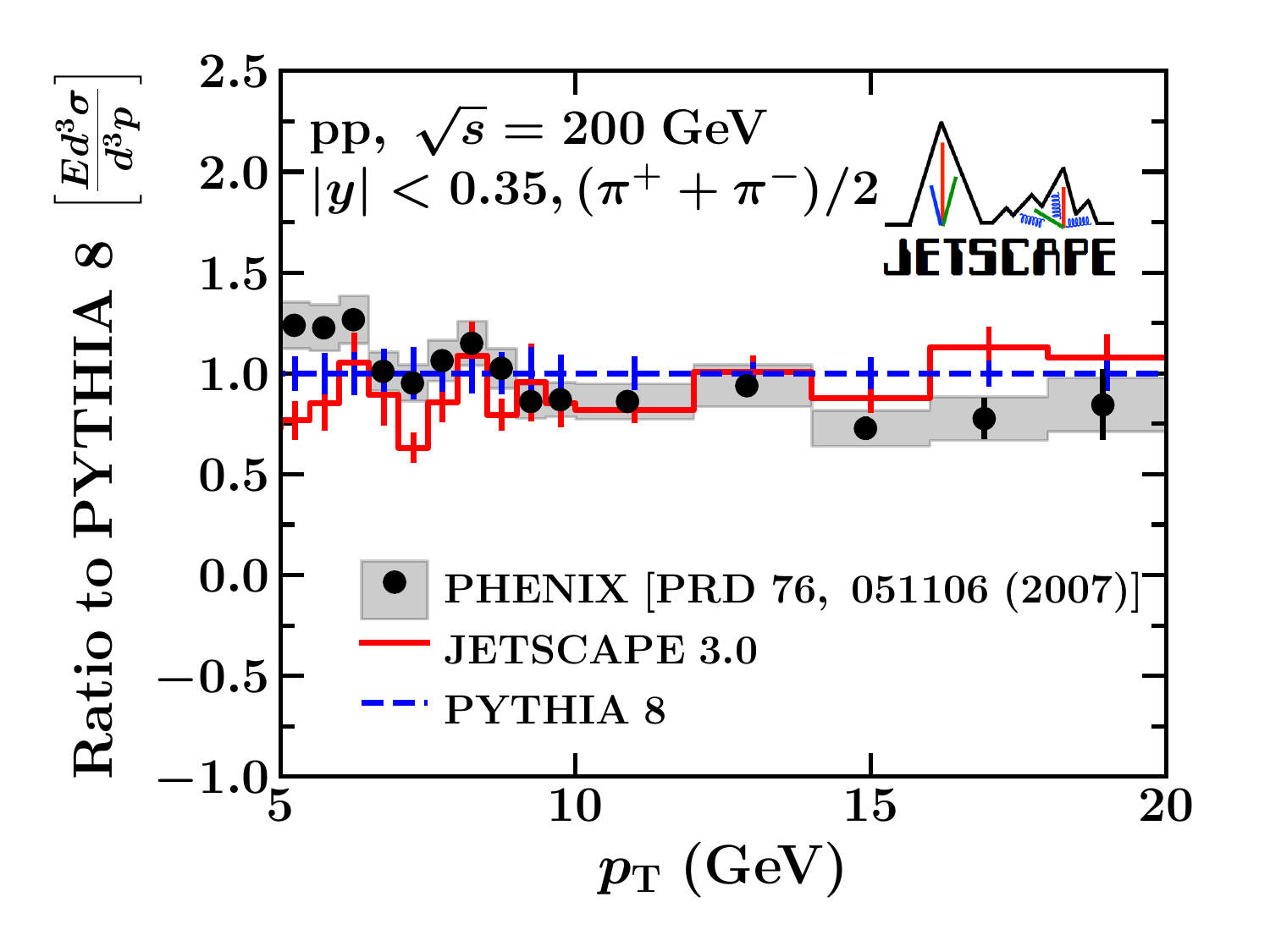}
\caption{Ratio of differential cross-section for inclusive charged-particle at mid-rapidity in $p+p$ collisions. The ratio is taken w.r.t. the default \textsc{pythia} 8. 
The solid red lines and dashed blue lines show the results from \textsc{jetscape} and \textsc{pythia} 8, respectively. 
Statistical errors (black error bars) and systematic uncertainties (grey bands) are plotted with the experimental data. 
Top panel: 
Results for inclusive charged particle with $\eta < 1.0$ at $\sqrt{s}=5.02$~TeV,  compared to CMS data~\cite{CMS:2016xef}. 
Middle panel: 
Results for inclusive charged particle with $\eta < 1.0$ at $\sqrt{s}=2.76$~TeV, 
compared to CMS data~\cite{CMS:2012aa}. 
Bottom panel: 
Results for charged pion with $y<0.35$ at $\sqrt{s}=200$~GeV, compared to PHENIX data~\cite{PHENIX:2007kqm}.}
\label{fig:ratio_plot_chargedParticleYield_2760GeV_200GeV_CMS_RHIC}
\end{figure}
In Fig.~\ref{fig:ratio_plot_chargedParticleYield_2760GeV_200GeV_CMS_RHIC}, we show the ratio of charged particle cross-sections at mid-rapidity in $p+p$ collisions at various collision energies. 
Throughout all the collision energies and $p_{\mathrm{T}}$ range, \textsc{jetscape} results match with \textsc{pythia} 8 results within the statistical errors and describe the experimental data. 

\subsection{Comparison between different $\hat{q}\cdot f$ formulations}
In this section, we perform a multistage based jet energy loss calculation and explore three different forms of 
$\hat{q}\cdot f$ presented in Eqs.~(\ref{eq:type1-q-hatform})-(\ref{eq:q2-dep-qhat}). 
As described in Sec.~\ref{Subsection:Q2-dependentQhat}, in the Type 3 formulation, the reduction in the medium induced emission due to coherence effects~\cite{Armesto:2011ir,Kumar:2019uvu} can be incorporated with a scale-dependent reduction factor $f(Q^2)$ included with the HTL expression for $\hat{q}$. 
We will demonstrate that this form is essential for the simultaneous description of inclusive jet and charged-particle $R_{\mathrm{AA}}$.

\begin{figure}[htbp]
\includegraphics[width=0.47\textwidth]{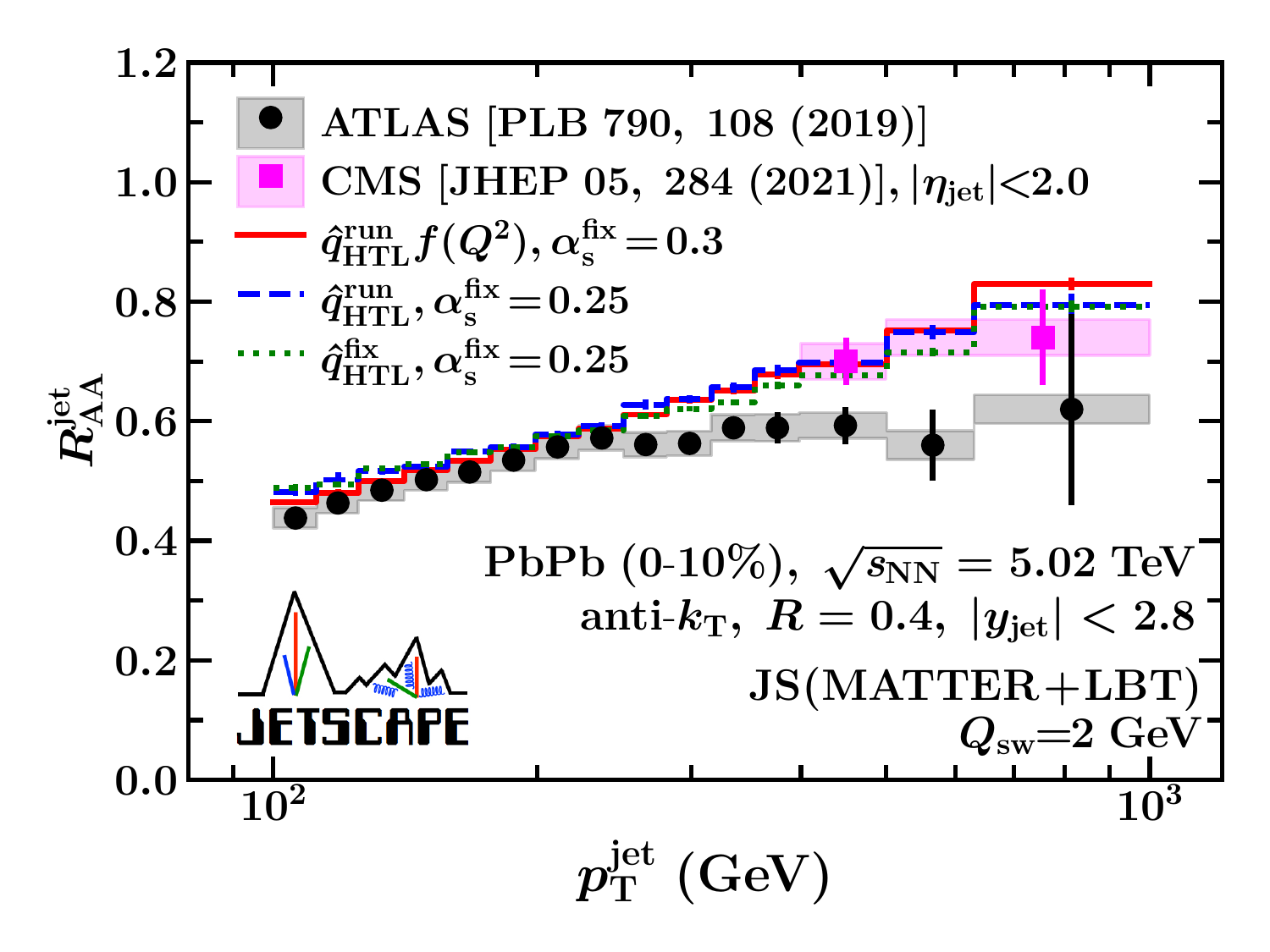}
\includegraphics[width=0.48\textwidth]{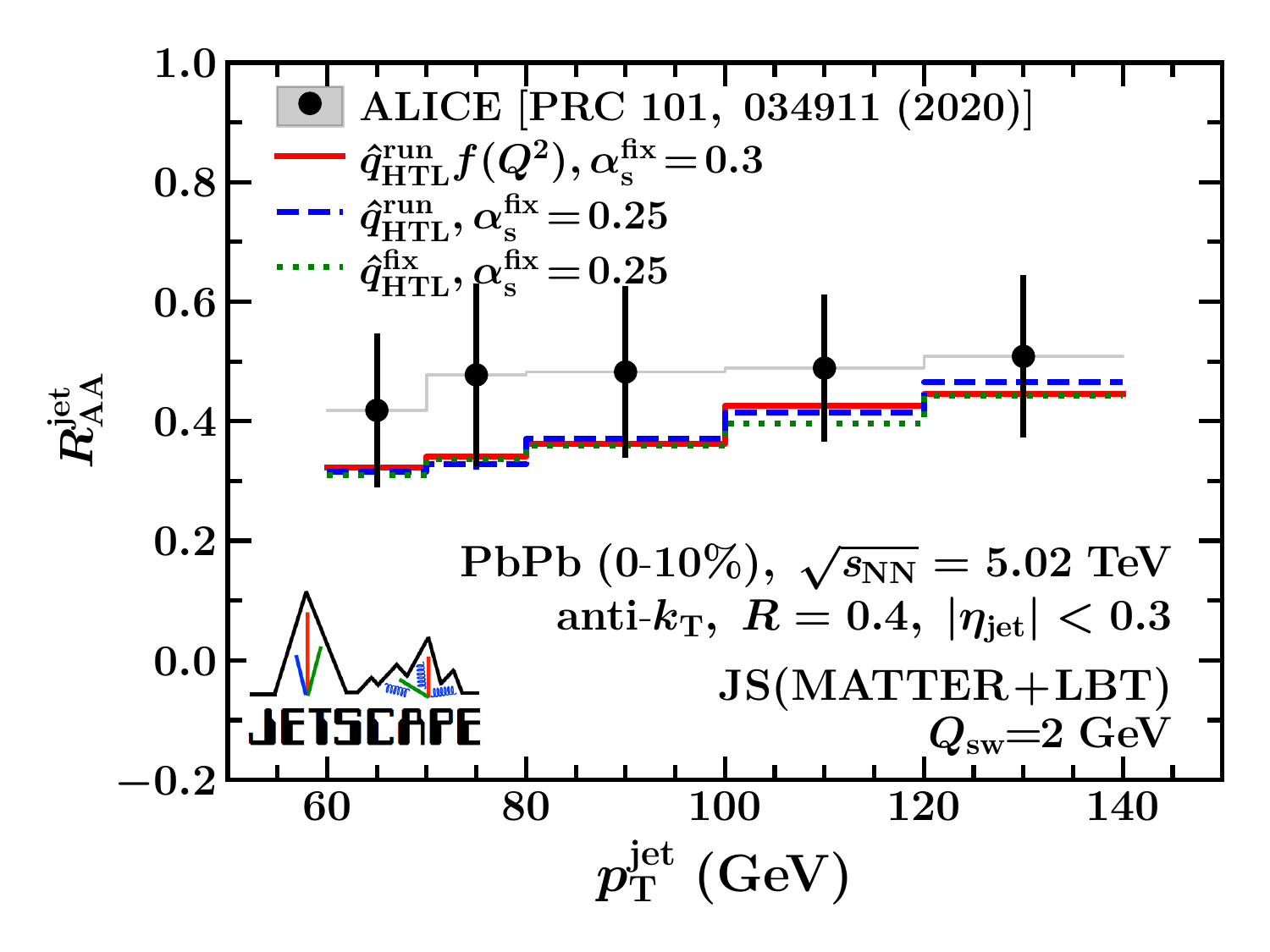}
\includegraphics[width=0.47\textwidth]{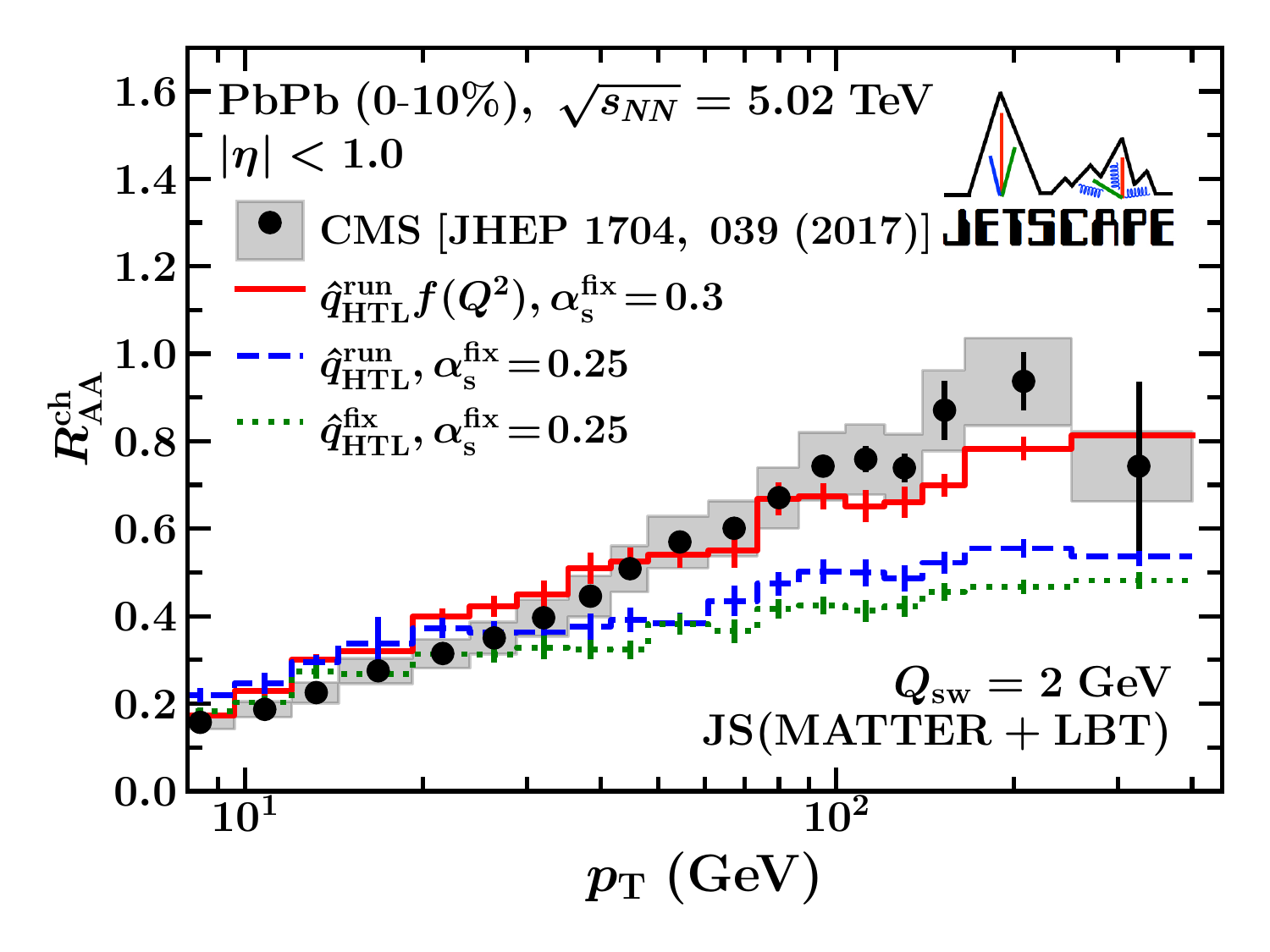}
\caption{Nuclear modification factor for inclusive jets and charged-particle in most central (0-10$\%$) Pb+Pb collisions at $\sqrt{s_{\mathrm{NN}}}=5.02$~TeV 
from \textsc{matter}+\textsc{lbt} simulations for three different $\hat{q}\cdot f$ formulations within the \textsc{jetscape} framework. 
Top panel: 
Results for inclusive jet $R_{\mathrm{AA}}$ 
with $R=0.4$ and $y_{\mathrm{jet}} < 2.8$, compared to ATLAS data~\cite{ATLAS:2018gwx} (black circles)
and 
CMS data for 
$\eta_{\mathrm{jet}} < 2.0$~\cite{CMS:2021vui}
(magenta squares). 
Middle panel: 
Results for inclusive jet $R_{\mathrm{AA}}$ with $R=0.4$, $y_{\mathrm{jet}} < 0.3$ and $p^{\mathrm{lead,\,ch}}_{\mathrm{T}} > 7\,\mathrm{GeV}$, compared to ALICE data~\cite{ALICE:2019qyj}.  
Bottom panel: 
Results for charged-particle $R_{\mathrm{AA}}$ with $\eta<1.0$, compared to CMS data~\cite{CMS:2016xef}.}
\label{fig:Effect_of_three_different_form_q-hat}
\end{figure}

In Fig.~\ref{fig:Effect_of_three_different_form_q-hat}, we present the nuclear modification factor for inclusive jets and charged-particles at most-central (0-10$\%$) Pb+Pb collisions at $\sqrt{s_\mathrm{NN}}=$ 5.02~TeV.
The calculations are carried out using the multistage jet quenching model (\textsc{matter}+\textsc{lbt}) for three different $\hat{q}\cdot f$ formulations and compared with the experimental data from ATLAS~\cite{ATLAS:2018gwx}, CMS~\cite{CMS:2021vui}, and ALICE~\cite{ALICE:2019qyj} for inclusive jets $R_{\mathrm{AA}}$ and from CMS ~\cite{CMS:2016xef} for charged-particles $R_{\mathrm{AA}}$. 
In the above calculation, the switching virtuality parameter is set to $Q_{\mathrm{sw}}=2$~GeV. 
The solid red lines show the results with the virtuality dependent factor $\hat{q}f(Q^2)$, i.e. Type 3 of Eq.~(\ref{eq:q2-dep-qhat}). 
The dashed blue lines and dotted green lines are for $\hat{q}$ with both couplings fixed [Type 1 of Eq.~(\ref{eq:type1-q-hatform})]
and $\hat{q}$ with running coupling [Type 2 of Eq.~(\ref{eq:type2-q-hatform})], respectively. 
In the experimental data, the statistical errors are represented by vertical lines (black color), and systematic uncertainties are shown in shaded rectangular boxes (grey color).

The best fit to inclusive jet $R_{\mathrm{AA}}$ and charged-particle $R_{\mathrm{AA}}$ yields   $\alpha^{\mathrm{fix}}_{\mathrm{s}}=0.3$ for Type 3 and $\alpha^{\mathrm{fix}}_{\mathrm{s}}=0.25$ for Type 1 and Type 2. 
All three formulations 
produce similar results for inclusive jet suppression given one needs to re-adjust the fixed coupling constant $\alpha^\mathrm{fix}_\mathrm{s}$ when going from Type 1/Type 2 
to Type 3.
Results for Type 1/Type 2 for other values of $\alpha^\mathrm{fix}_\mathrm{s}=0.3$ with $Q_{\mathrm{sw}}=2$~GeV are shown in the Appendices. (see Fig.~\ref{fig:type1_alphas_dependence} in Appendix~\ref{type1_param} for Type 1 and Fig.~\ref{fig:type2_alphas_dependence} in Appendix~\ref{type2_param} for Type 2). We note the inclusive jet $R_{\mathrm{AA}}$ presented in the top panel of Fig.~\ref{fig:Effect_of_three_different_form_q-hat} shows good agreement with the ATLAS data for jet $p_{\mathrm{T}}<$ 250~GeV. However, for jet $p_{\mathrm{T}}>$ 300~GeV, inclusive jet $R_{\mathrm{AA}}$ deviates ($\lesssim 10\%$) from the ATLAS data and strongly favors the CMS inclusive jet data. In addition to this, the inclusive jet $R_{\mathrm{AA}}$ for each $\hat{q}$ formulation are in agreement with the ALICE experimental data for all measured jet $p_{\mathrm{T}}$. 

In contrast to the reconstructed jet, 
the suppression of high-$p_{\mathrm{T}}$ charged-particle yields 
exhibits strong sensitivity to the 
virtuality dependence of coherence effects. The results from \textsc{matter}+\textsc{lbt} 
with a virtuality dependent modulation factor $f(Q^2)$ (Type 3) reproduces the experimental data from the CMS Collaboration quite well throughout the entire $p_{\mathrm{T}}$ range. 
The other two formulations without the $Q^2$ dependence (Type 1 and 2)
agree with the data and Type 3 at low $p_{\mathrm{T}}$,  
but show significant over suppression at high $p_{\mathrm{T}}$. 

It should be noted that this behavior of Type 1 and 2 formulations cannot be improved by changing the parameters (see Appendices~\ref{type1_param} and \ref{type2_param}). 
This strongly indicates that the virtuality dependence in energy loss (the gradual onset of coherence effects) is essential to describe the $p_{\mathrm{T}}$ dependence of the $R_{\mathrm{AA}}$ of charged particles. 

The insensitivity in jet $R_{\mathrm{AA}}$ and sensitivity in charged particle $R_{\mathrm{AA}}$ to the virtuality dependence in parton energy loss can be interpreted as follows. 
The charged particle distribution is dominated by the contribution of hadrons from leading partons in jet shower. 
Leading partons with large $p_{\mathrm{T}}$ are more likely to have large virtuality during the early stage of the energy loss. 
When partons undergo energy loss based on the virtuality dependent formulation in Eq.~(\ref{eq:q2-dep-qhat}), the strength of interaction with the medium is diminished for the high-$p_{\mathrm{T}}$ leading partons due to their large virtuality. 
This results in the weaker suppression of larger-$p_{\mathrm{T}}$ charged particles. 

On the other hand, the jet energy loss is mainly brought by the medium effects on partons at larger angles, which causes energy outflow from the jet cone. 
Through successive interactions with the medium, the large-angle region of a jet is dominated by 
soft daughter partons in the low-virtuality phase and becomes less relevant to the inner core structure directly radiated from the leading parton. 
Thus, jet suppression has small sensitivity to the details of the leading-parton energy loss, in particular for large jet cone sizes.  

Although the difference between the formulations is not visible in the $p^{\mathrm{jet}}_{\mathrm{T}}$ dependence of jet $R_{\mathrm{AA}}$, it can be seen in the inner structures of jets, particularly in the core region dominated by the leading parton; Jets are more likely to have particles with a larger-$p_{\mathrm{T}}$ fraction in their core for the case with the virtuality-dependent formulation. 
Thus, one may see more details in the energy loss of high-virtuality partons by studying the medium modification of jet substructure observables, e.g. jet fragmentation function, which will be discussed in an upcoming effort.

\subsection{Exploring A+A model parameters}
\label{type_3_param}
In this section, we explore the sensitivity of the free parameters in the multistage jet quenching model (\textsc{matter}+\textsc{lbt}). 
We shall employ the virtuality dependent formulation (Type 3) as it gives the best simultaneous description of the data. 
The free parameters in the jet quenching model are summarized in Tab.~\ref{tab:aa_parameter_set}. 
By changing the parameters from the default values, we present the inclusive jet $R_{\mathrm{AA}}$ and charged-particle $R_{\mathrm{AA}}$ at most central (0-10$\%$) collision at $\sqrt{s_\mathrm{NN}}=$ 5.02~TeV and show the parameter dependence in our model.

\subsubsection{Sensitivity to jet-medium coupling $\alpha^\mathrm{fix}_\mathrm{s}$}
\label{section:sensitivity_alphas}
\begin{figure}[htbp]
\centering
\includegraphics[width=0.45\textwidth]{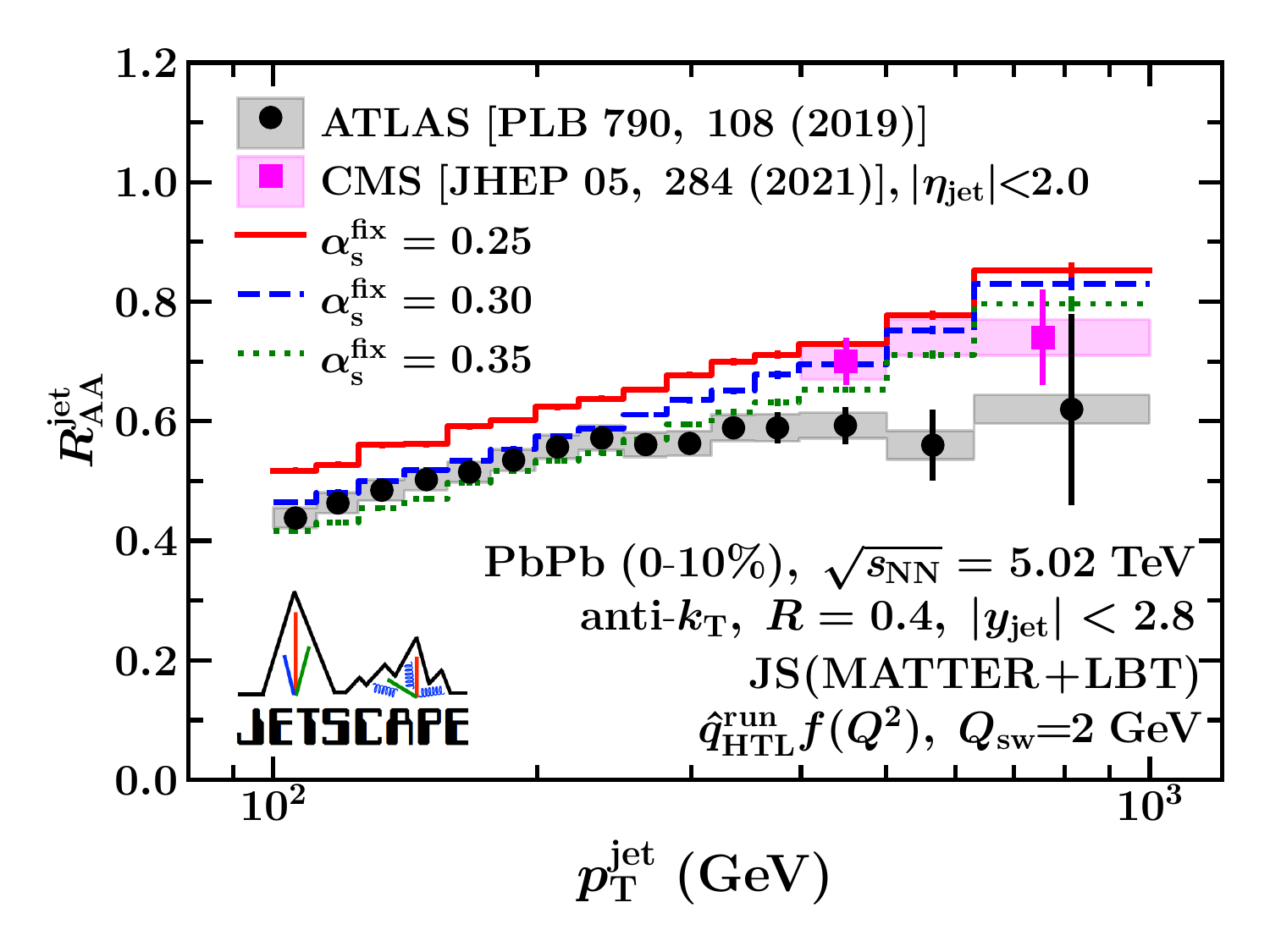}
\includegraphics[width=0.45\textwidth]{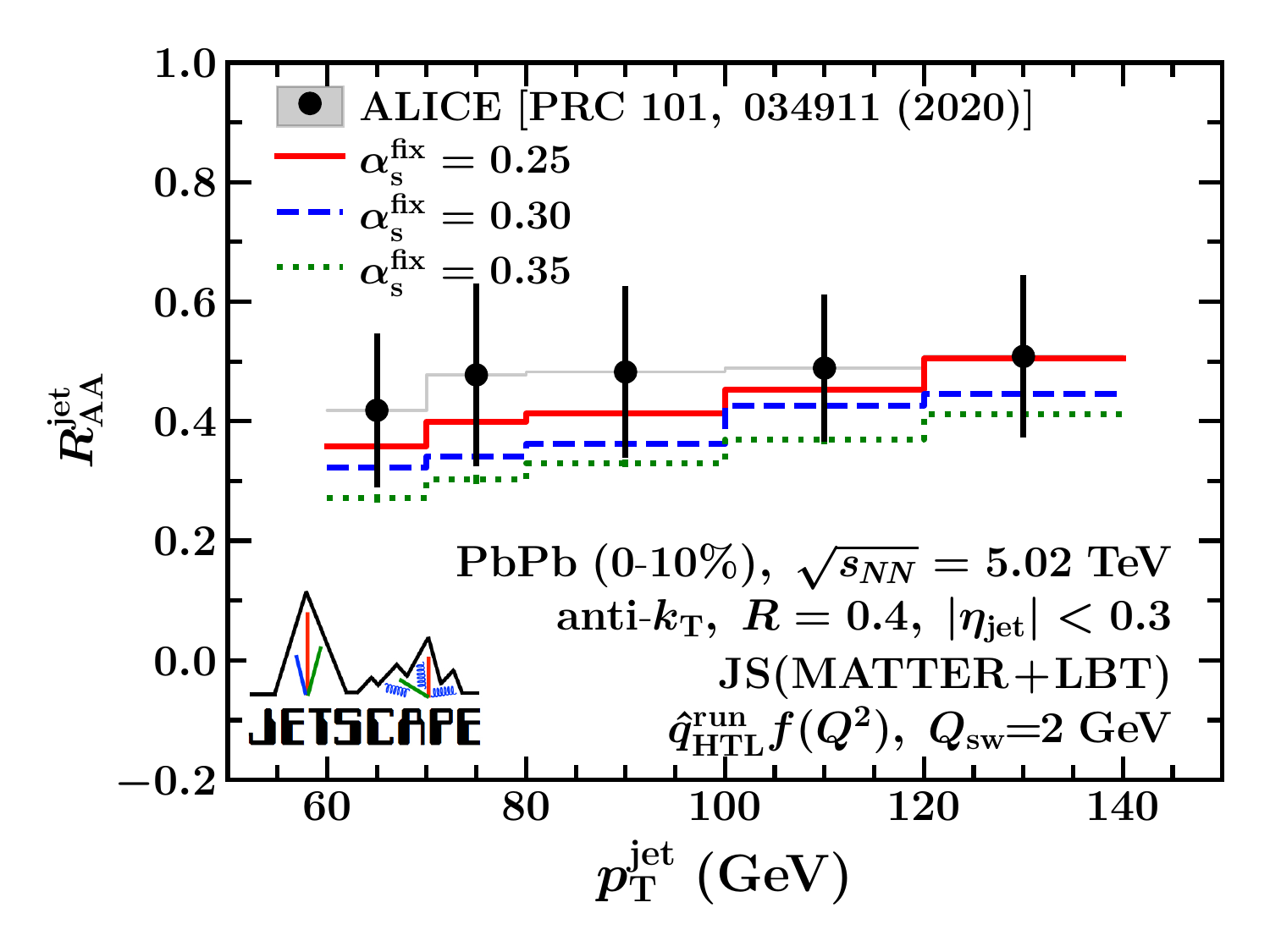}
\includegraphics[width=0.45\textwidth]{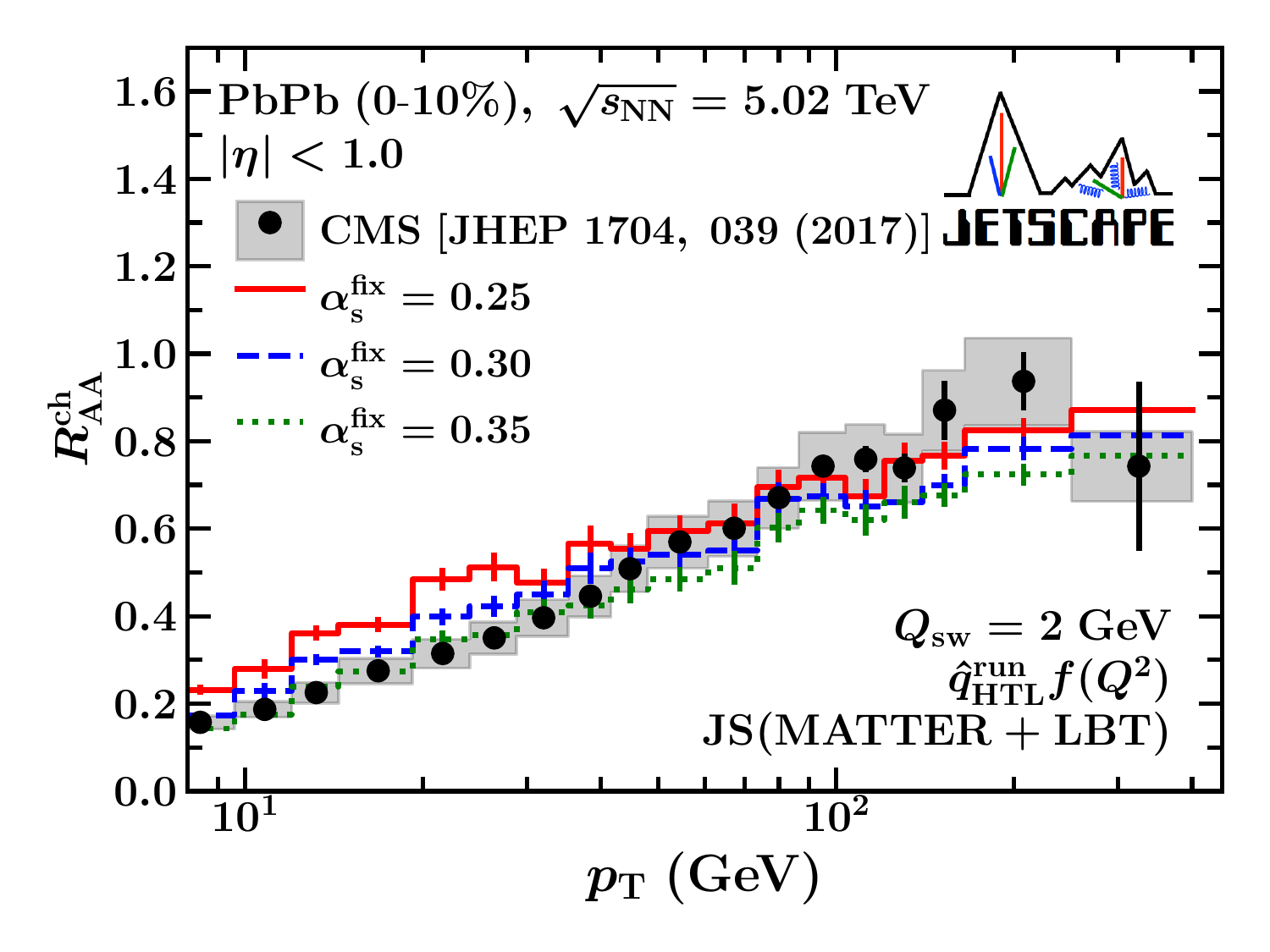}
\caption{Same as Fig.~\ref{fig:Effect_of_three_different_form_q-hat}. 
The solid red, dashed blue, and dotted green lines show results with virtuality dependence (Type 3) 
for $\alpha^{\mathrm{fix}}_{\mathrm{s}}=0.25,\,0.3,\,\mbox{and }0.35$, respectively.
}
\label{fig:Type3-q-hat-effect_fixed_alphas}
\end{figure}
In Fig.~\ref{fig:Type3-q-hat-effect_fixed_alphas}, we present the nuclear modification factor for inclusive jets and charged-particles for a jet-medium coupling parameter $\alpha^\mathrm{fix}_\mathrm{s}=$ 0.25, 0.3 and 0.35. 
The \textsc{jetscape} results are obtained using the multistage jet energy loss approach (\textsc{matter}+\textsc{lbt}) where a virtuality dependent factor $f(Q^2)$ given 
in Eq.~(\ref{eq:qhatSuppressionFactor}) 
is employed. 
Increasing $\alpha^\mathrm{fix}_\mathrm{s}$ from 0.25 to 0.35 leads to suppression ($\approx10 \%$) in the inclusive jet $R_{\mathrm{AA}}$ for all jet $p_{\mathrm{T}}$. 
A similar trend is observed in the charged-particle $R_{\mathrm{AA}}$ as well. Since the variation of $Q_{\mathrm{sw}}$ from 1 to 3 GeV increases the effective length of the LBT energy loss stage, the parton at low-$p_{\mathrm{T}}$ undergoes significant energy loss. This leads to an overall suppression of $R_{\mathrm{AA}}$ for inclusive jets and charged particles. 

Comparison of both inclusive jet $R_{\mathrm{AA}}$ with ATLAS/CMS data and charged-particle $R_{\mathrm{AA}}$ with CMS data favors $\alpha^\mathrm{fix}_\mathrm{s}$ to be between $0.3$ and 0.35, 
whereas ALICE data favors $\alpha^\mathrm{fix}_\mathrm{s} \lesssim 0.3$. 
Overall, the optimized value of $\alpha^\mathrm{fix}_\mathrm{s}$ from the above comparisons comes out to be 0.3. 
For $N_{f}=3$ flavors and $T \in [150,400]$~MeV, 
we estimate the Debye mass in Eq.~(\ref{eq:DebyMassEquation}) to be $m_\mathrm{D} \in [357,951]$~MeV.

\subsubsection{Sensitivity to switching virtuality parameter $Q_\mathrm{sw}$}
\begin{figure}[htbp]
\centering
\includegraphics[width=0.45\textwidth]{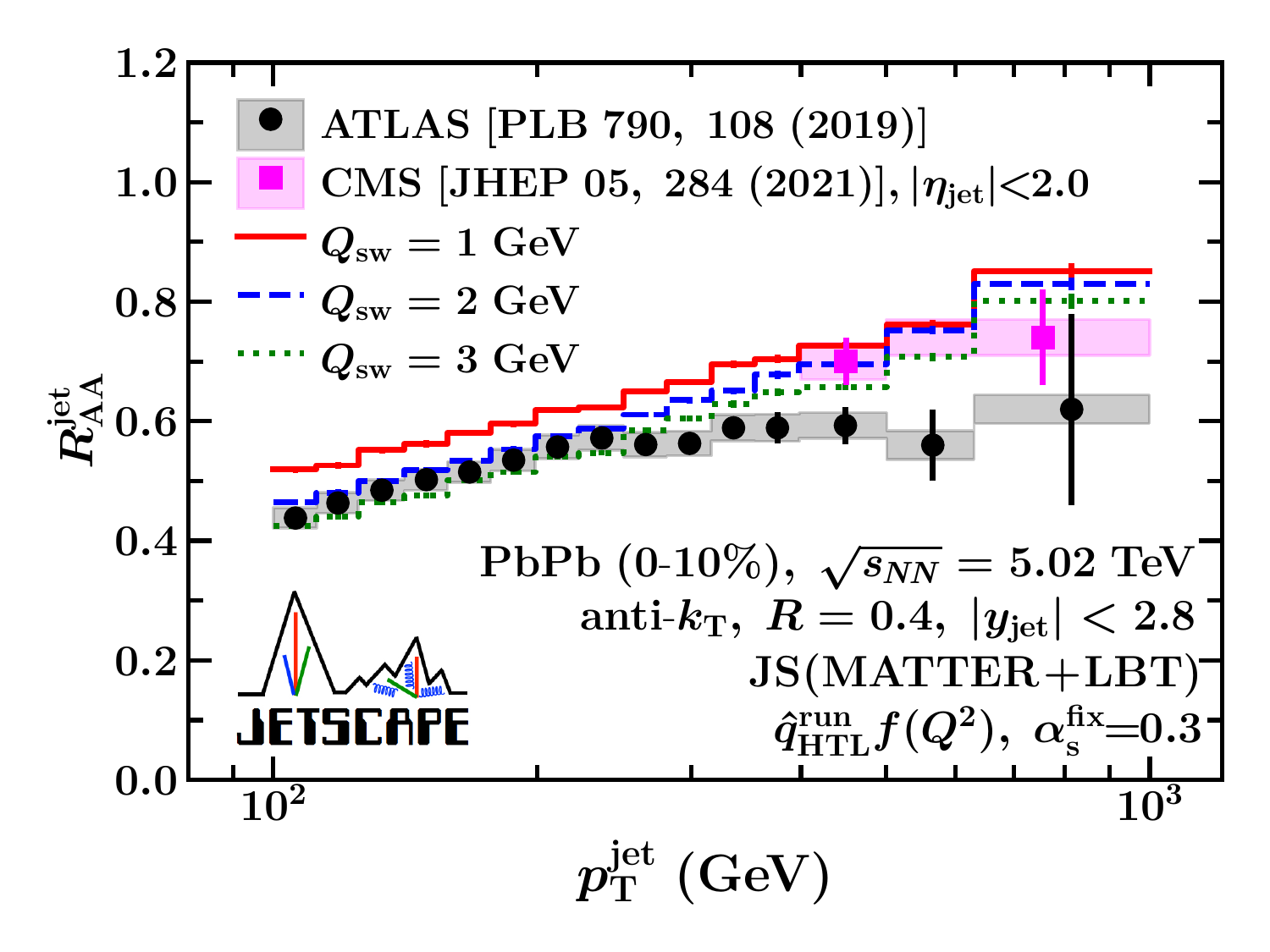}
\includegraphics[width=0.45\textwidth]{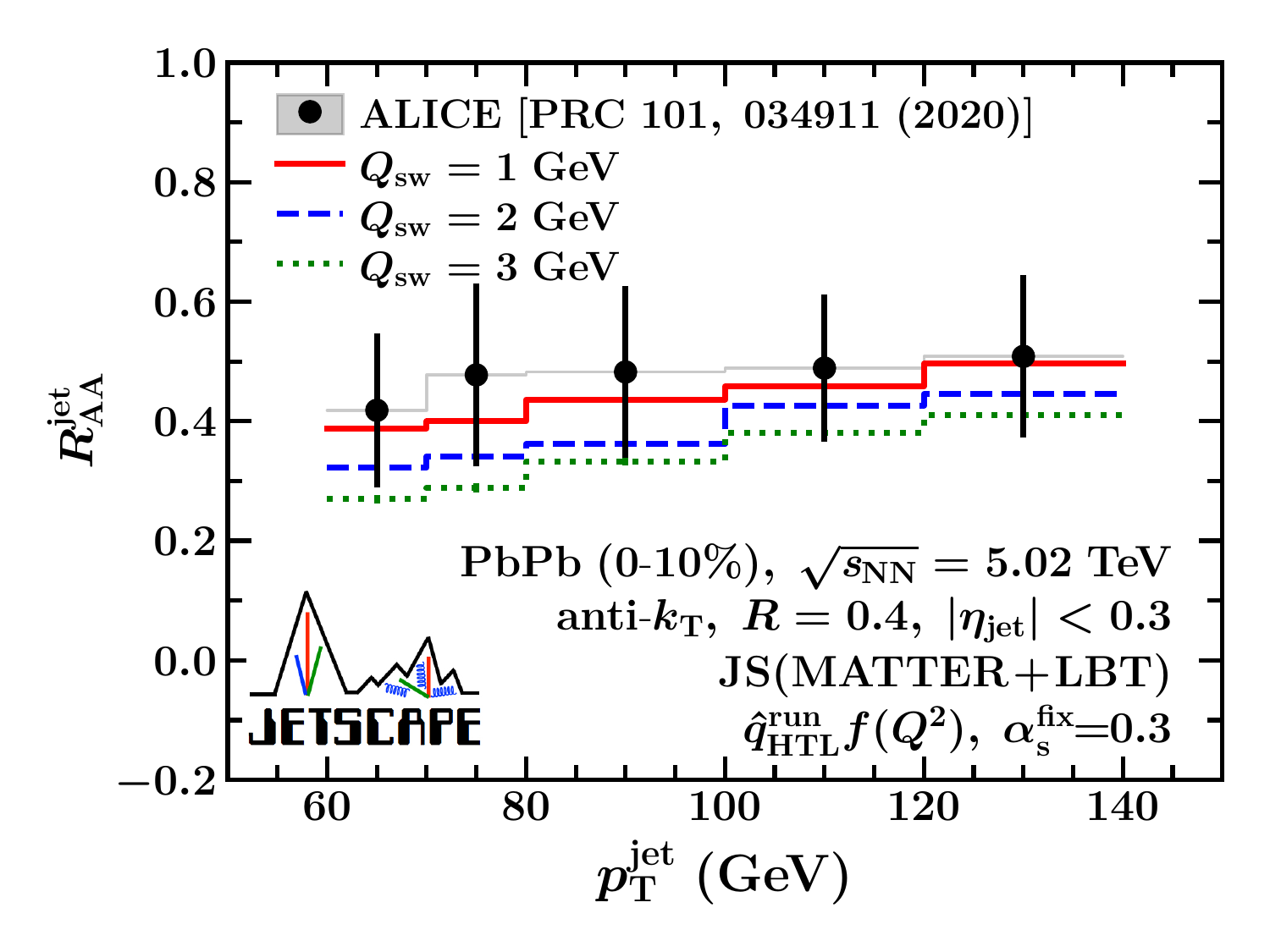}
\includegraphics[width=0.45\textwidth]{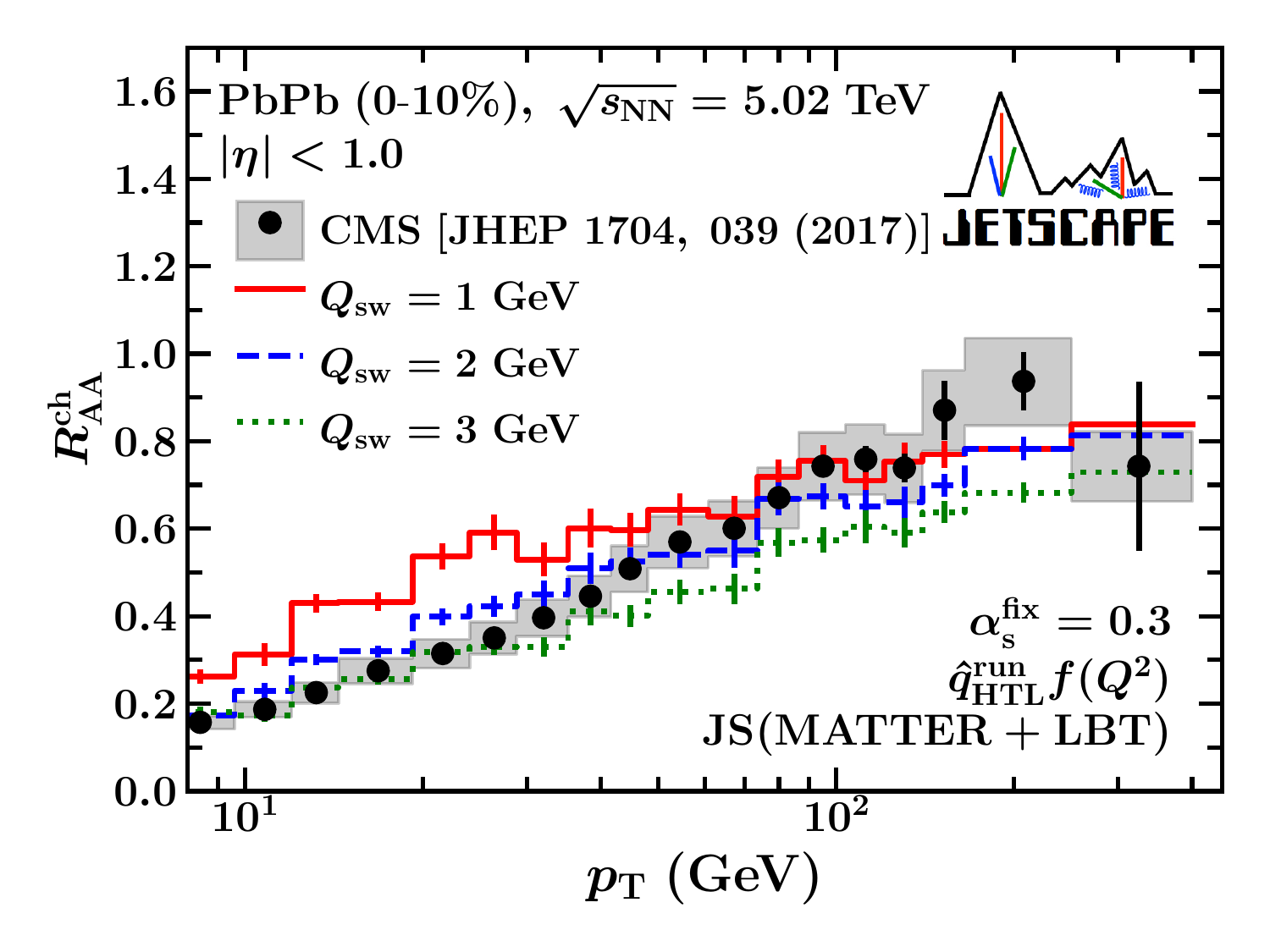}
\caption{Same as Fig.~\ref{fig:Effect_of_three_different_form_q-hat}. The solid red, dashed blue, and dotted green lines show results with virtuality dependence (Type 3) 
for $Q_{\mathrm{sw}}=1,\,2,\,\mbox{and }3$~GeV,  respectively. Here we set $\alpha^{\mathrm{fix}}_{\mathrm{s}}=0.3$.}
\label{fig:Type3-q-hat-Effect_of_switching_virtuality}
\end{figure}
In Fig.~\ref{fig:Type3-q-hat-Effect_of_switching_virtuality}, we present inclusive jet $R_{\mathrm{AA}}$ and charged-particle $R_{\mathrm{AA}}$ for the switching virtuality parameter $Q_\mathrm{sw}=1$~GeV, 2~GeV and 3~GeV.
The \textsc{jetscape} results are obtained using the multistage jet energy loss approach (\textsc{matter}+\textsc{lbt}) where a virtuality dependent factor $f(Q^2)$ given 
in Eq.~(\ref{eq:qhatSuppressionFactor}) 
is employed. 
Increasing $Q_\mathrm{sw}$ from 1~GeV to 3~GeV leads to suppression ($\approx 10 \%$) in the inclusive jet $R_{\mathrm{AA}}$ for all jet $p_{\mathrm{T}}$. Similar trend is observed in the charged-particle $R_{\mathrm{AA}}$ as well. Since the variation of $Q_\mathrm{sw}$ from 1~GeV to 3~GeV increases the effective length of \textsc{lbt} energy loss stage, 
where there is no virtuality-driven suppression of the medium effect. 
This leads to an overall suppression of $R_{\mathrm{AA}}$ for inclusive jets and charged particles.

Comparison of inclusive jet $R_{\mathrm{AA}}$ with ATLAS and CMS data indicate $Q_\mathrm{sw}$ to be between 2~GeV and 3~GeV, whereas ALICE data favors $Q_\mathrm{sw} \lesssim 2$~GeV. Moreover, CMS data for the charged-particle $R_{\mathrm{AA}}$ disfavors $Q_\mathrm{sw}=1$~GeV in the low-$p_{\mathrm{T}}$ region and $Q_\mathrm{sw}=3$~GeV in the high-$p_{\mathrm{T}}$ region.
Overall, the optimized value of $Q_\mathrm{sw}$ from the above comparison comes out to be about 2~GeV.

\subsubsection{Sensitivity to start time of jet energy loss parameter $\tau_{0}$}
\begin{figure}[htbp]
\centering
\includegraphics[width=0.45\textwidth]{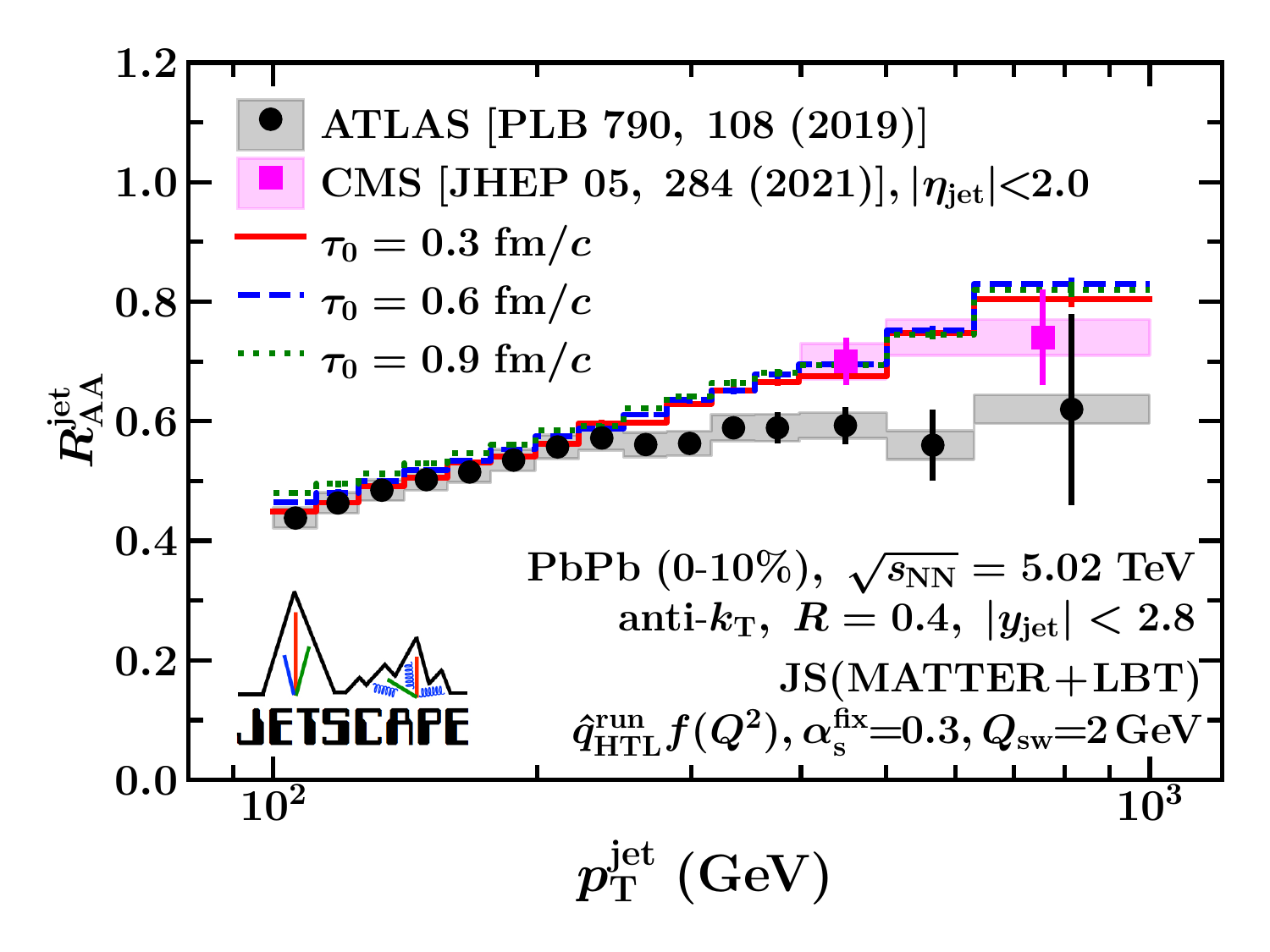}
\includegraphics[width=0.45\textwidth]{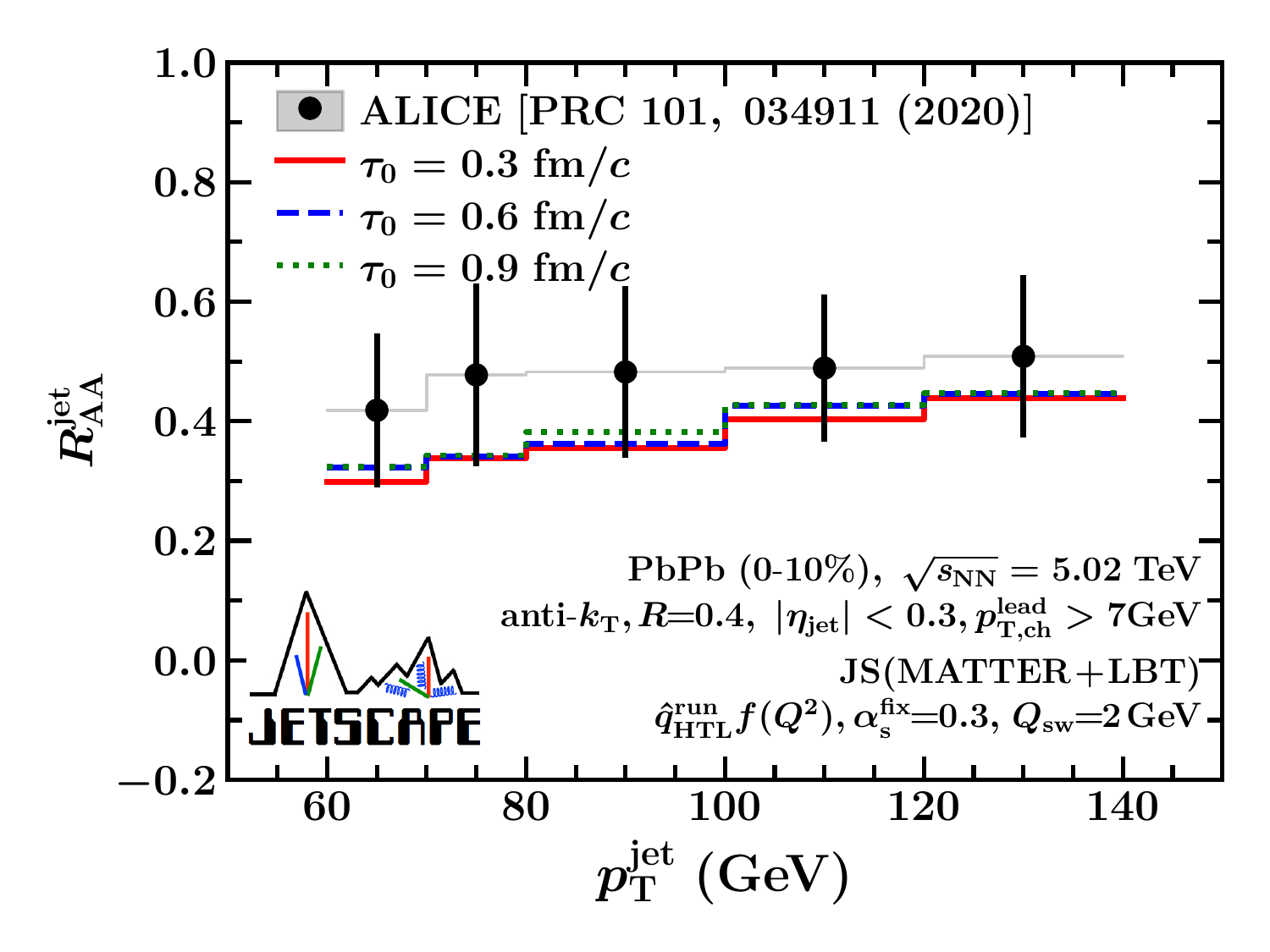}
\includegraphics[width=0.45\textwidth]{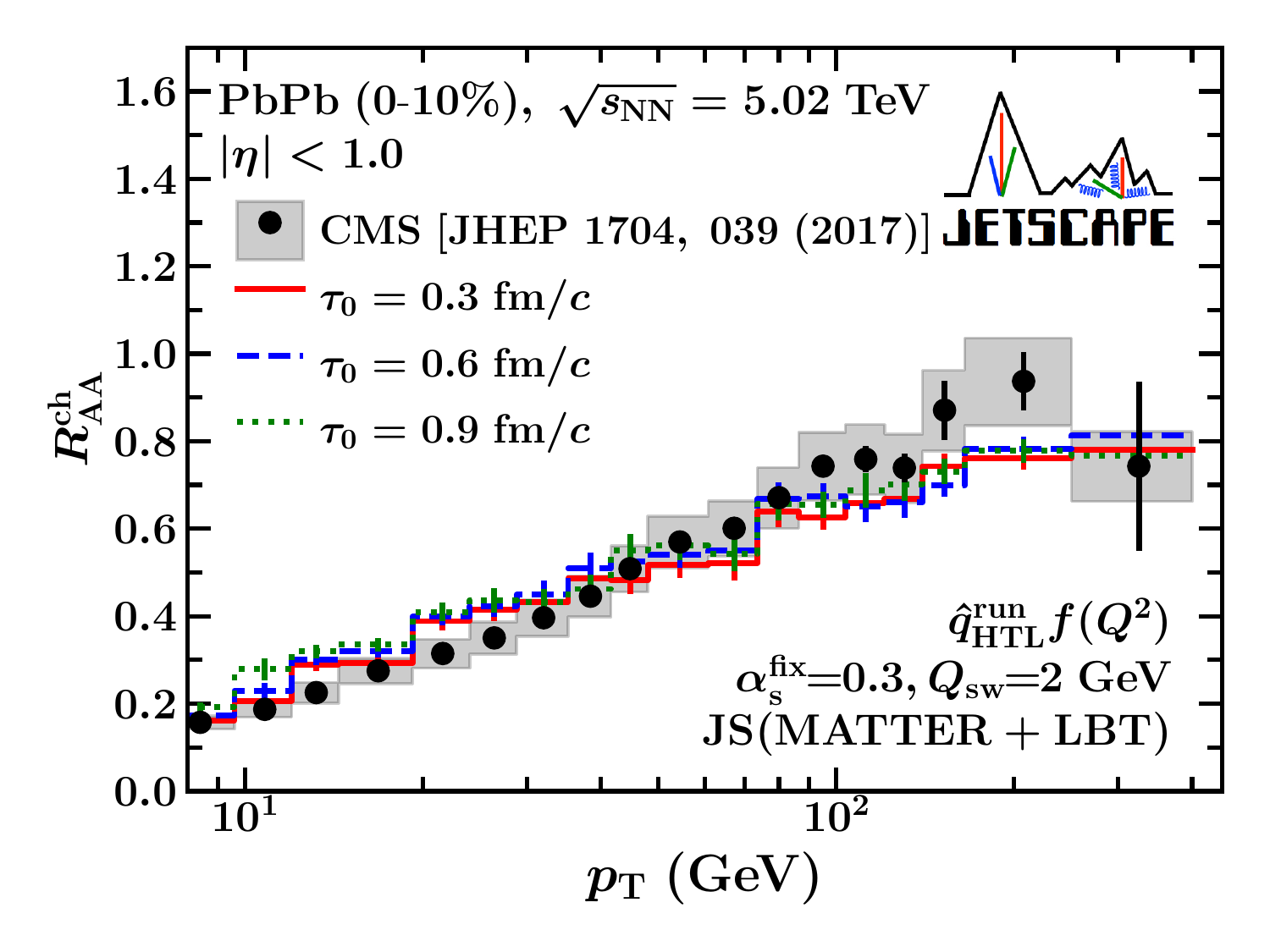}
\caption{
Same as Fig.~\ref{fig:Effect_of_three_different_form_q-hat}. 
The solid red, dashed blue, and dotted green lines show results with virtuality dependence (Type 3) 
for  
starting longitudinal proper times for in-medium jet energy loss 
$\tau_0=0.3,\,0.6,\,\mbox{and }0.9$ fm, respectively.
}
\label{fig:Type3-q-hat-Effect_of_tau0}
\end{figure}
In Fig.~\ref{fig:Type3-q-hat-Effect_of_tau0}, we present inclusive jet $R_{\mathrm{AA}}$ and charged-particle $R_{\mathrm{AA}}$ for the jet quenching start time parameter $\tau_{0}=0.3$~fm/$c$, 0.6~fm/$c$ and 0.9~fm/$c$.
The \textsc{jetscape} results are obtained using the multistage jet energy loss approach (\textsc{matter}+\textsc{lbt}) 
where a virtuality dependent factor $f(Q^2)$ given 
in Eq.~(\ref{eq:qhatSuppressionFactor}) 
is employed. 

Increasing $\tau_{0}$ from 0.3~fm/$c$ to 0.9~fm/$c$ does not seem to affect the inclusive jet $R_{\mathrm{AA}}$ or the charged-particle $R_{\mathrm{AA}}$. 
The effect of the starting time of jet-medium interaction 
as studied in Refs.~\cite{Andres:2016iys,Andres:2019eus} is not visible 
in the modifications of inclusive jet and charged-particle spectra within our model.
This is primarily due to the fact that the jet initiating parton is highly virtual, and the virtuality-dependence in Eq.~(\ref{eq:q2-dep-qhat}) highly suppresses the medium effect in the early stage. It is still possible, though unlikely that the choice of $\tau_0$ will affect the azimuthal anisotropy. This will be explored in a future effort. 

\subsubsection{Sensitivity to  temperature parameter $T_{\mathrm{c}}$ in jet energy loss}
\begin{figure}[htbp]
\centering
\includegraphics[width=0.45\textwidth]{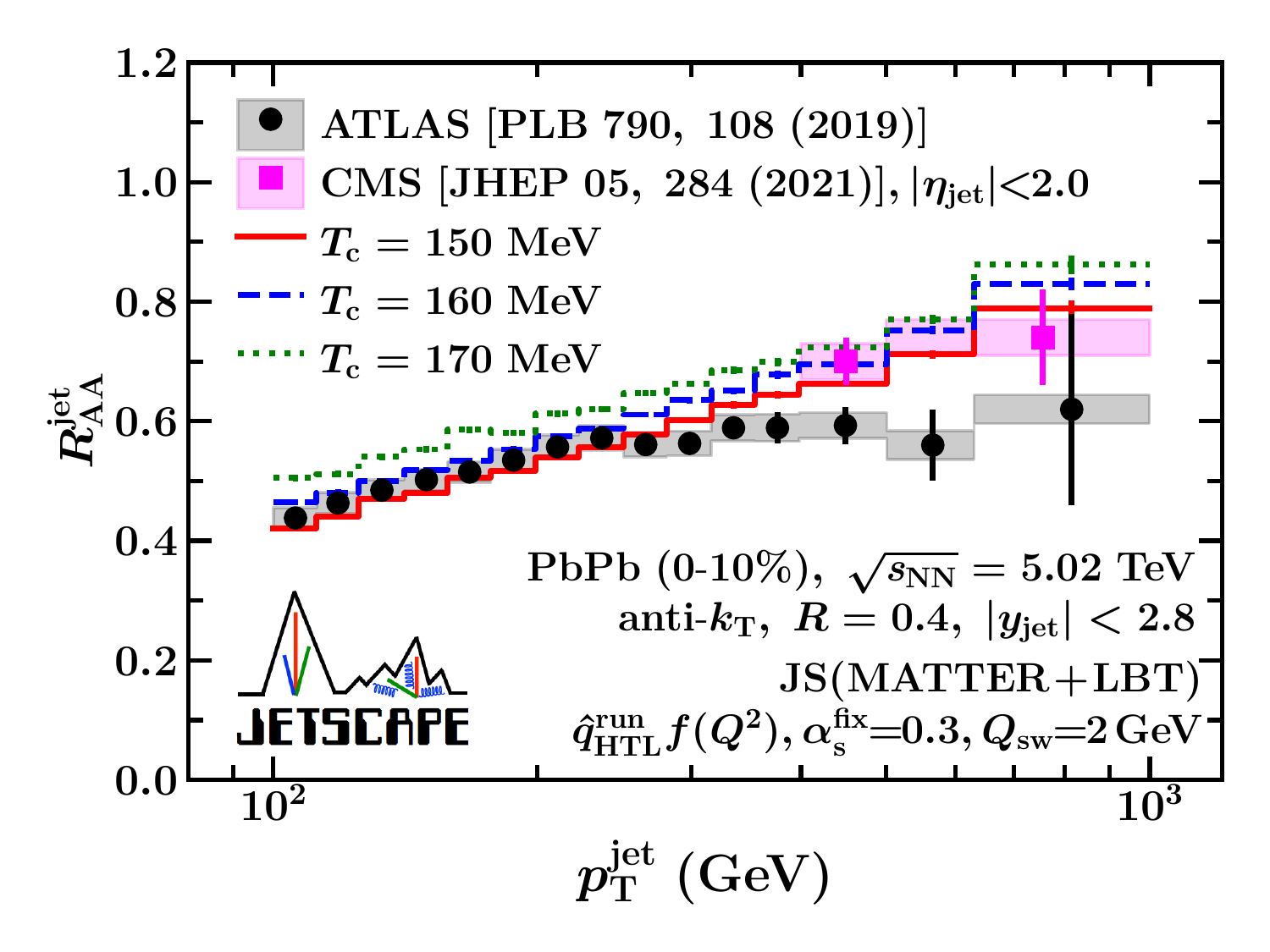}
\includegraphics[width=0.45\textwidth]{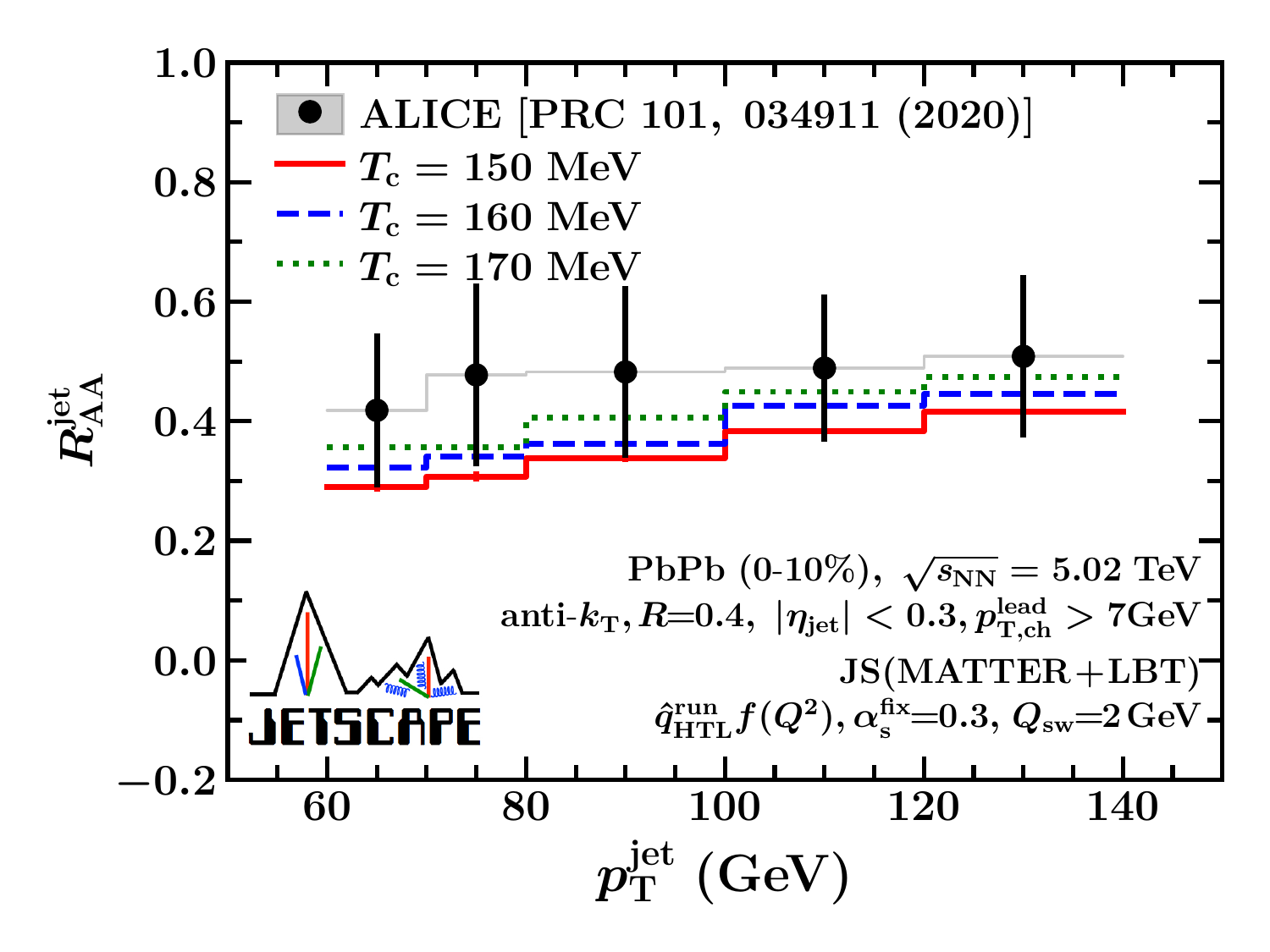}
\includegraphics[width=0.45\textwidth]{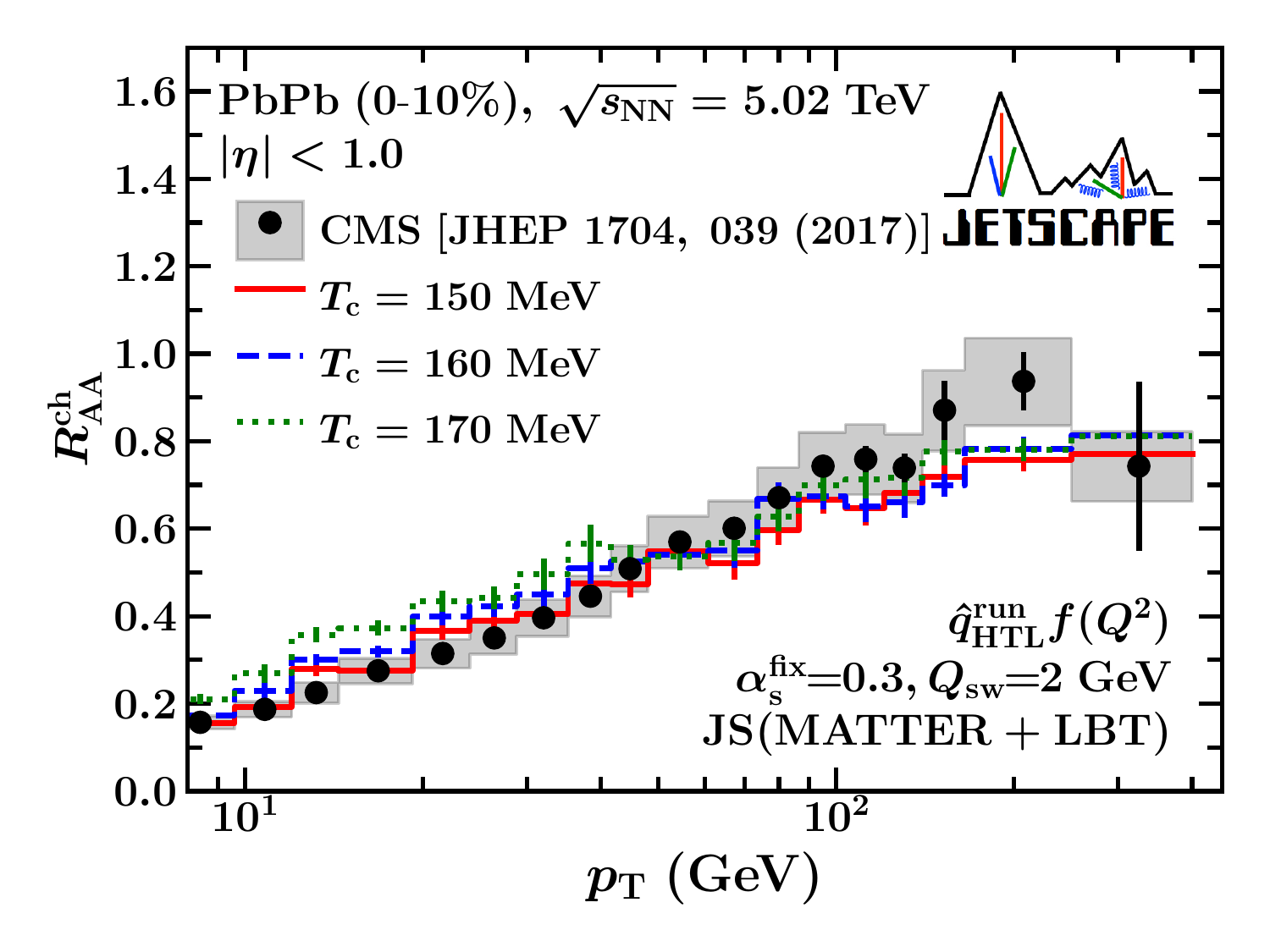}
\caption{
Same as Fig.~\ref{fig:Effect_of_three_different_form_q-hat}. 
The solid red, dashed blue, and dotted green lines show results with virtuality dependence (Type 3) 
for the energy loss termination temperatures  
$T_{\mathrm{c}}=150,\,160,\,\mbox{and }170$~MeV, respectively.
}
\label{fig:Type3-q-hat-Effect_of_Tc}
\end{figure}
We vary the temperature cut-off parameter for jet energy loss $T_{\mathrm{c}}=150$~MeV, 160~MeV and 170~MeV, and present results for inclusive jet $R_{\mathrm{AA}}$ and charged-particle $R_{\mathrm{AA}}$ in Fig~\ref{fig:Type3-q-hat-Effect_of_Tc}. 
The \textsc{jetscape} results are obtained using the multistage jet energy loss approach (\textsc{matter}+\textsc{lbt}) where a virtuality dependent factor $f(Q^2)$ given 
in Eq.~(\ref{eq:qhatSuppressionFactor}) 
is employed. 
All other parameters are set to their default value.
Increasing $T_{\mathrm{c}}$ from 150~MeV to 170~MeV decreases the in-medium portion of the jet energy loss. Since, $T\in [150,170]$~MeV corresponds to the late time dynamics of jet energy loss, the relevant energy loss stage is \textsc{lbt}. The decrease in the effective length of the \textsc{lbt} energy loss stage enhances the low $p_{\mathrm{T}}$ region of the charged-particle $R_{\mathrm{AA}}$ significantly,  compared to high-$p_{\mathrm{T}}$ region. This leads to an overall enhancement ($\approx 10 \%$) in the inclusive jet $R_{\mathrm{AA}}$ at all jet $p_{\mathrm{T}}$.

Comparison of inclusive jet $R_{\mathrm{AA}}$ with ATLAS and CMS data indicate $T_{\mathrm{c}}$ to be between 150~MeV and 160~MeV, whereas ALICE data favors $T_{\mathrm{c}} \gtrsim 160$~MeV. Moreover, CMS data for the charged-particle $R_{\mathrm{AA}}$ favors $T_{\mathrm{c}} \in [160,170]$~MeV.  Overall, the optimized value of $T_{\mathrm{c}}$ from the above comparison comes out to be 160~MeV.

\subsection{Inclusive jet and hadron suppression at semi-peripheral collisions}
In this section, we use the optimized value of the free parameters listed in Table \ref{tab:aa_parameter_set} and present comparisons for centrality dependence of inclusive-jet $R_{\mathrm{AA}}$ and charged-particle $R_{\mathrm{AA}}$ at $\sqrt{s_\mathrm{NN}}$=5.02~TeV. 
The \textsc{jetscape} results are obtained using the multistage jet energy loss (\textsc{matter}+\textsc{lbt}), where we employ a virtuality dependent factor $f(Q^2)$ along with the $\hat{q}$ in Eq.~(\ref{eq:q2-dep-qhat}), to account for a reduction in the rate of medium induced emission in the high virtuality phase. 

\begin{figure*}[htbp]
\centering
\includegraphics[width=0.45\textwidth]{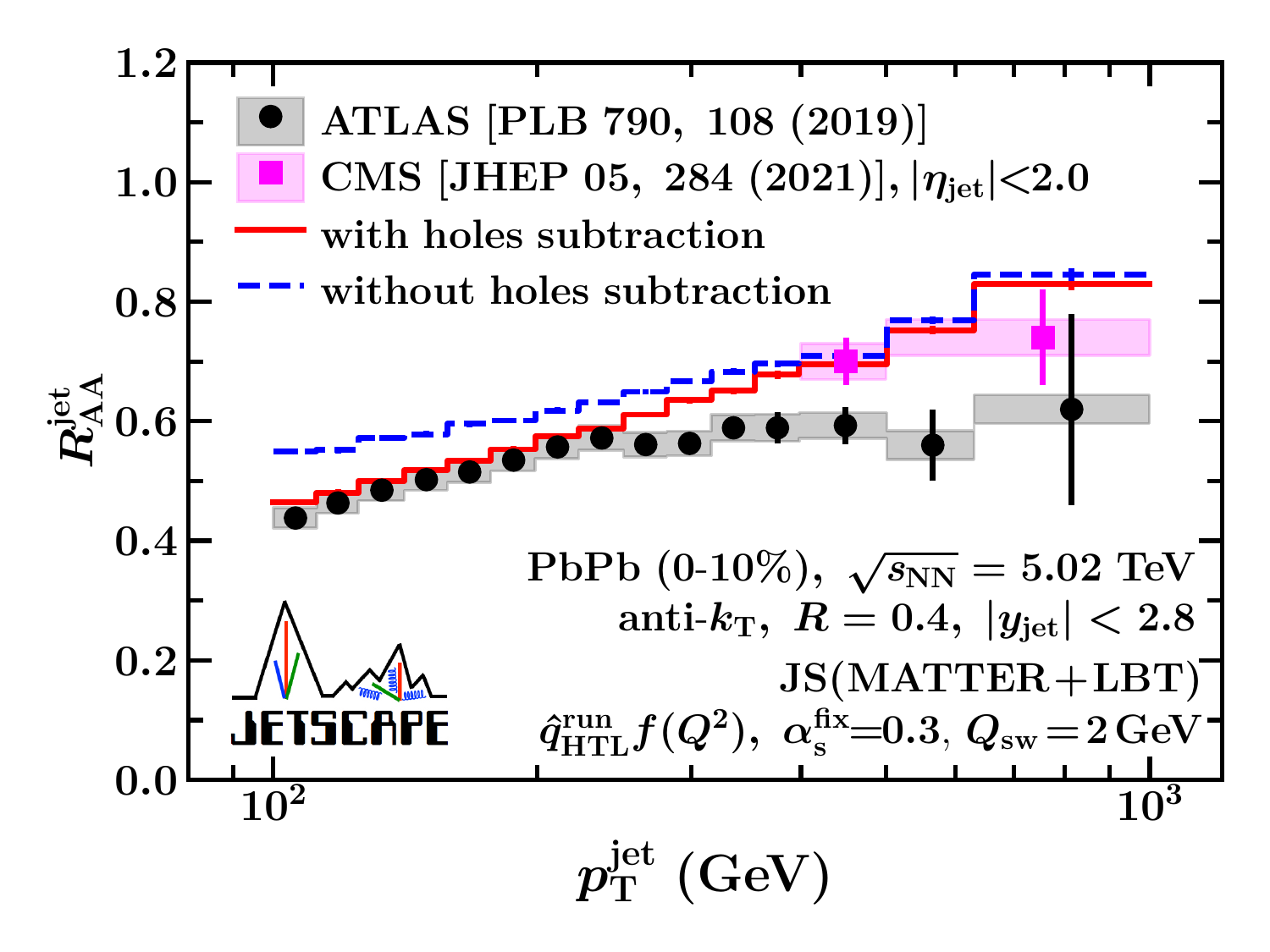}
\includegraphics[width=0.45\textwidth]{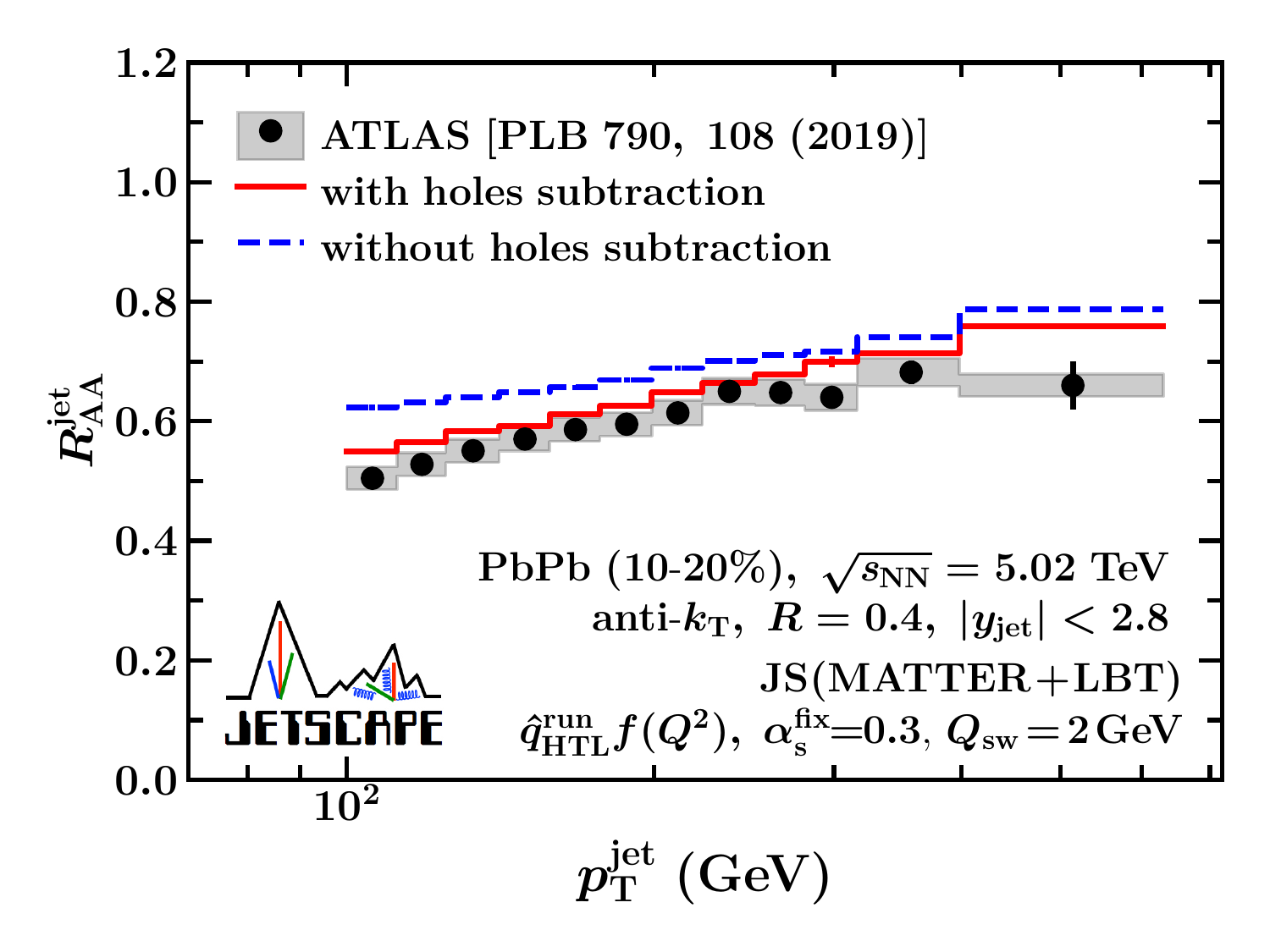}
\includegraphics[width=0.45\textwidth]{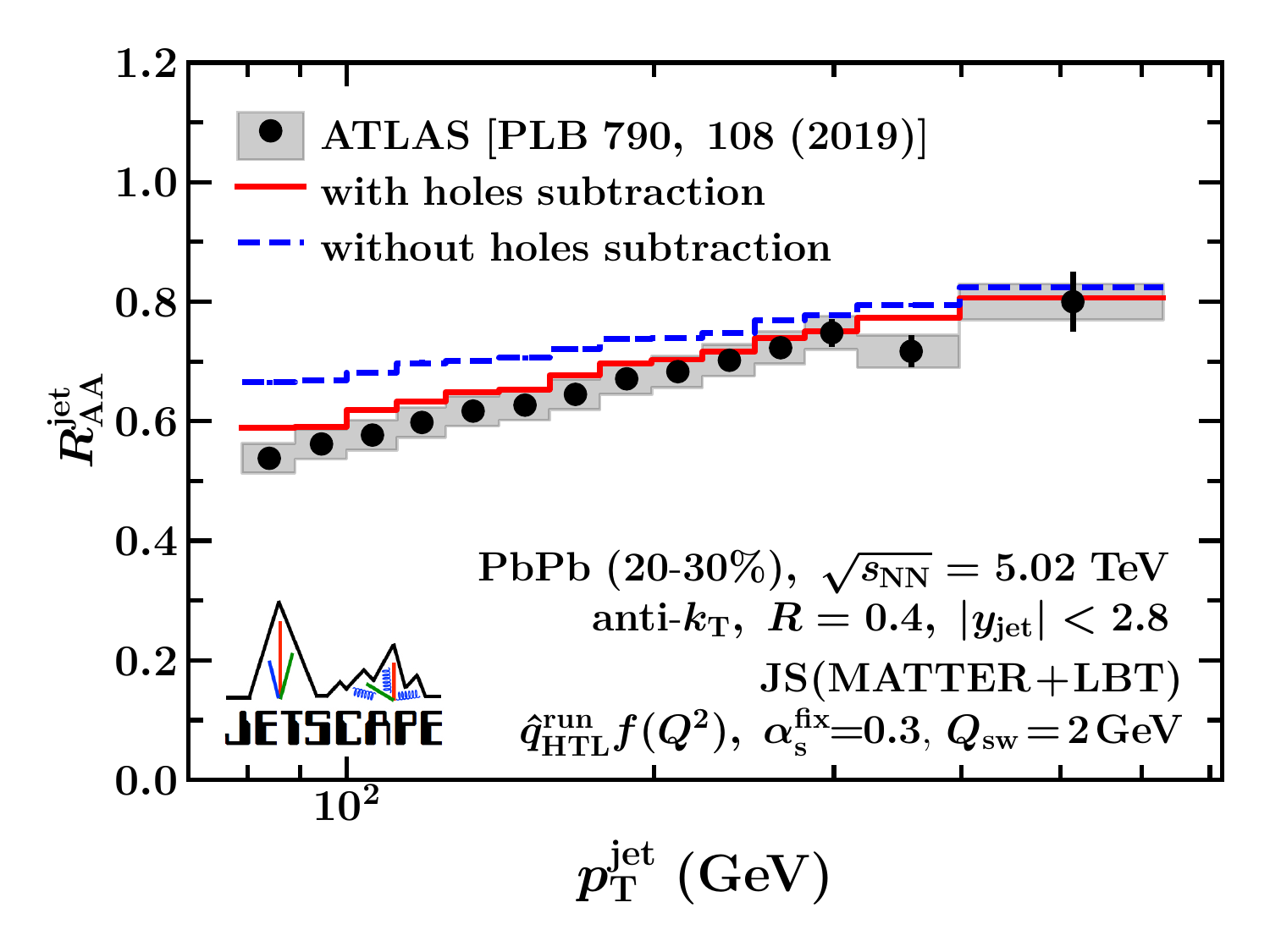}
\includegraphics[width=0.45\textwidth]{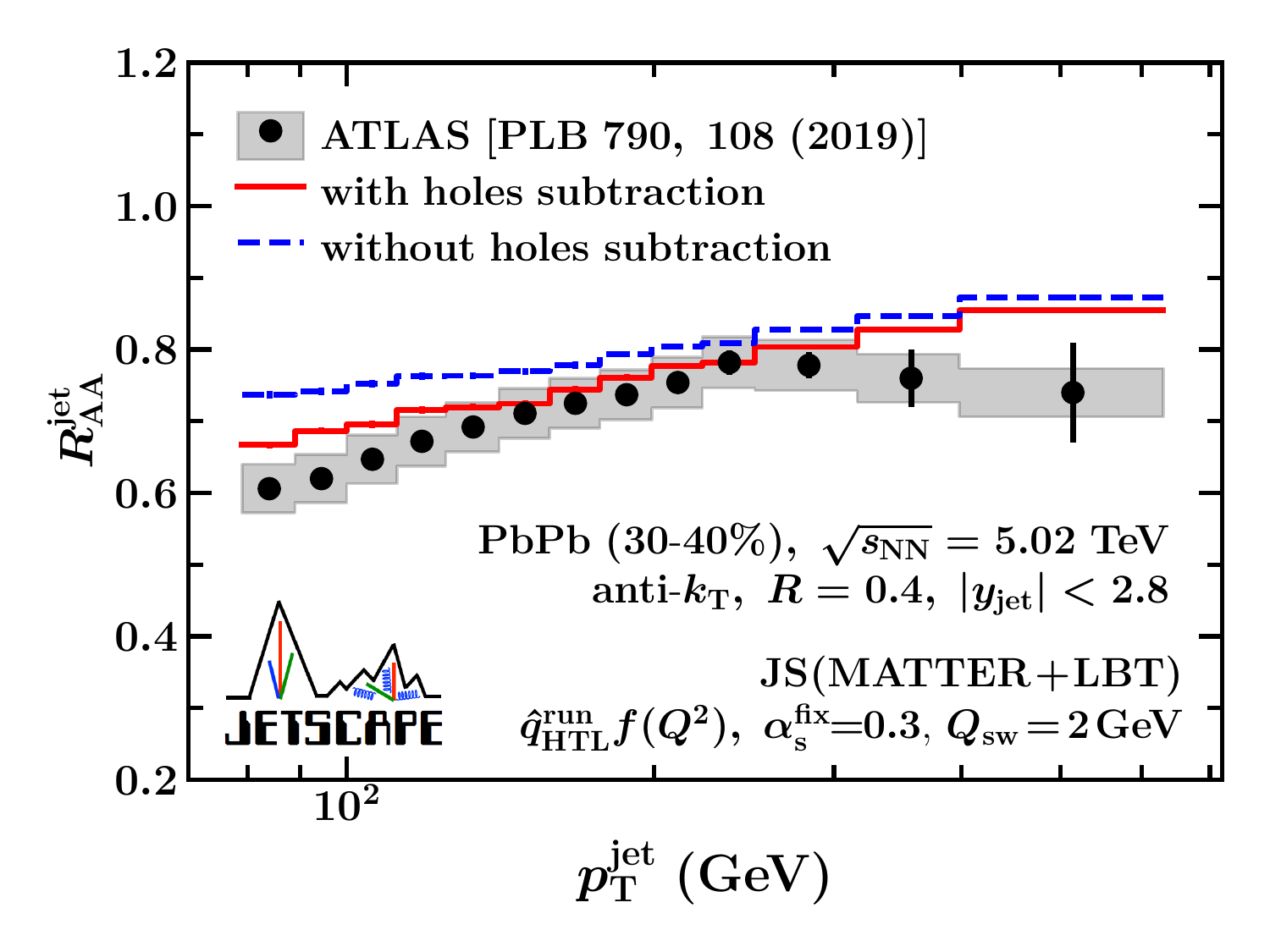}
\includegraphics[width=0.45\textwidth]{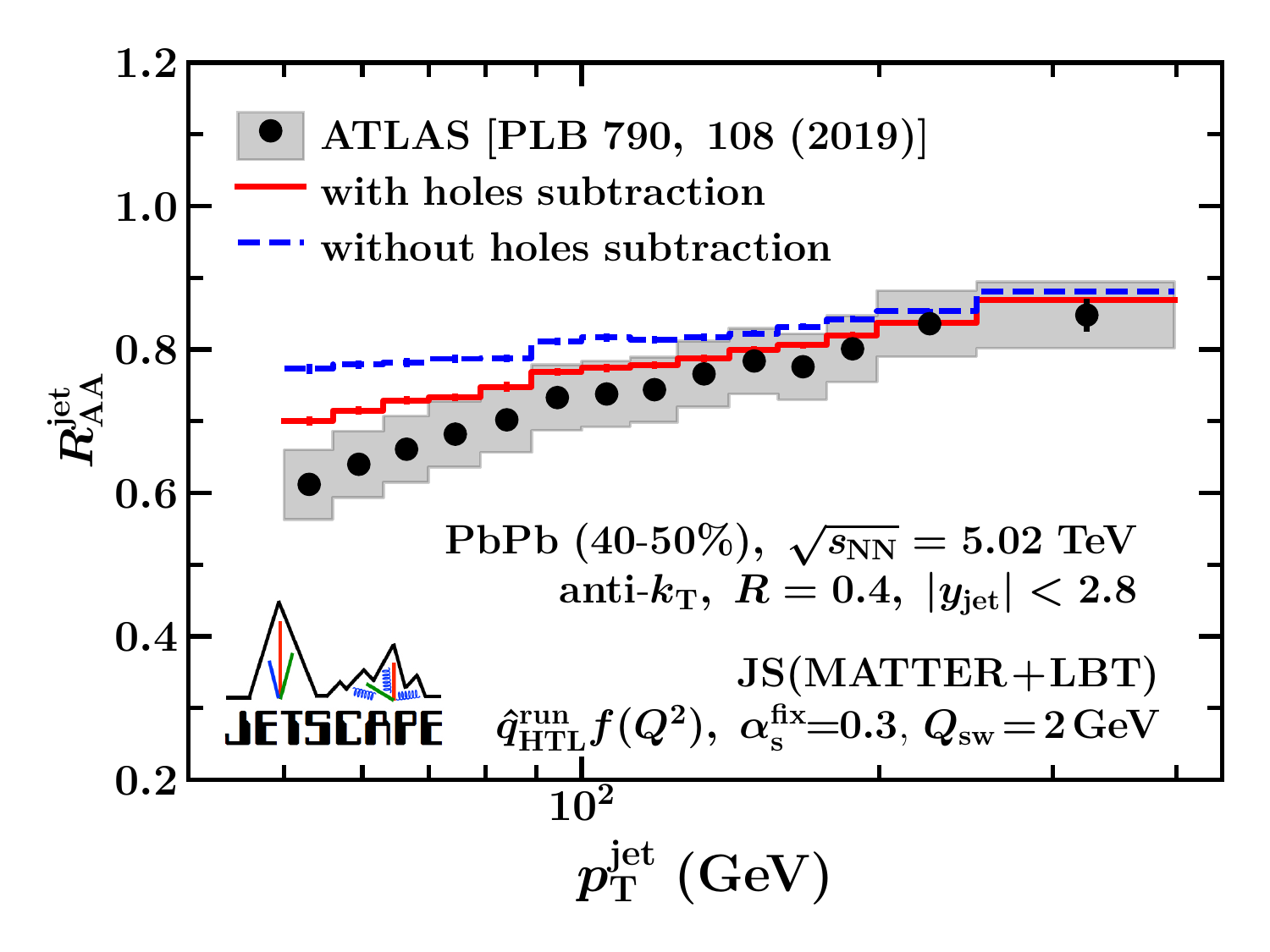}
\includegraphics[width=0.45\textwidth]{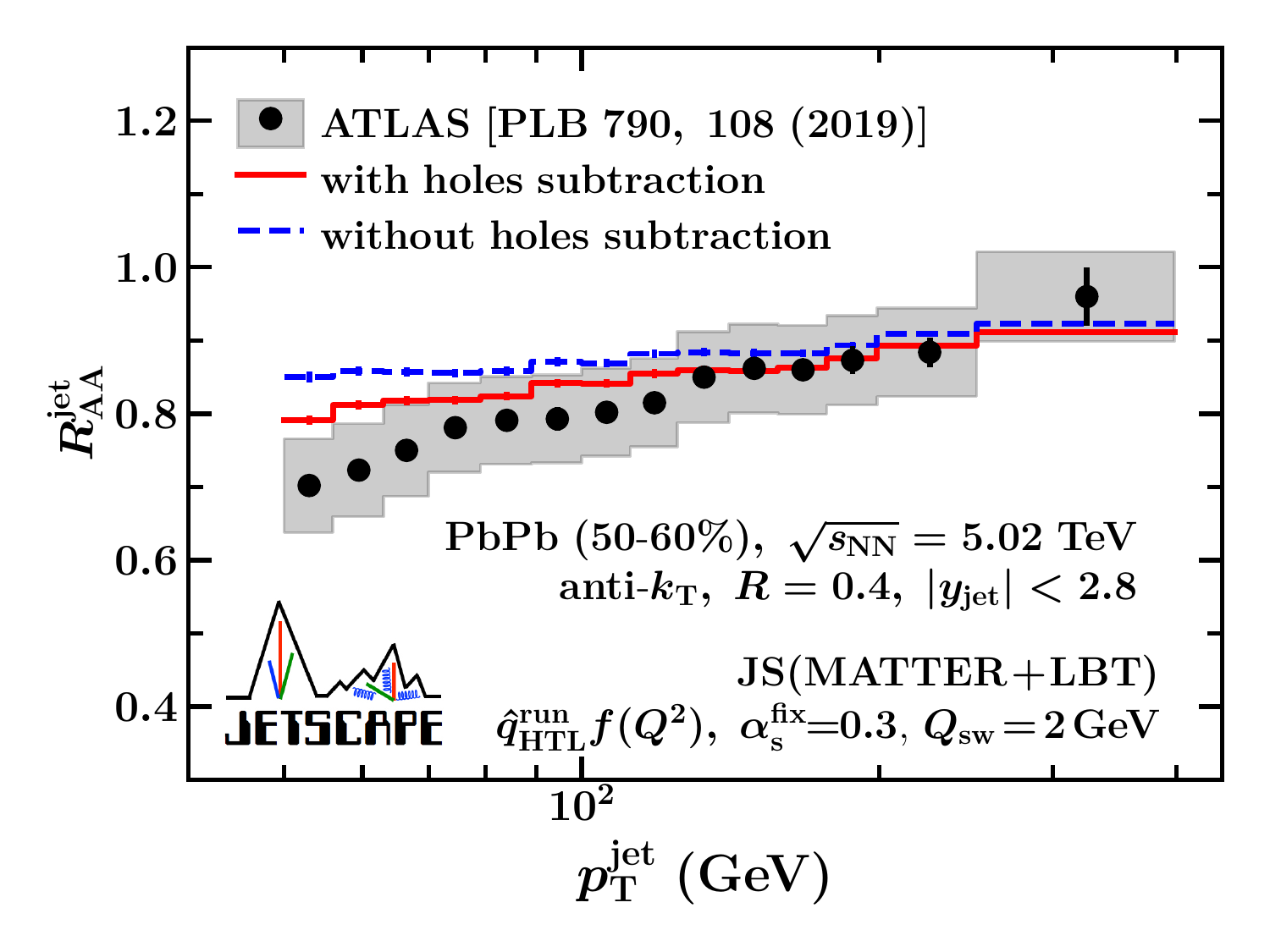}
\caption{Centrality dependence of inclusive jet $R_{\mathrm{AA}}$ 
with $R=0.4$ and $y_{\mathrm{jet}} < 2.8$ 
at $\sqrt{s_\mathrm{NN}}$=5.02~TeV. The calculation is performed using a multistage jet quenching model (\textsc{matter}+\textsc{lbt}). Jet transport coefficient multiplied by virtuality dependent factor [$\hat{q}^{\mathrm{run}}_{\mathrm{HTL}}f(Q^2)$] is used. The free parameters employed in the jet quenching model are extracted from simultaneous fit to inclusive jet $R_{\mathrm{AA}}$ and charged-particle $R_{\mathrm{AA}}$ at most central (0-10$\%$, $\sqrt{s_\mathrm{NN}}$=5.02~TeV) Pb+Pb  collisions (top left plot for jets) and no further re-tuning has been performed.
Also shown in the dashed blue lines, is the effect of not subtracting the holes. 
Results are compared to ATLAS data~\cite{ATLAS:2018gwx} (black circles) in all centrality cases and CMS data for $\eta_{\mathrm{jet}} < 2.0$~\cite{CMS:2021vui} (magenta square) in only the 0-10$\%$ case.
}
\label{fig:semi_peripheral_collisions_5TeV_jets}
\end{figure*}

\begin{figure*}[htbp]
\centering
\includegraphics[width=0.49\textwidth]{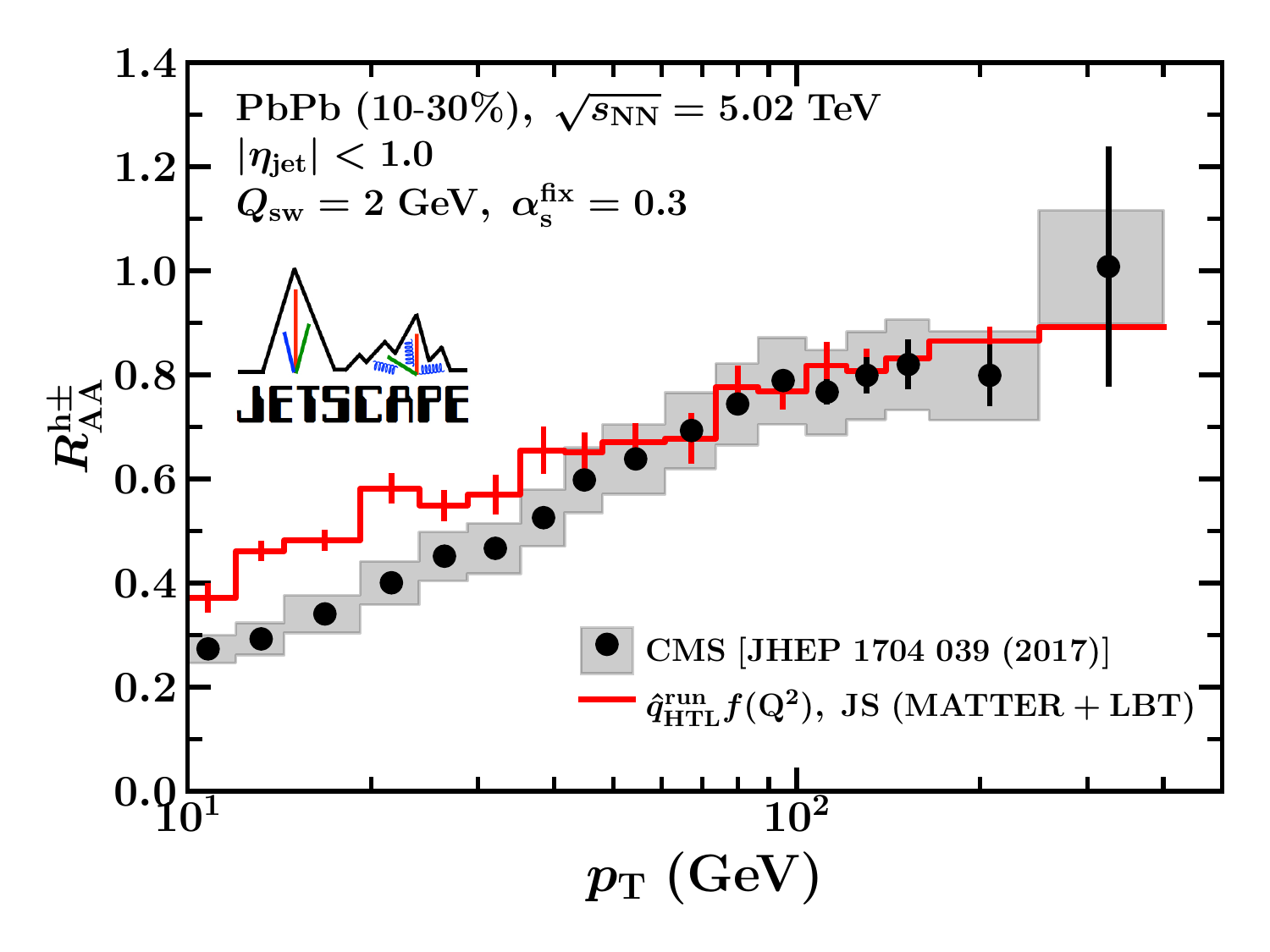}
\includegraphics[width=0.49\textwidth]{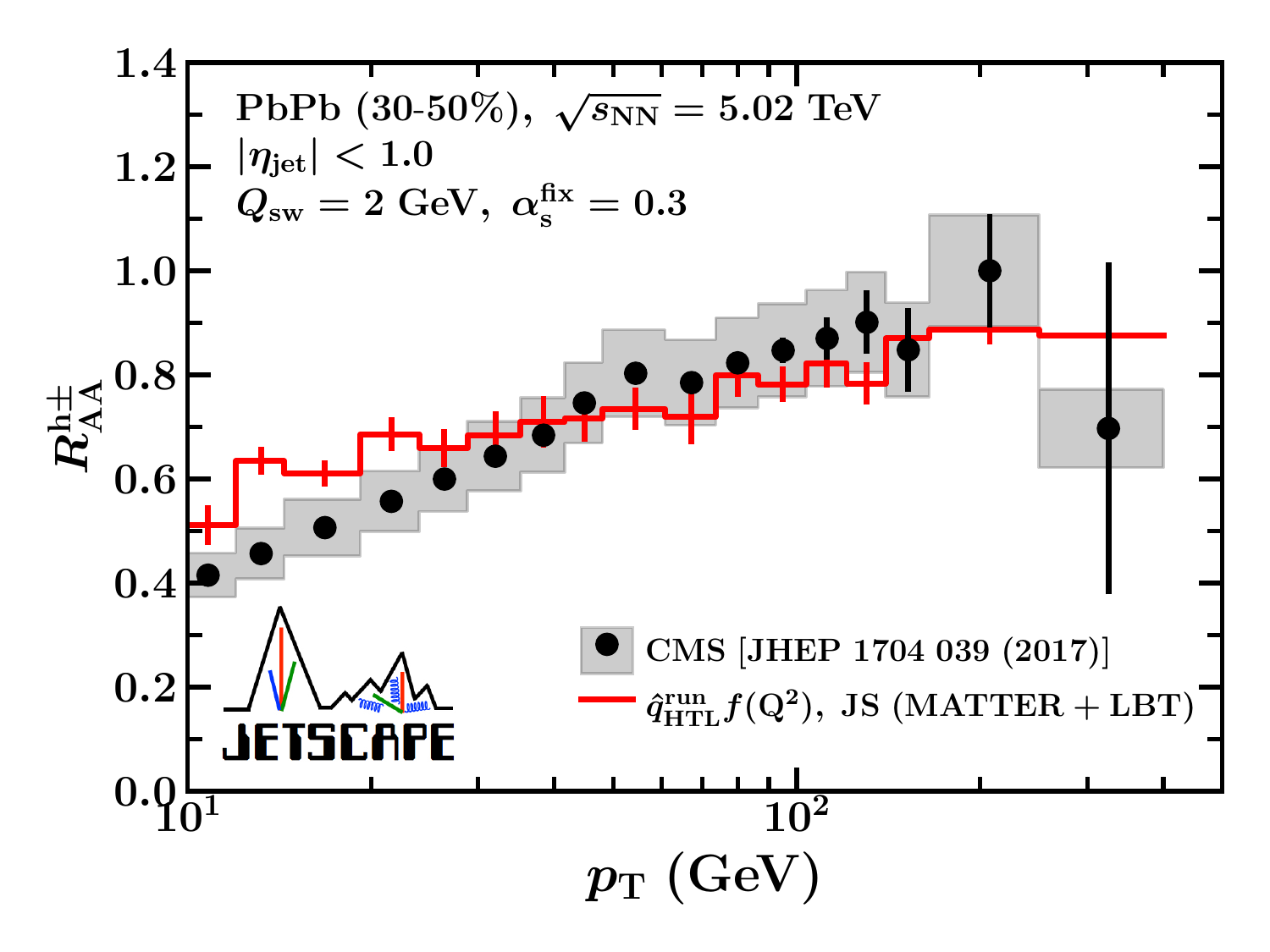}
\caption{Centrality dependence of inclusive charged hadron $R_{\mathrm{AA}}$ for $\eta<1.0$ at $\sqrt{s_\mathrm{NN}}$=5.02~TeV. The calculation is performed using a multistage jet quenching model (\textsc{matter}+\textsc{lbt}). Jet transport coefficient multiplied by virtuality dependent factor [$\hat{q}^{\mathrm{run}}_{\mathrm{HTL}}f(Q^2)$] is used. The free parameters employed in the jet quenching model are extracted from simultaneous fit to inclusive jet $R_{\mathrm{AA}}$ and charged-particle $R_{\mathrm{AA}}$ at most central (0-10$\%$, $\sqrt{s_\mathrm{NN}}$=5.02~TeV) Pb+Pb  collisions and no further re-tuning has been performed. 
Results are compared to CMS data~\cite{CMS:2016xef} for both centralities. 
}
\label{fig:semi_peripheral_collisions_5TeV_hadrons}
\end{figure*}

Figure~\ref{fig:semi_peripheral_collisions_5TeV_jets} shows 
our inclusive-jet $R_{\mathrm{AA}}$ results 
for different centrality classes at $\sqrt{s_\mathrm{NN}}$=5.02~TeV, 
compared with the experimental data from ATLAS~\cite{ATLAS:2018gwx}. 
Our full results (solid red lines) are typically consistent within the uncertainties of the ATLAS data 
for centralities 20-30$\%$, 40-50$\%$, and 50-60$\%$ for all jet $p_{\mathrm{T}}$, while we observe a deviation of about  $5\%$ in highest jet $p_{\mathrm{T}}$ bin at 400-600~GeV for centrality bins 10-20$\%$ and 30-40$\%$. 
For the case of charged-particle $R_{\mathrm{AA}}$ shown in  Fig.~\ref{fig:semi_peripheral_collisions_5TeV_hadrons}, our multistage jet quenching model can describe CMS data~\cite{CMS:2016xef} in the high-$p_{\mathrm{T}}$ region very well with a deviation $~10\%$ in the low-$p_{\mathrm{T}}$ region.

The deviation from the data in charged-particle $R_{\mathrm{AA}}$ at $p_{\mathrm{T}}<30$~GeV, for more peripheral events, is mainly due to the absence of the jet energy loss in the hadronic phase. As one moves away from central collisions, the expectation is that the hadronic phase will be a larger fraction of the entire system and hence play a non-negligible role in the quenching of jets in semi-peripheral and peripheral collisions.
We emphasize that the calculation presented here employed the same free parameters as the calculation for most-central (0-10$\%$) collisions at $\sqrt{s_\mathrm{NN}}$=5.02~TeV, and no further re-tuning of the free parameters has been performed.

\subsection{Effects of medium response and hole subtraction on inclusive jets}
In this section, we highlight the importance of the recoil-hole formalism and demonstrate their effect on inclusive jet $R_{\mathrm{AA}}$ at $\sqrt{s_\mathrm{NN}}$=5.02~TeV.
We use the optimized value of the free parameters listed in Tab.~\ref{tab:aa_parameter_set} and present the effect of hole subtraction in inclusive jets at different centralities (dashed blue lines) in Fig.~\ref{fig:semi_peripheral_collisions_5TeV_jets}. The \textsc{jetscape} results are obtained using the multistage jet energy loss (\textsc{matter}+\textsc{lbt}), where we employ a virtuality dependent factor $f(Q^2)$ that modulates the effective value of $\hat{q}$ in the high virtuality (\textsc{matter}) stage to account for a reduction in the medium induced gluon radiation rate due to coherence effects [Eq.~\eqref{eq:q2-dep-qhat}].

We remind the reader that the hole hadrons are the product of hadronization of the thermal partons sampled from the medium during the jet-medium interaction. The \textsc{jetscape} framework keeps track of such partons, which are then used to determine the background correlated to the jet. For inclusive jets, we subtract the hole hadrons according to the criteria discussed in Eq.~(\ref{eq:neg_sub}). 
 
The effects of holes subtraction are roughly $\lesssim 5 \%$ at jet $p_{\mathrm{T}}>400$~GeV and becomes gradually larger as one goes to lower value of jet $p_{\mathrm{T}}$.
The calculation demonstrates that, for inclusive jets, hole subtraction correctly reproduces the jet $p_{\mathrm{T}}$ dependence for all centralities, except the deviation of roughly $5\%$ at the highest jet $p_{\mathrm{T}}$ bins for centrality 10-20$\%$ and 30-40$\%$. 
The deviation from the data at low jet $p_{\mathrm{T}}$ can be attributed to the fact that we do not have jet energy loss in the hadronic phase.

\subsection{Nuclear modification of jets and leading hadron at $\sqrt{s_\mathrm{NN}}=2.76$~TeV and $\sqrt{s_\mathrm{NN}}=200$~GeV }

In this section, we present the comparisons for nuclear modification factor for inclusive jets and charged-particle $R_{\mathrm{AA}}$ at lower collision energies and demonstrate that the multistage jet quenching model, with a recoil-hole formalism and a virtuality dependent factor $f(Q^2)$ that modulates the effective value of $\hat{q}$ in the high virtuality phase, captures the essential aspects of the parton energy loss in the QGP.

\begin{figure}[htbp]
\centering
\includegraphics[width=0.45\textwidth]{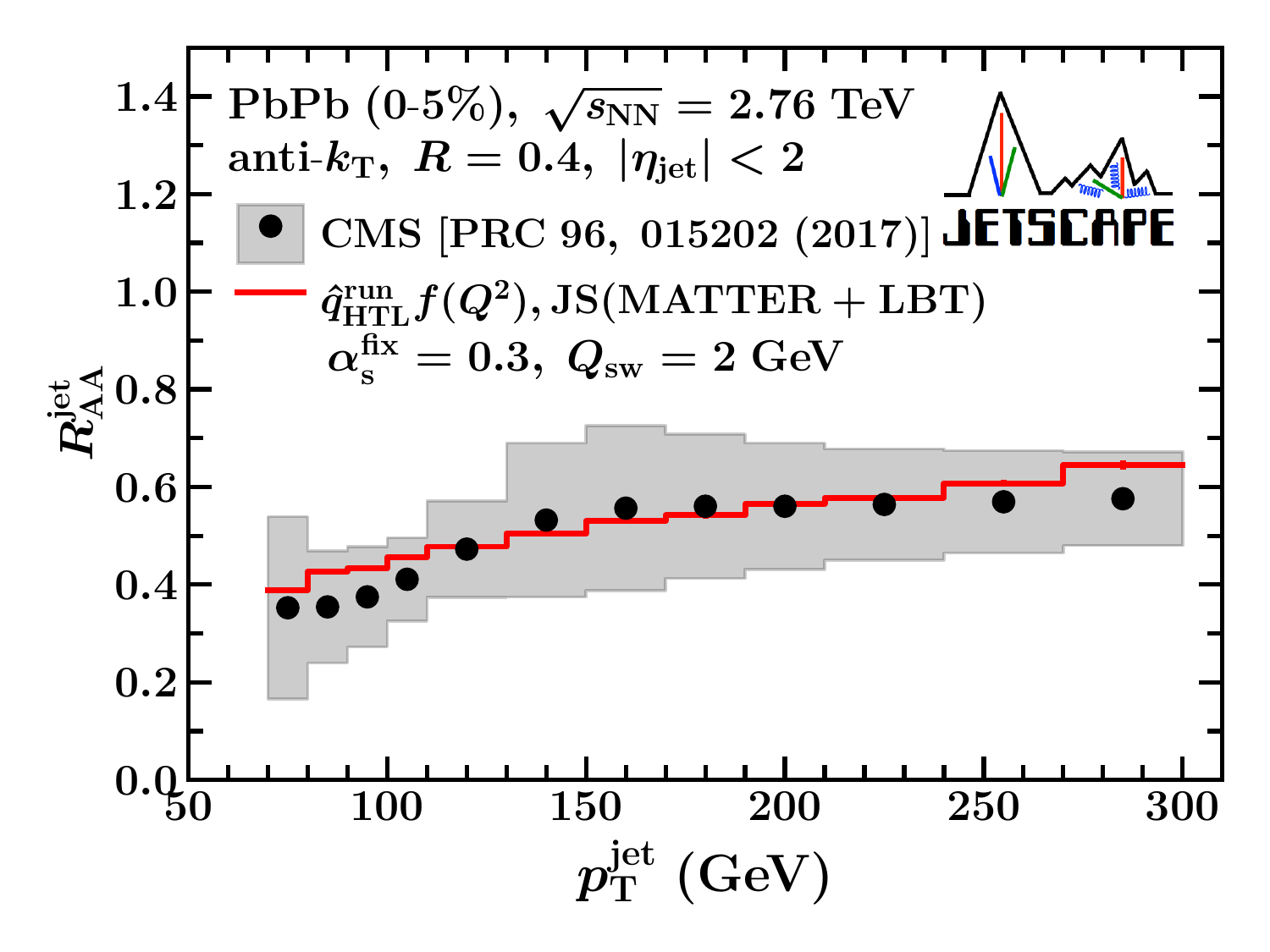}
\includegraphics[width=0.45\textwidth]{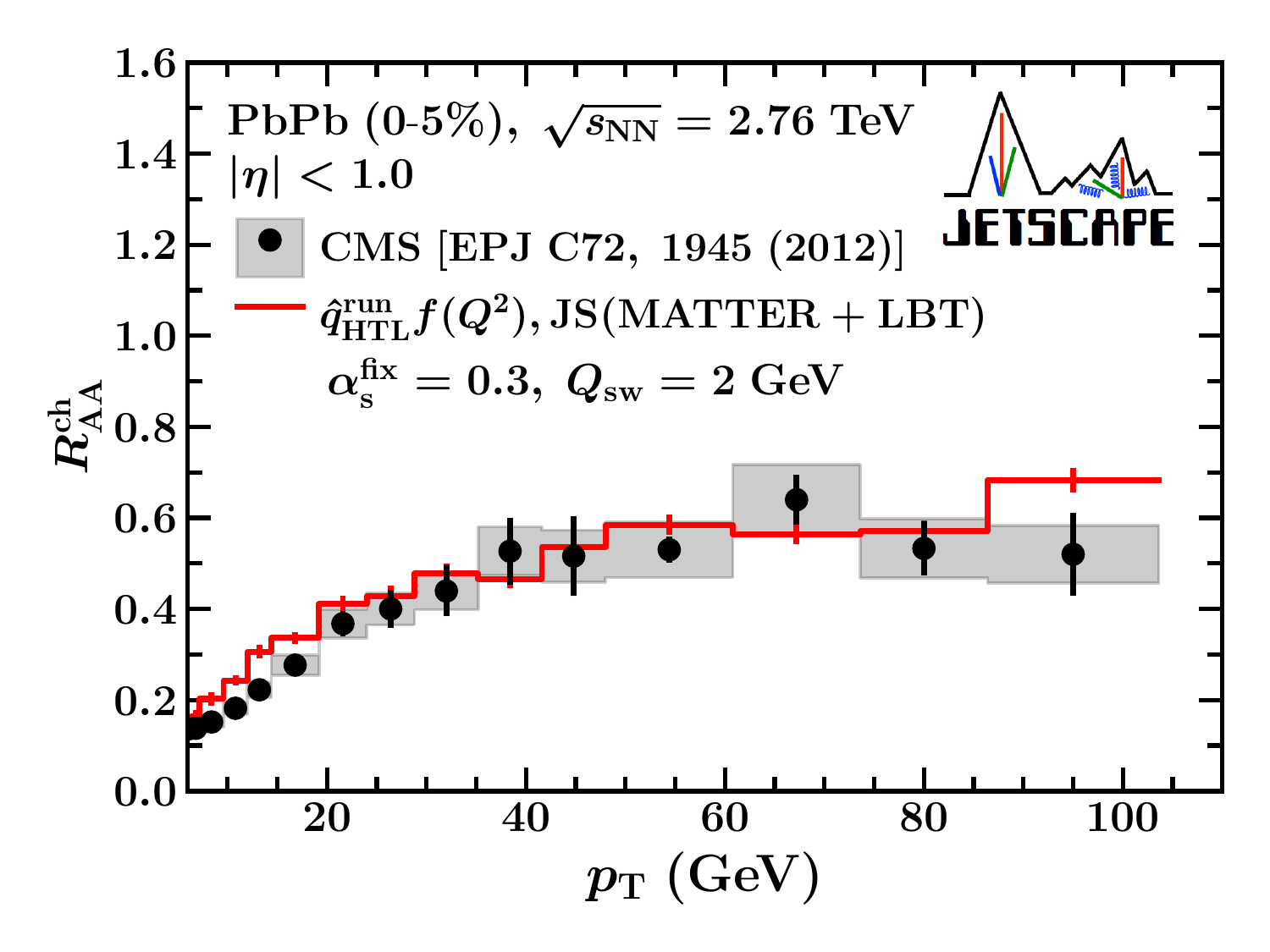}
\caption{The inclusive jet $R_{\mathrm{AA}}$ and charged-particle $R_{\mathrm{AA}}$ at most central (0-5$\%$) Pb+Pb collisions at $\sqrt{s_\mathrm{NN}}$=2.76~TeV. The calculation is performed using the multistage jet quenching model (\textsc{matter}+\textsc{lbt}) with virtuality dependence (Type 3). The free parameters employed in the jet quenching model are extracted from simultaneous fit to inclusive jet $R_{\mathrm{AA}}$ and charged-particle $R_{\mathrm{AA}}$ at most central (0-10$\%$, $\sqrt{s_\mathrm{NN}}$=5.02~TeV) Pb+Pb  collisions and no further re-tuning has been performed.
Top panel: 
Results for inclusive jet $R_{\mathrm{AA}}$ 
with $R=0.4$ and $\eta_{\mathrm{jet}} < 2$, compared to CMS data~\cite{CMS:2016uxf}. 
Bottom panel: 
Results for inclusive charged-particle $R_{\mathrm{AA}}$ with $\eta<1.0$, compared to CMS data~\cite{CMS:2012aa}.
}
\label{fig:InclusiveJetRAA_Charged-ParticleRAA_276GeV}
\end{figure}

First, we present the inclusive jet $R_{\mathrm{AA}}$ for most central (0-10$\%$) PbPb collisions at $\sqrt{s_\mathrm{NN}}$=2.76~TeV in the top panel of Fig. \ref{fig:InclusiveJetRAA_Charged-ParticleRAA_276GeV}. The inclusive jets are constructed using anti-$k_{\mathrm{T}}$ algorithm with cone size $R=0.4$ and $|\eta_\mathrm{jet}|<2$, and compared with CMS data~\cite{CMS:2016uxf}. 
The theory calculation shows a very good agreement with the experimental data for all jet $p_{\mathrm{T}}$. In the bottom panel of Fig.~\ref{fig:InclusiveJetRAA_Charged-ParticleRAA_276GeV}, we present the charged-particle $R_{\mathrm{AA}}$ for most central (0-5$\%$) collision at $\sqrt{s_\mathrm{NN}}$=2.76~TeV. The comparison of the theory calculation with the CMS data~\cite{CMS:2012aa} is quite remarkable. 

\begin{figure*}[htbp]
\centering
\includegraphics[width=0.45\textwidth]{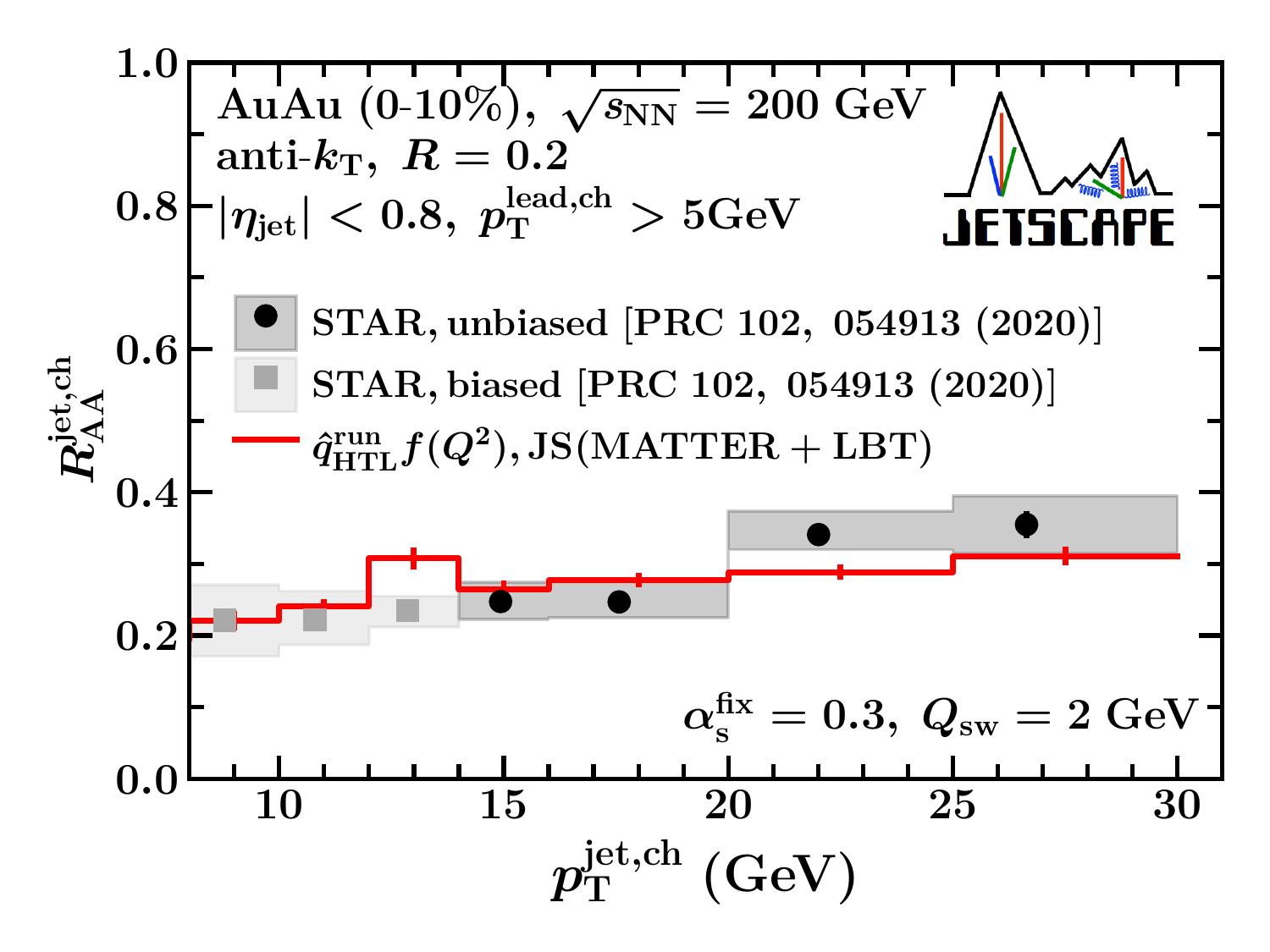}
\includegraphics[width=0.45\textwidth]{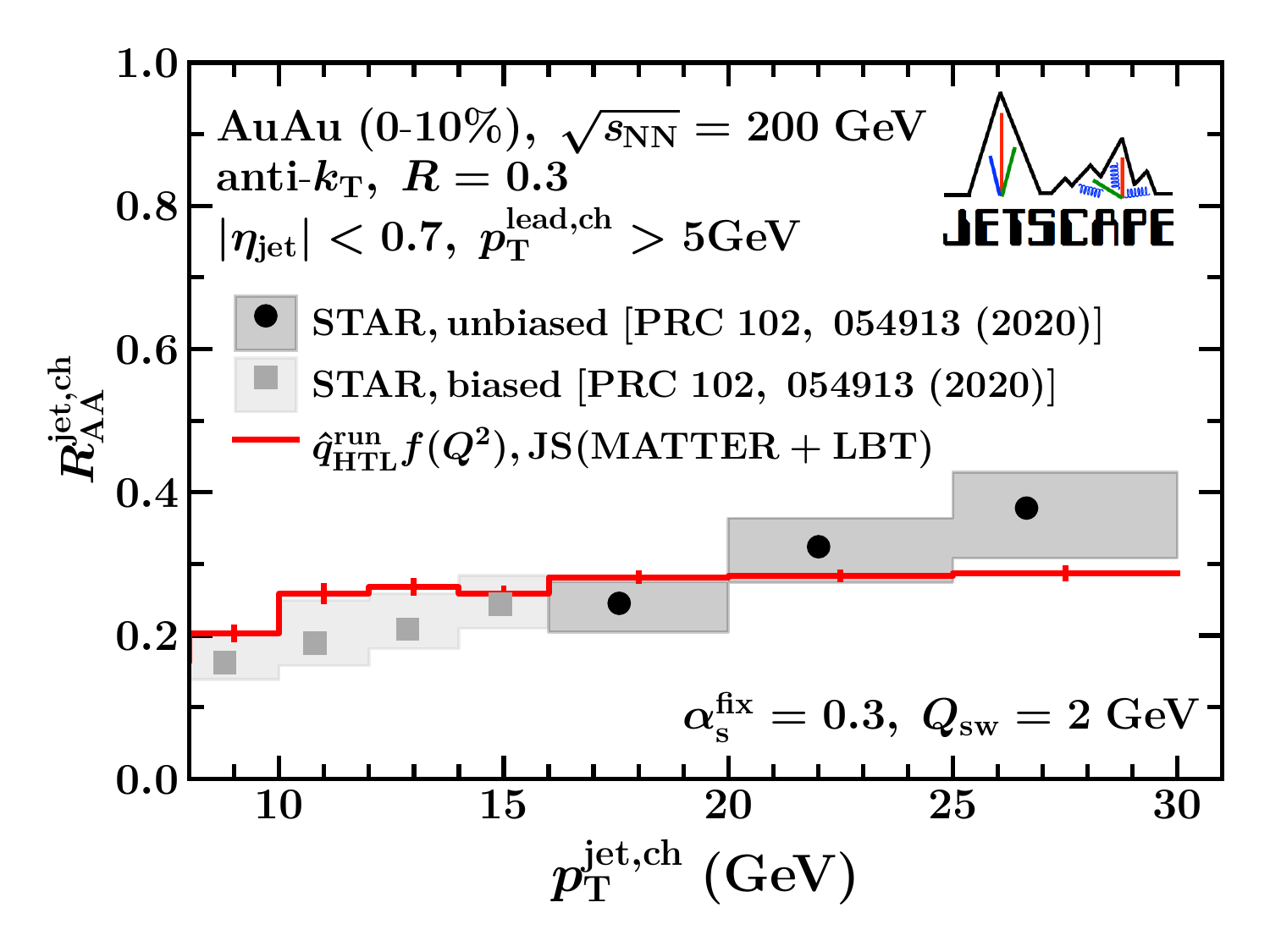}
\includegraphics[width=0.45\textwidth]{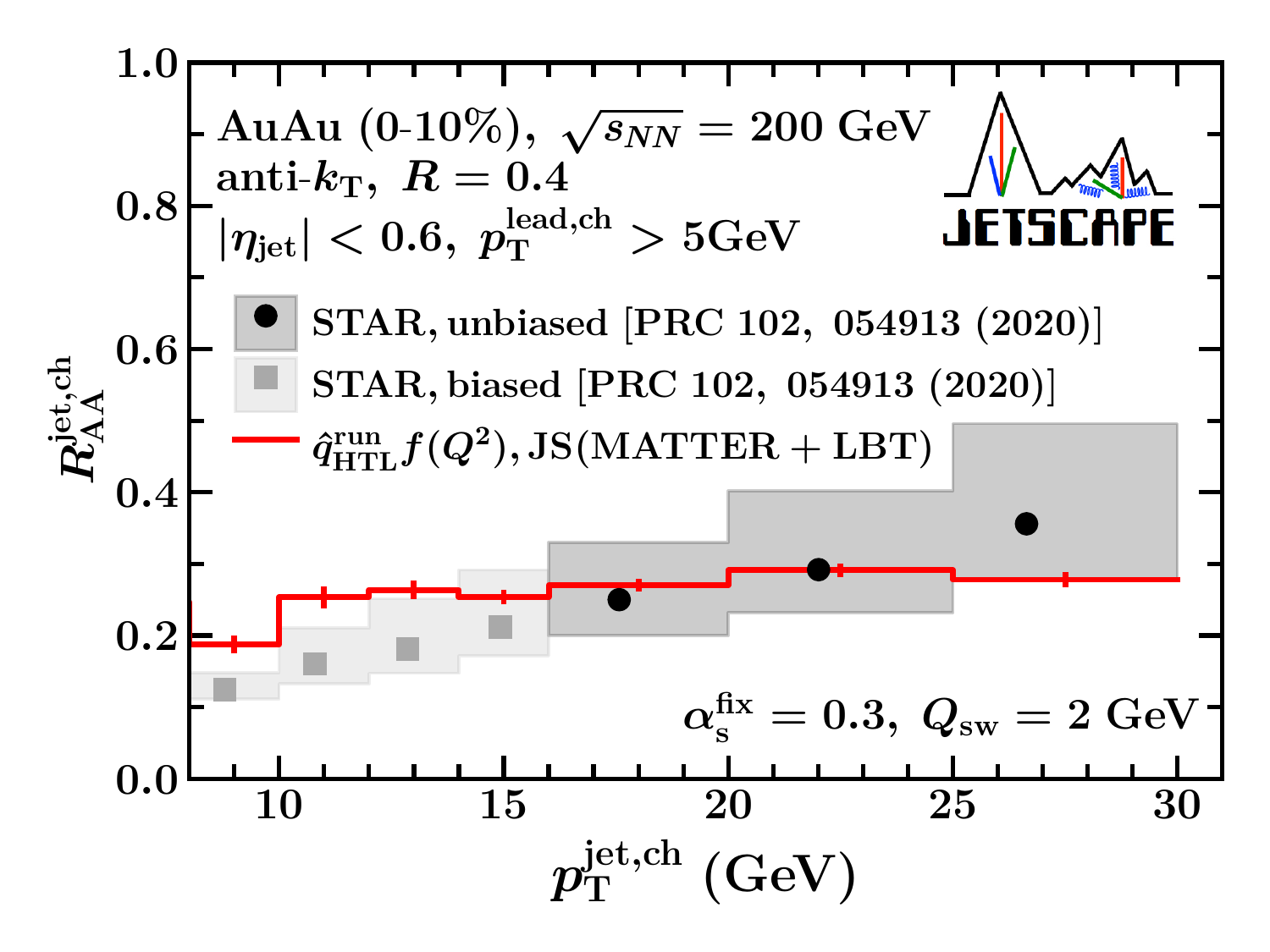}
\includegraphics[width=0.45\textwidth]{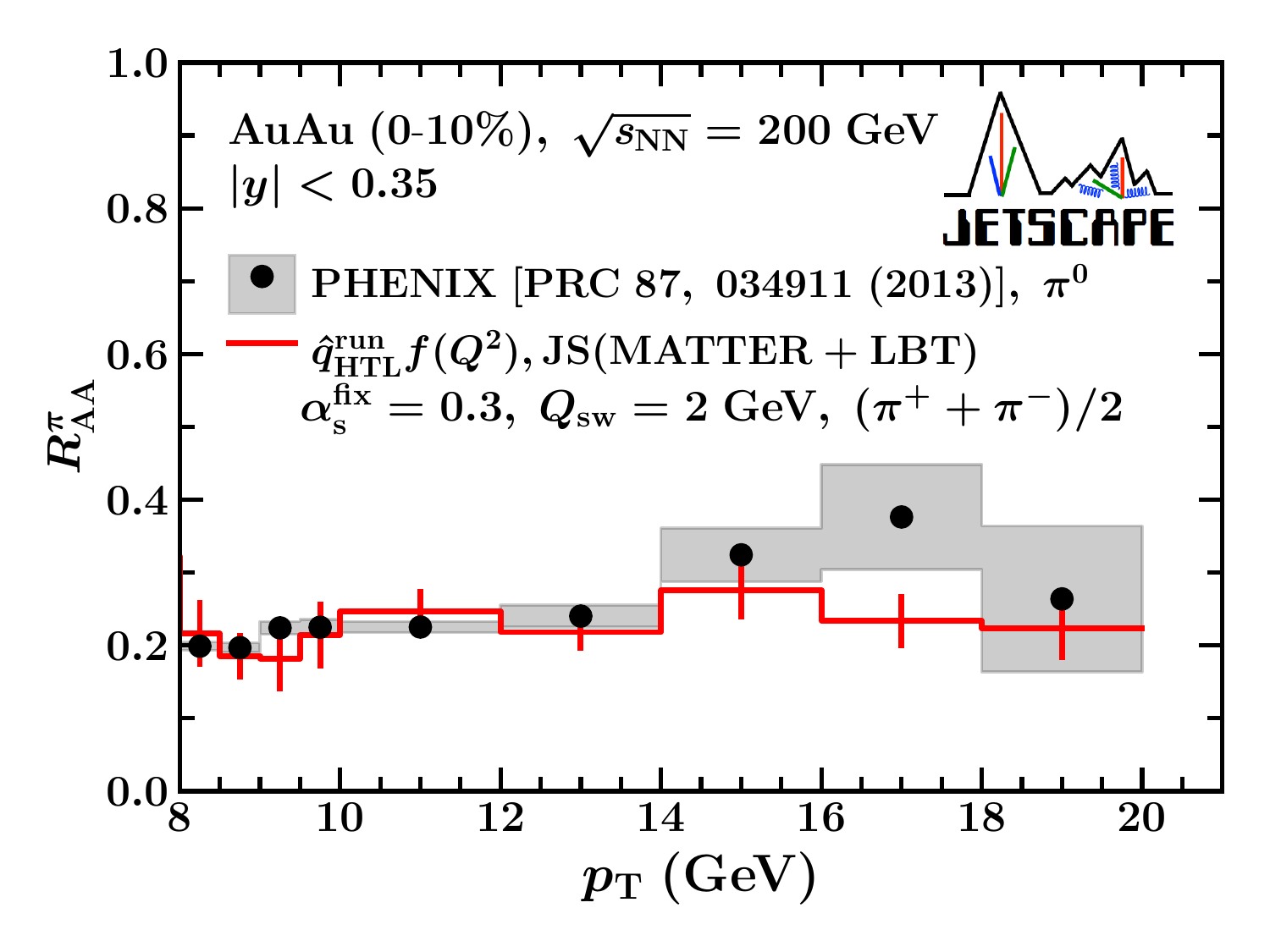}
\caption{The charged-particle jet $R_{\mathrm{AA}}$ and charged-pion $R_{\mathrm{AA}}$ at most central (0-10$\%$) Au+Au collisions at $\sqrt{s_\mathrm{NN}}$=200~GeV. 
The top two-panels and the bottom left panel are 
charged-jet $R_{\mathrm{AA}}$ for cone sizes 
$R$=0.2 ($\eta_{\mathrm{jet}}<0.8$), 
0.3 ($\eta_{\mathrm{jet}}<0.7$), 
and 0.4 ($\eta_{\mathrm{jet}}<0.6$), compared to STAR data~\cite{STAR:2020xiv}. 
The bottom right panel is charged-pion $R_{\mathrm{AA}}$ compared to 
PHENIX data of neutral pion $R_{\mathrm{AA}}$~\cite{PHENIX:2012jha}. 
The calculation is performed using the multistage jet quenching model (\textsc{matter}+\textsc{lbt}) with virtuality dependence (Type 3). The free parameters employed in the jet quenching model are extracted from simultaneous fit to inclusive jet $R_{\mathrm{AA}}$ and charged-particle $R_{\mathrm{AA}}$ at most central (0-10$\%$, $\sqrt{s_\mathrm{NN}}$=5.02~TeV) Pb+Pb  collisions and no further re-tuning has been performed. 
}
\label{fig:ChargedJetRAA_Charged-PionRAA_200GeV}
\end{figure*}

Second, we present in Fig.~\ref{fig:ChargedJetRAA_Charged-PionRAA_200GeV} the comparisons for the charged-particle jet $R_{\mathrm{AA}}$ and charged-pion $R_{\mathrm{AA}}$ at RHIC collision energy $\sqrt{s_\mathrm{NN}}$=200~GeV. The charged-particle jets are constructed using anti-$k_{\mathrm{T}}$ algorithm for cone sizes $R=$0.2, 0.3 and 0.4 with kinematic cut $|\eta_\mathrm{jet}|<(1.0-R)$. 
We impose the leading charged-particle trigger bias $p^{\mathrm{lead,ch}}_{\mathrm{T}}>$ 5~GeV to select true jets in the same way as done in the experiment and compare the results with STAR data~\cite{STAR:2020xiv} at most central (0-10$\%$) Au+Au collisions. The charged-jet $R_{\mathrm{AA}}$ for all three jet cone sizes show a good agreement with the STAR measurements. 
In the bottom-right panel (Fig.~\ref{fig:ChargedJetRAA_Charged-PionRAA_200GeV}), we show comparisons of the charged-pion $R_{\mathrm{AA}}$ at most central (0-10$\%$) RHIC collision energy $\sqrt{s_\mathrm{NN}}$=200~GeV. The theoretical calculations are compared to PHENIX data~\cite{PHENIX:2012jha}, where we only compare with data available at top RHIC energy that covers hadron $p_{\mathrm{T}}$ up to 20~GeV. It should be pointed out that such comparisons are meaningful due to the iso-spin symmetry between charged pions and neutral pions. Here again, the theoretical calculations are in good agreement with the experimental data at all hadron $p_{\mathrm{T}}$.

\begin{figure*}[htbp]
\centering
\includegraphics[width=0.45\textwidth]{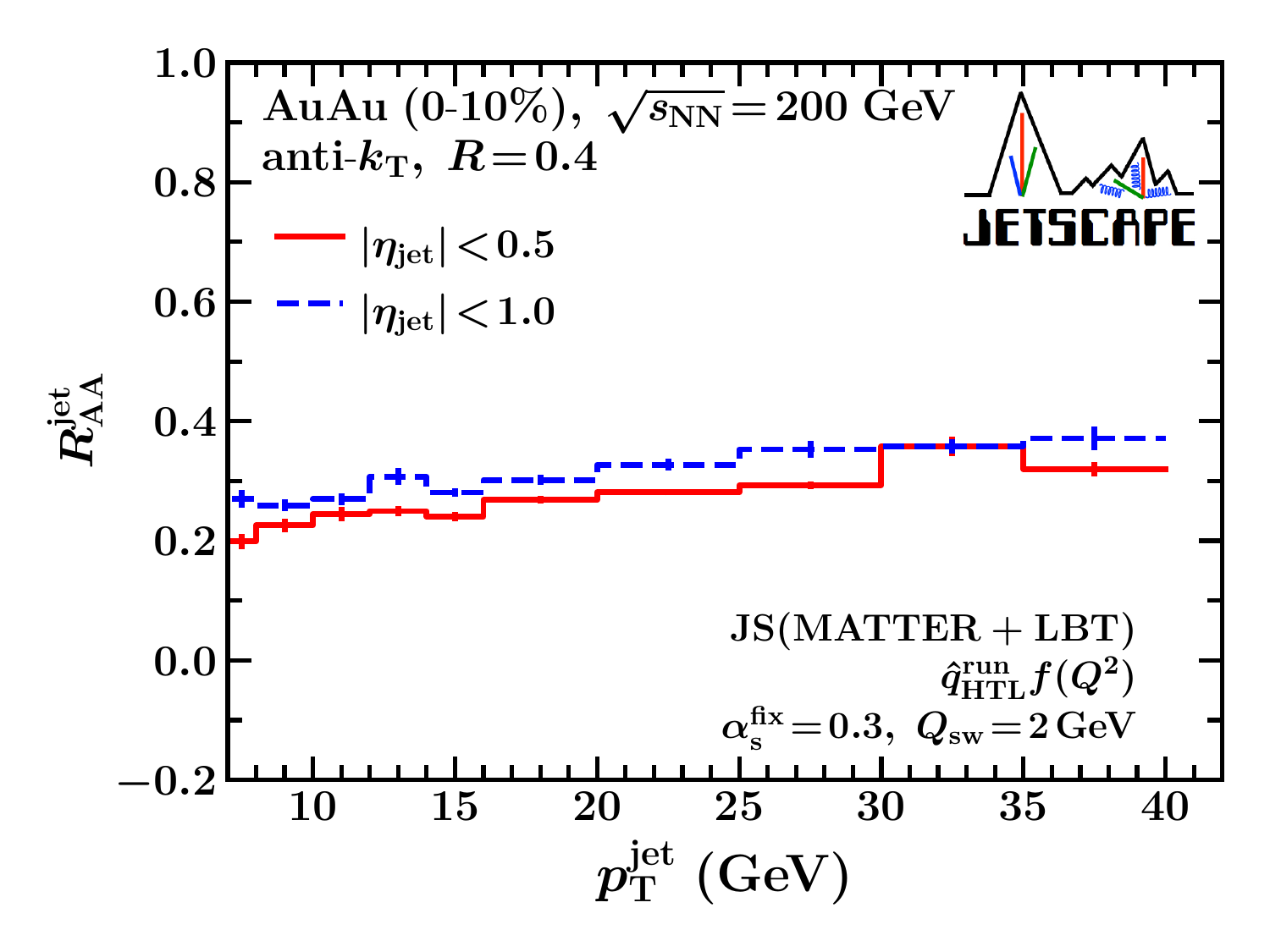}
\caption{Predictions for the nuclear modification factor for inclusive jets  for most central (0-10$\%$) Au+Au collisions at $\sqrt{s_\mathrm{NN}}$=200~GeV. The inclusive jets are constructed with the cone size $R=$0.4 for two different kinematic cuts: $|\eta_{\mathrm{jet}}|<0.5$ and $|\eta_{\mathrm{jet}}|<1.0$. The calculation is performed using the multistage jet quenching model (\textsc{matter}+\textsc{lbt}) with virtuality dependence (Type 3). The free parameters employed in the jet quenching model are extracted from simultaneous fit to inclusive jet $R_{\mathrm{AA}}$ and charged-particle $R_{\mathrm{AA}}$ at most central (0-10$\%$, $\sqrt{s_\mathrm{NN}}$=5.02~TeV) Pb+Pb  collisions.}
\label{fig:InclusiveJetRAA_200GeV}
\end{figure*}

Next, we present the prediction of inclusive jet $R_{\mathrm{AA}}$ in Fig.~\ref{fig:InclusiveJetRAA_200GeV} for most-central (0-10$\%$) Au+Au collisions at $\sqrt{s_\mathrm{NN}}=200$~GeV. 
The calculation is performed using the multistage jet quenching model (\textsc{matter}+\textsc{lbt}) with virtuality dependence (Type 3) for the default values of free parameters presented in Tab.~\ref{tab:aa_parameter_set}. We have shown inclusive jet $R_{\mathrm{AA}}$ with the jet cone size $R=$ 0.4 for kinematic cuts $|\eta_\mathrm{jet}|<0.5$ (solid red line) and $|\eta_\mathrm{jet}|<1.0$ (dashed blue line). The jets show significant suppression, but a weak jet $p_{\mathrm{T}}$ dependence. 

\section{Summary and Discussion}
\label{Section:Summary}
In this manuscript, we have presented a multistage [\textsc{matter}+\textsc{lbt}\,(recoils on)\,+\,\textsc{colorless hadronization}] jet quenching model within the \textsc{jetscape} framework and demonstrated, for the first time, a simultaneous description of the nuclear modification factor for inclusive jets and single hadrons from the top RHIC to the top LHC collision energies. 
We covered three collision energies $\sqrt{s_{\rm NN}}$=5.02~TeV, 2.76~TeV, and 200~GeV and performed model-to-data comparison for selected data sets from ALICE, ATLAS, CMS, PHENIX, and STAR experiments.

Event-by-event bulk medium simulations, without jets, were carried out first and calibrated to data \cite{Bernhard:2019bmu}. Binary collision profiles extracted from these individual simulations are sampled to yield locations of hard scattering. The \textsc{pythia} generator with ISR and MPI turned on is used to simulate hard scatterings that produce final state partons, without any final state shower. These are transferred to the \textsc{matter} generator, where they are imbued with a timelike virtuality $Q$, which depends on their transverse momentum $p_{\mathrm{T}}$.

To incorporate the multiscale dynamics of jet energy loss within the \textsc{jetscape} framework an effective parton evolution was set up, in which we encoded the space-time profile of the QGP, obtained from the bulk simulations, within the parton energy loss process. The initial high virtuality stage is modeled by the \textsc{matter} event generator, followed by the low virtuality stage, modeled by the \textsc{lbt} event generator. The switching of jet energy loss stage from high virtuality to low virtuality is carried out on a parton-by-parton level, depending on the virtuality of the parton: Those with $Q>Q_{\rm sw}$ remain in \textsc{matter}, while those partons, whose virtuality drops below $Q_{\rm sw}$, in the process of multiple emission, are transferred to \textsc{lbt}. Further medium-induced emission within \textsc{lbt} is assumed to maintain the virtuality at or below $Q_{\rm sw}$. Partons that escape the medium and still have virtuality larger than $Q_0 = 1$~GeV undergo vacuum evolution using the \textsc{matter} module. A weakly-coupled description of the medium in terms of thermal partons (recoils/holes) is used to include the medium response to the jet.

We systematically explored the three functional forms of the parameter dominating the jet-medium interaction strength. 
We demonstrated that 
the inclusion of a virtuality dependent factor [$f(Q^2)$] which modulates the effective value of $\hat{q}$ to account for a reduced medium-induced emission in the high virtuality phase, due to coherence effects~\cite{Armesto:2011ir,Kumar:2019uvu}, is essential for a simultaneous description of inclusive jet $R_\mathrm{AA}$ and charged-particle $R_\mathrm{AA}$ at LHC and RHIC collision energies. 

The success of this approach, in comparison with inclusive jet and hadron data, at all energies and centralities (at both RHIC and LHC), may imply that $\hat{q}/T^3$ does \emph{not} have a cusplike behavior at 300~MeV$<T<$ 400~MeV. Moreover, the effective reduction in $\hat{q}/T^3$, at LHC collision energies compared to RHIC, is mainly due to the fact that the energy of the jets produced at the LHC is an order of magnitude higher, compared to those produced at RHIC energies. The jets emanating from the hard parton, at LHC collision energies, start out with significantly higher virtuality, and hence, experience a significantly smaller stimulated emission rate, compared to RHIC energies. Alternatively, as the virtuality of the hard parton increases, the transverse size of the dipole formed by the hard parton and the emitted gluon decreases, due to which the plasma starts to appear more dilute (one often states that the partons in the medium cannot resolve the different parts of the dipole for stimulated emission). 
While coherence effects have been included in the high virtuality phase, effects such as the rise in the effective $\hat{q}$ in radiative processes compared to purely scattering processes, at lower virtualities, have not~\cite{Liou:2013qya,Blaizot:2014bha}. However, these effects can be approximately applied after extraction of $\hat{q}$ has been made.

We explored A+A model parameters: $\alpha^\mathrm{fix}_\mathrm{s}$, $Q_\mathrm{sw}$, $\tau_{0}$, and $T_\mathrm{c}$, and studied their effects on inclusive jets and charged-particle $R_\mathrm{AA}$. These explorations were carried out using a virtuality dependent factor $f(Q^2)$ effectively modulating $\hat{q}$ in the \textsc{matter} phase, as this gives the best simultaneous description of the inclusive jet and charged-particle $R_\mathrm{AA}$ data. The study suggests that the  $\tau_{0}$ parameter does not seem to have much effect on inclusive jets and charged-particle spectra, indicating that the medium effects on the jet energy loss may be highly suppressed during the early stage. The jet interaction termination temperature $T_{\mathrm{c}}$ seems to have a noticeable effect on both jet and hadron spectra, particularly at low $p_{\mathrm{T}}$. The neglect of jet energy loss in the hadronic phase is the most probable reason for the reduced suppression for $p_{\mathrm{T}} < 30$\,GeV, particularly at more peripheral collisions, which have a proportionately larger hadronic phase. In spite of these minor offsets, the model proposed in this paper presents a good comparison with a wide swath of experimental data.

Most of the results presented in Sec.~\ref{Section:Results} included simulations with a modulating factor $f(Q^2)$ that reduces the effective $\hat{q}$ in the \textsc{matter} phase to account for the reduction in the medium induced radiation. We have also carried out similar simulations without this modulating factor. These results are presented in Appendix~\ref{type1_param} and Appendix~\ref{type2_param}. As the reader will note, neither of these formulations can describe the inclusive charged particle $R_{\mathrm{AA}}$ and the inclusive jet $R_{\mathrm{AA}}$ simultaneously. The emerging picture from these simulations is that the modification of jets is dominated by the migration of the softer wider-angle components beyond the jet cone, with only minor modifications of the core region. Thus a considerable portion of jet energy loss is possible in the nonperturbative region. This is discussed in Appendix~\ref{parton_jet}, and will be further explored in the upcoming effort on jet medium correlations. Along with these studies, other efforts will focus on the heavy-quark sector, jet substructure, and azimuthal anisotropies. 

Our simulations have increased the number of parameters typically invoked in jet simulations. Some of this increase is simply due to the increased sophistication of a multistage jet modification framework. There is currently no well-established theory for how to transition from the higher virtuality phase simulated by \textsc{matter} to a lower virtuality phase simulated by \textsc{lbt}. While we have used an energy scale $Q_{\rm sw}$, in an effort to get a direct handle on the average scale of the transition, a more apt method may use $Q_{\rm sw} = C \sqrt{2 \hat{q} E }$. 
The case for $C=1$ was already explored in Ref.~\cite{JETSCAPE:2017eso}, however, no comparison to data was carried out. 
The parameters in $f(Q^2)$ are a fit to the form derived in Ref.~\cite{Kumar:2019uvu}, however, this could also be further parametrized as more observables are included in a more global fit. We point out that another user of the \textsc{jetscape} framework may easily choose to replace either one or both of the energy loss modules that we used with any number of their own modules, with different criteria for transition. This may lead to a different collection of parameters. 

More parameters are expected to arise in the introduction and simulation of the nonperturbative wake of the jet~\cite{Tachibana:2020mtb}, which is required for the simulation of the jet medium interactions and jets at large angles~\cite{CMS:2021vui}. The physics of soft energy-momentum spreading away from a jet is inherently nonperturbative, and thus, the inclusion of new parameters is unavoidable. The current effort evades the need for many of these parameters by limiting its scope to only jets and single hadrons. As each new tranche of parameters is introduced, a new Bayesian analysis should be carried out to constrain and correlate these against the existing parameters.

\acknowledgments
\label{Ack}

This work was supported in part by the National Science Foundation (NSF) within the framework of the JETSCAPE collaboration, under grant number OAC-2004571 (CSSI:X-SCAPE). It was also supported under ACI-1550172 (Y.C. and G.R.), ACI-1550221 (R.J.F., F.G., and M.K.), ACI-1550223 (D.E., U.H., L.D., and D.L.), ACI-1550225 (S.A.B., T.D., W.F., R.W.), ACI-1550228 (J.M., B.J., P.J., X.-N.W.), and ACI-1550300 (S.C., A.K., A.M., C.N., A.S., J.P., L.S., C.Si., R.A.S. and G.V.); by PHY-1516590 and PHY-1812431 (R.J.F., F.G., M.K.), by PHY-2012922 (C.S.); it was supported in part by NSF CSSI grant number \rm{OAC-2004601} (BAND; D.L. and U.H.); it was supported in part by the US Department of Energy, Office of Science, Office of Nuclear Physics under grant numbers \rm{DE-AC02-05CH11231} (X.-N.W.), \rm{DE-FG02-00ER41132} (D.O), \rm{DE-AC52-07NA27344} (A.A., R.A.S.), \rm{DE-SC0013460} (S.C., A.K., A.M., C.S. and C.Si.), \rm{DE-SC0021969} (C.S.), \rm{DE-SC0004286} (L.D., D.E., U.H. and D. L.), \rm{DE-SC0012704} (B.S.), \rm{DE-FG02-92ER40713} (J.P.) and \rm{DE-FG02-05ER41367} (T.D., J.-F.P., D. S. and S.A.B.). The work was also supported in part by the National Science Foundation of China (NSFC) under grant numbers 11935007, 11861131009 and 11890714 (Y.H. and X.-N.W.), under grant numbers 12175122 and 2021-867 (S.C.), by the Natural Sciences and Engineering Research Council of Canada (C.G., M.H., S.J., and G.V.), by the Office of the Vice President for Research (OVPR) at Wayne State University (Y.T.), by JSPS KAKENHI Grant No.~22K14041 (Y.T.), by the S\~{a}o Paulo Research Foundation (FAPESP) under projects 2016/24029-6, 2017/05685-2 and 2018/24720-6 (A. L. and  M.L.), and by the University of California, Berkeley - Central China Normal University Collaboration Grant (W.K.). U.H. would like to acknowledge support by the Alexander von Humboldt Foundation through a Humboldt Research Award. C.S. acknowledges a DOE Office of Science Early Career Award. Allocation of super-computing resources (Project: PHY180035) were obtained in part through the Extreme Science and Engineering Discovery Environment (XSEDE), which is supported by National Science Foundation grant number ACI-1548562. Calculation were performed in part on Stampede2 compute nodes, generously funded by the National Science Foundation (NSF) through award ACI-1134872, within the Texas Advanced Computing Center (TACC) at the University of Texas at Austin \cite{TACC}, and in part on the Ohio Supercomputer \cite{OhioSupercomputerCenter1987} (Project PAS0254). Computations were also carried out on the Wayne State Grid funded by the Wayne State OVPR. The bulk medium simulations were done using resources provided by the Open Science Grid (OSG) \cite{Pordes:2007zzb, Sfiligoi:2009cct}, which is supported by the National Science Foundation award \#2030508. Data storage was provided in part by the OSIRIS project supported by the National Science Foundation under grant number OAC-1541335.

\appendix
\label{Appen}
\section{Parameter dependence for Type 1: HTL $\hat{q}$ with fixed coupling}
\label{type1_param}
\begin{figure}[htb]
\centering
\includegraphics[width=0.45\textwidth]{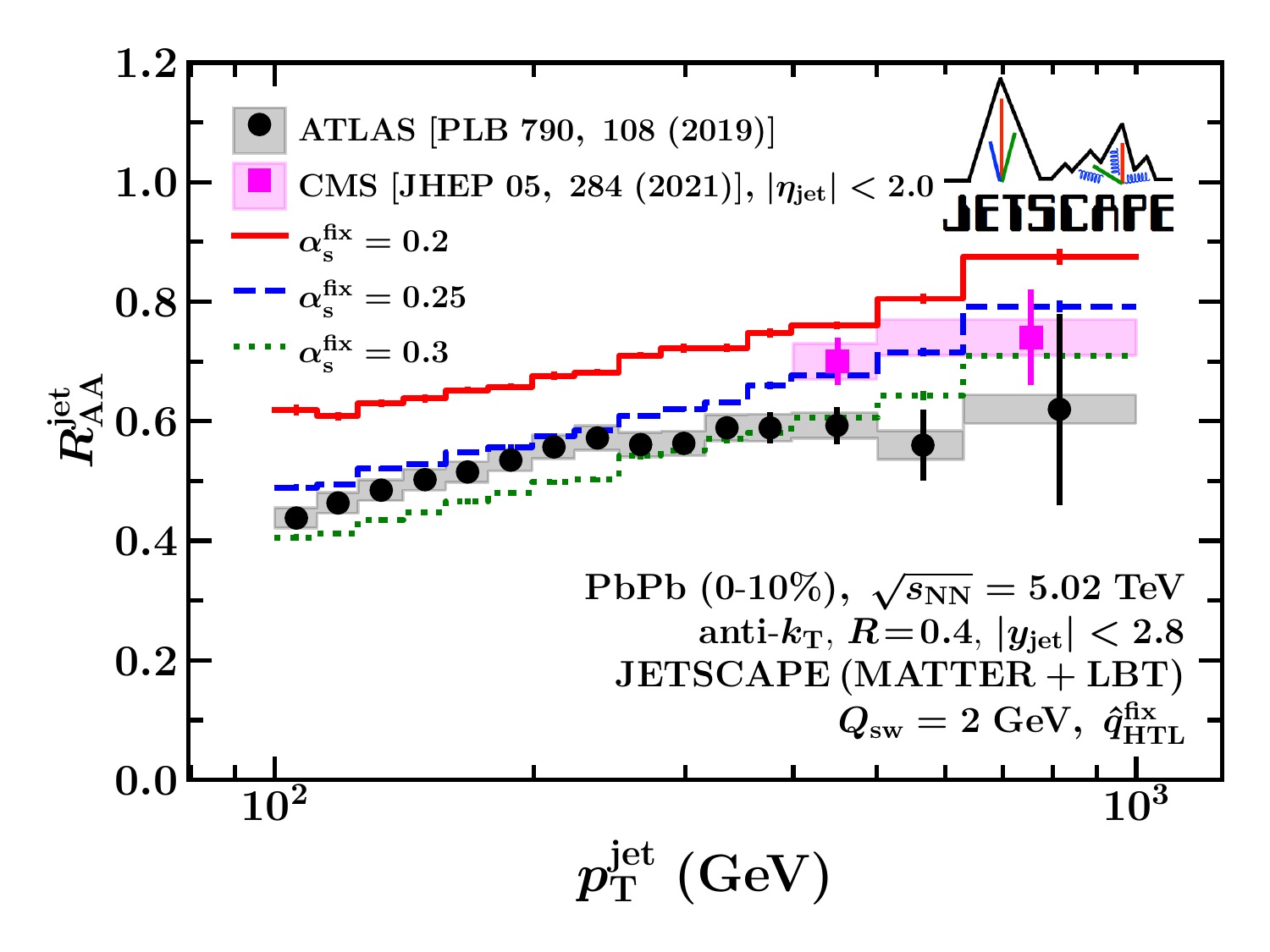}
\includegraphics[width=0.45\textwidth]{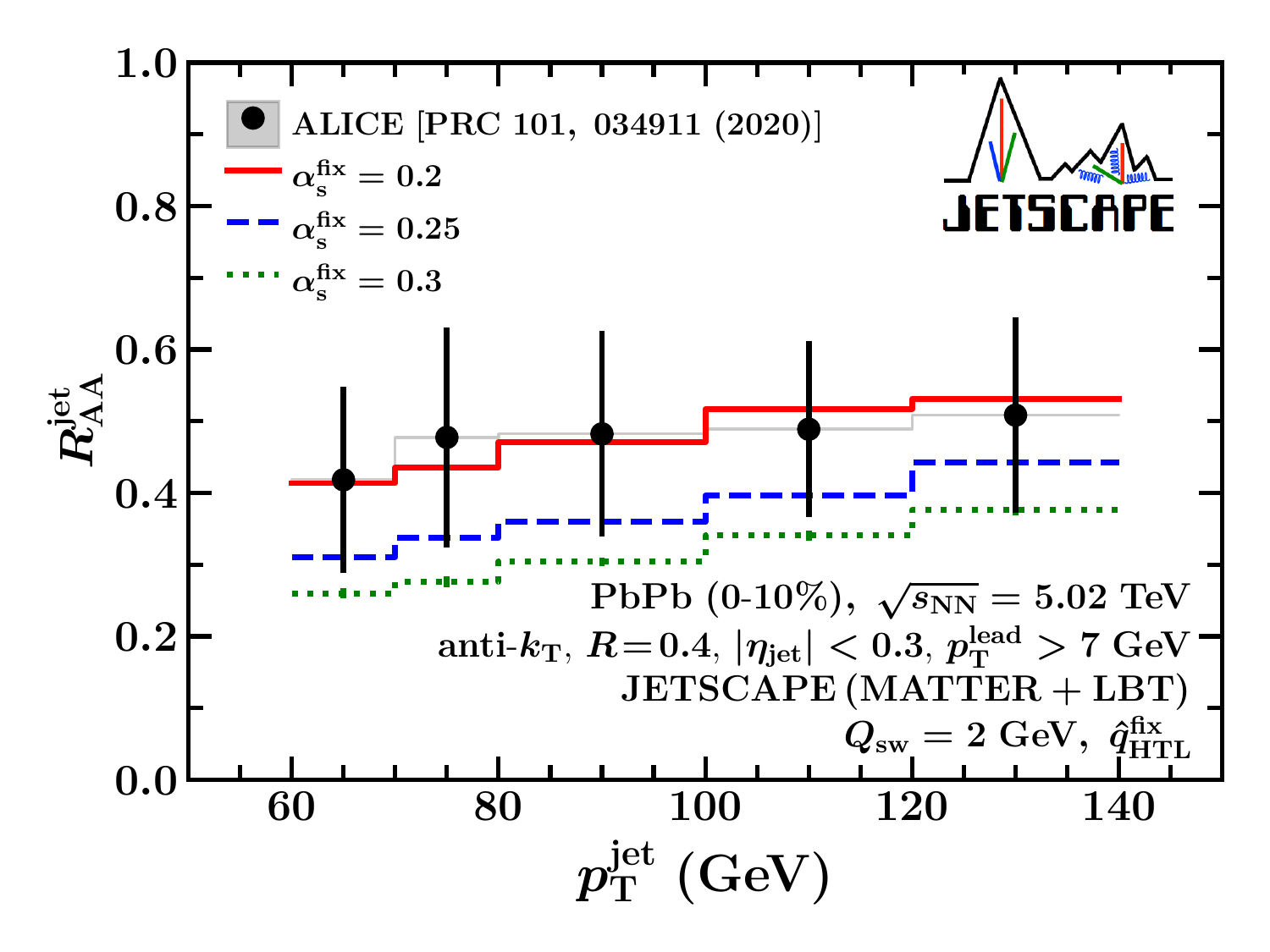}
\includegraphics[width=0.45\textwidth]{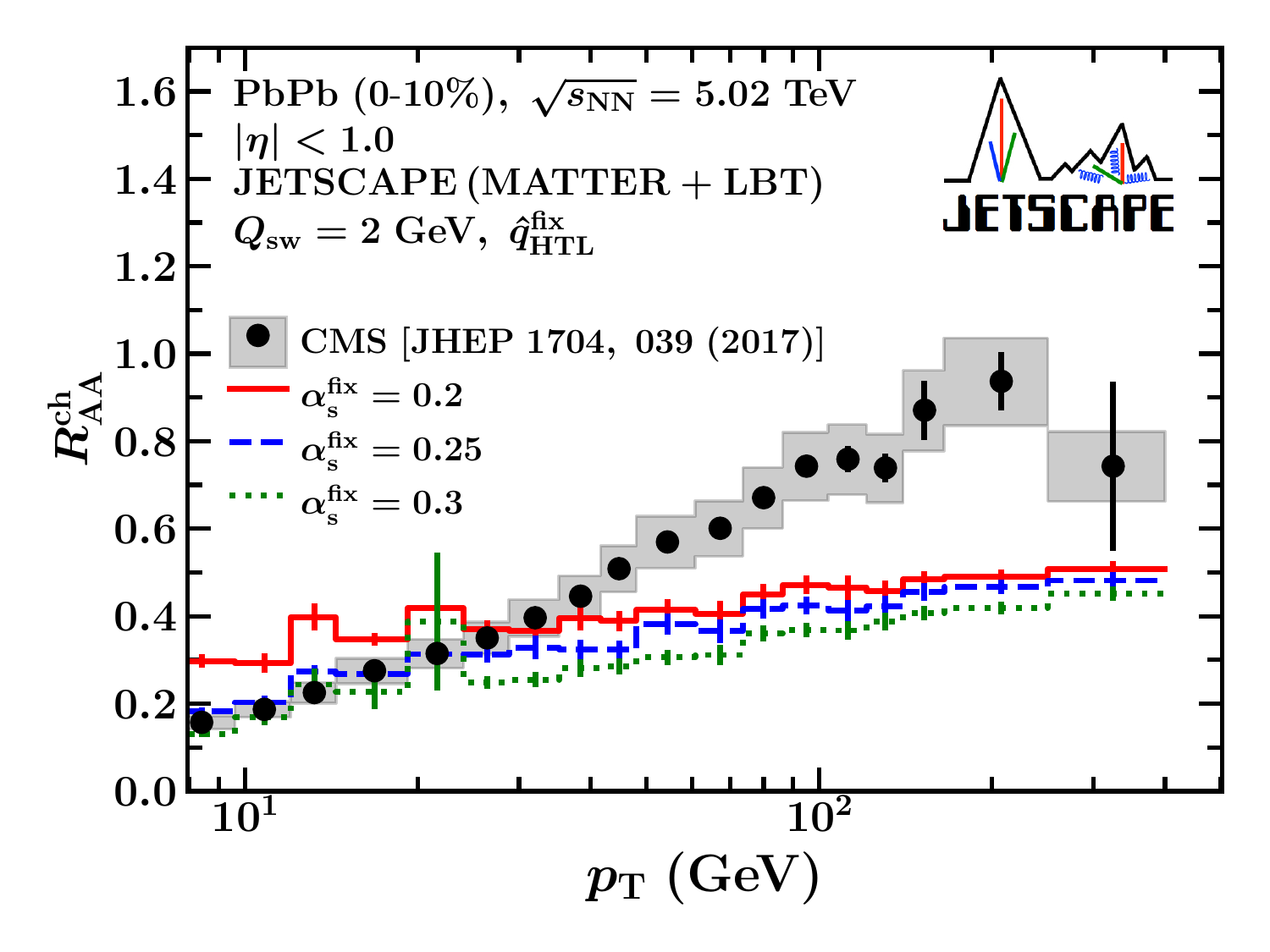}
\caption{Same as Fig.~\ref{fig:Effect_of_three_different_form_q-hat}. 
For the $\hat{q}$ formulation, the HTL with the fixed coupling (Type 1) is employed. 
The solid red, dashed blue, and dotted green lines show results 
with $\alpha^{\mathrm{fix}}_{\mathrm{s}}=0.2,\,0.25$ and $0.3$, respectively. 
Here we set $Q_{\mathrm{sw}}=2$~GeV.}
\label{fig:type1_alphas_dependence}
\end{figure}
In Fig.~\ref{fig:type1_alphas_dependence}, 
we present results of 
the nuclear modification factor of inclusive jet with $R=0.4$ and 
charged particle in Pb+Pb collisions at $\sqrt{s_{\mathrm{NN}}} = 5.02$~TeV 
from \textsc{jetscape} (\textsc{matter}+\textsc{lbt}) calculations with 
the fixed-coupling HTL $\hat{q}$ (Type 1) 
for different coupling strengths $\alpha_\mathrm{s}^\mathrm{fix}=0.2,0.25$, and $0.3$. 
The parameters other than $\alpha_\mathrm{s}^\mathrm{fix}$ are set to their default values shown in Tab.~\ref{tab:aa_parameter_set}. 

The trend of stronger suppression with increasing $\alpha^{\mathrm{fix}}_{\mathrm{s}}$ is visible for both jets and charged particles throughout almost the entire $p_{\mathrm{T}}$ range. 
Most notably, the $\alpha_\mathrm{s}^\mathrm{fix}$ dependence in jet suppression for the case with the fixed-coupling $\hat{q}$ is much larger than that for the case with virtuality-dependent formulation (top and middle panels in Fig.~\ref{fig:Type3-q-hat-effect_fixed_alphas}). 
Since a larger number of daughter partons, whose interactions with the medium give the main contribution to the jet energy loss, are branched off the leading partons, the stronger sensitivity is seen in the fixed-coupling $\hat{q}$ case. 
This strongly implies that the modification pattern of the inner jet structure can be very different between the formulations, Type 1 and 3, even when we tune the coupling strength to make their jet energy loss the same.

\begin{figure}[htb]
\centering
\includegraphics[width=0.45\textwidth]{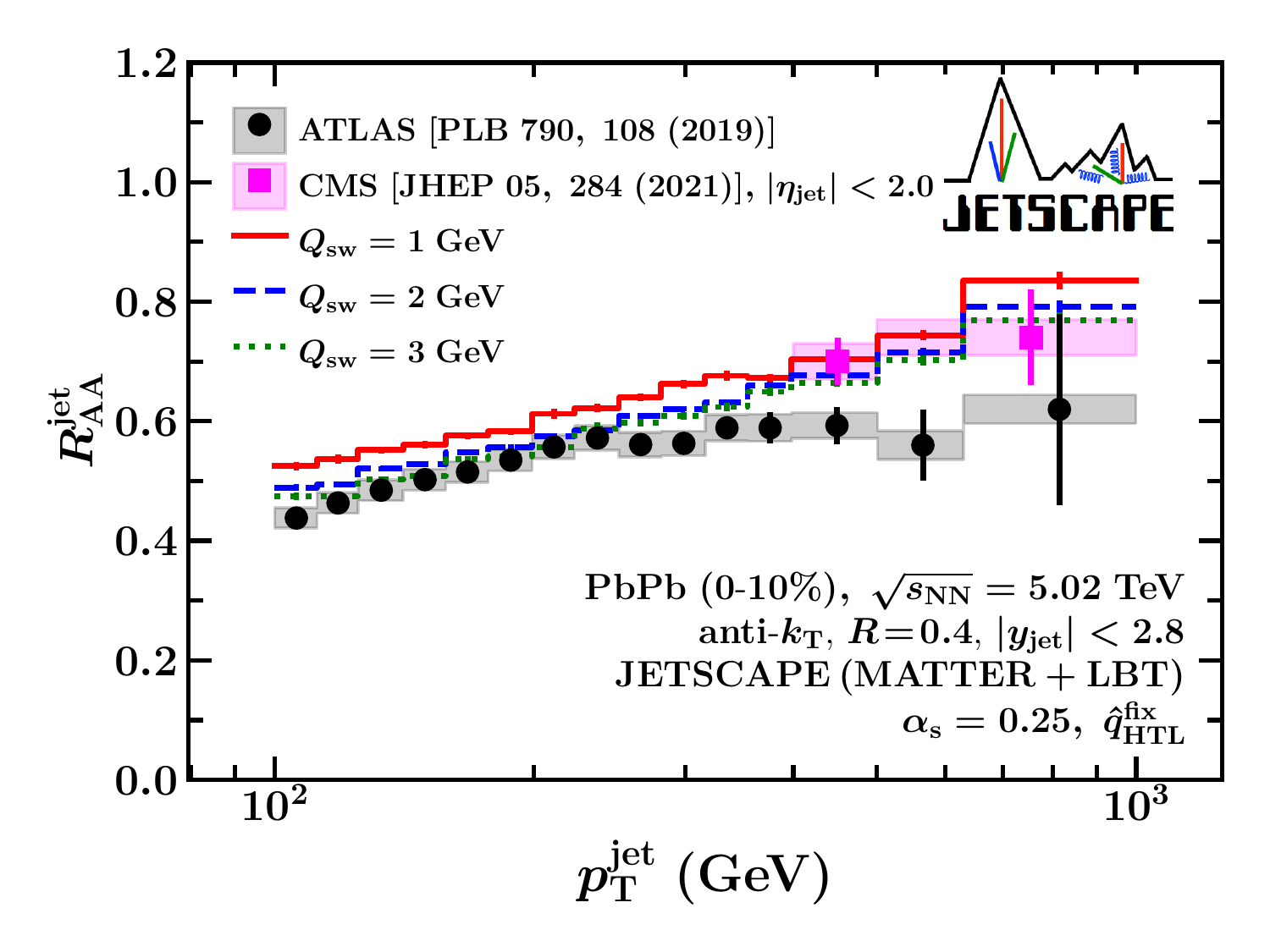}
\includegraphics[width=0.45\textwidth]{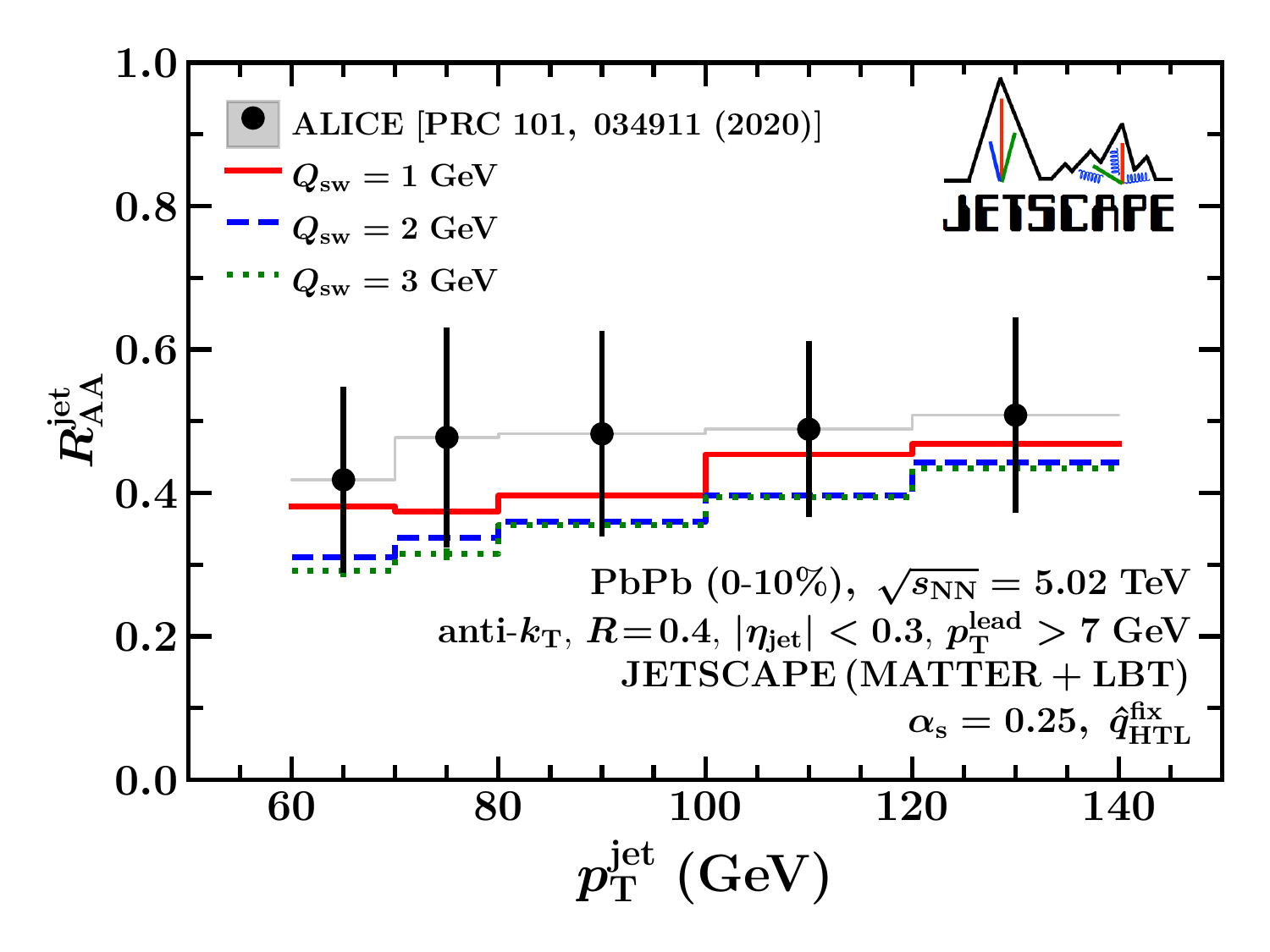}
\includegraphics[width=0.45\textwidth]{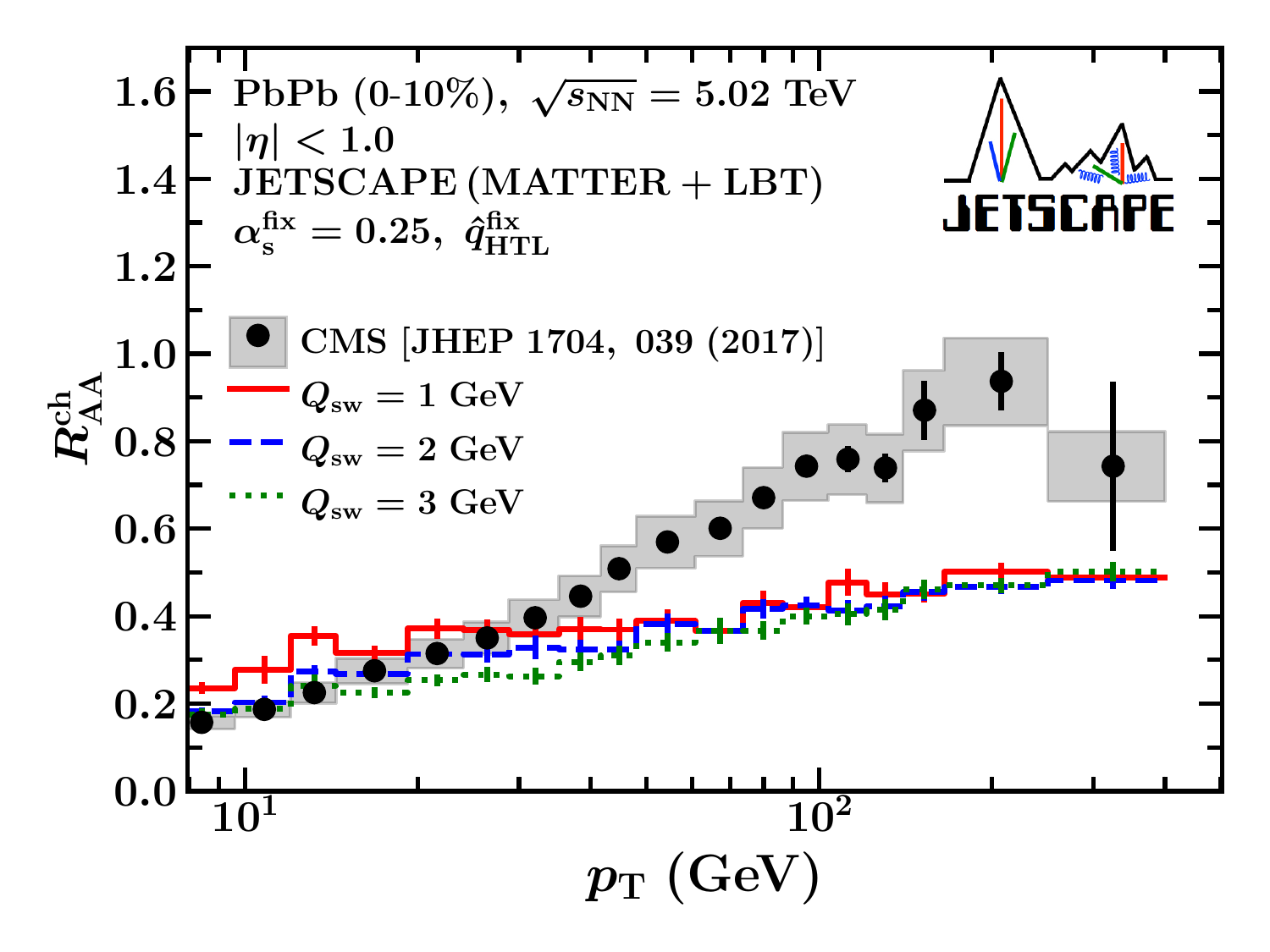}
\caption{Same as Fig.~\ref{fig:Effect_of_three_different_form_q-hat}. 
For the $\hat{q}$ formulation, the HTL with the fixed coupling (Type 1) is employed. 
The solid red, dashed blue, and dotted green lines show results 
with $Q_{\mathrm{sw}}=1,\,2$ and $3$~GeV, respectively. 
Here we set $\alpha^{\mathrm{fix}}_{\mathrm{s}}=0.25$.
}
\label{fig:type1_qsw_dependence}
\end{figure}
Figure~\ref{fig:type1_qsw_dependence}
shows results of 
the nuclear modification factor of inclusive jet with $R=0.4$ and 
charged particle in Pb+Pb collisions at $\sqrt{s_{\mathrm{NN}}} = 5.02$~TeV 
from \textsc{jetscape} (\textsc{matter}+\textsc{lbt}) calculations with 
the fixed-coupling HTL $\hat{q}$ (Type 1) 
for different switching virtuality $Q_\mathrm{sw}=1,2$, and $3$~GeV. 
The other free parameters are set to their default values shown in Tab.~\ref{tab:aa_parameter_set}. 
The same trend as that from the virtuality-dependent formulation (Fig.~\ref{fig:Type3-q-hat-Effect_of_switching_virtuality}) 
is shown here: 
stronger suppression with increasing $Q_\mathrm{sw}$
in both inclusive jet and charged particle $R_\mathrm{AA}$. 

Here we emphasize that 
the rising behavior of the data in the charged particle $R_\mathrm{AA}$ as a function of $p_{\mathrm{T}}$
cannot be reproduced with any values of the main free parameters $\alpha_{\mathrm{s}}$ and $Q_{\mathrm{sw}}$ 
when we employ the formulation of the fixed-coupling HTL $\hat{q}$ (Type 1). 

\section{Parameter dependence for Type 2: HTL $\hat{q}$ with running coupling}
\label{type2_param}
\begin{figure}[!htb]
\centering
\includegraphics[width=0.45\textwidth]{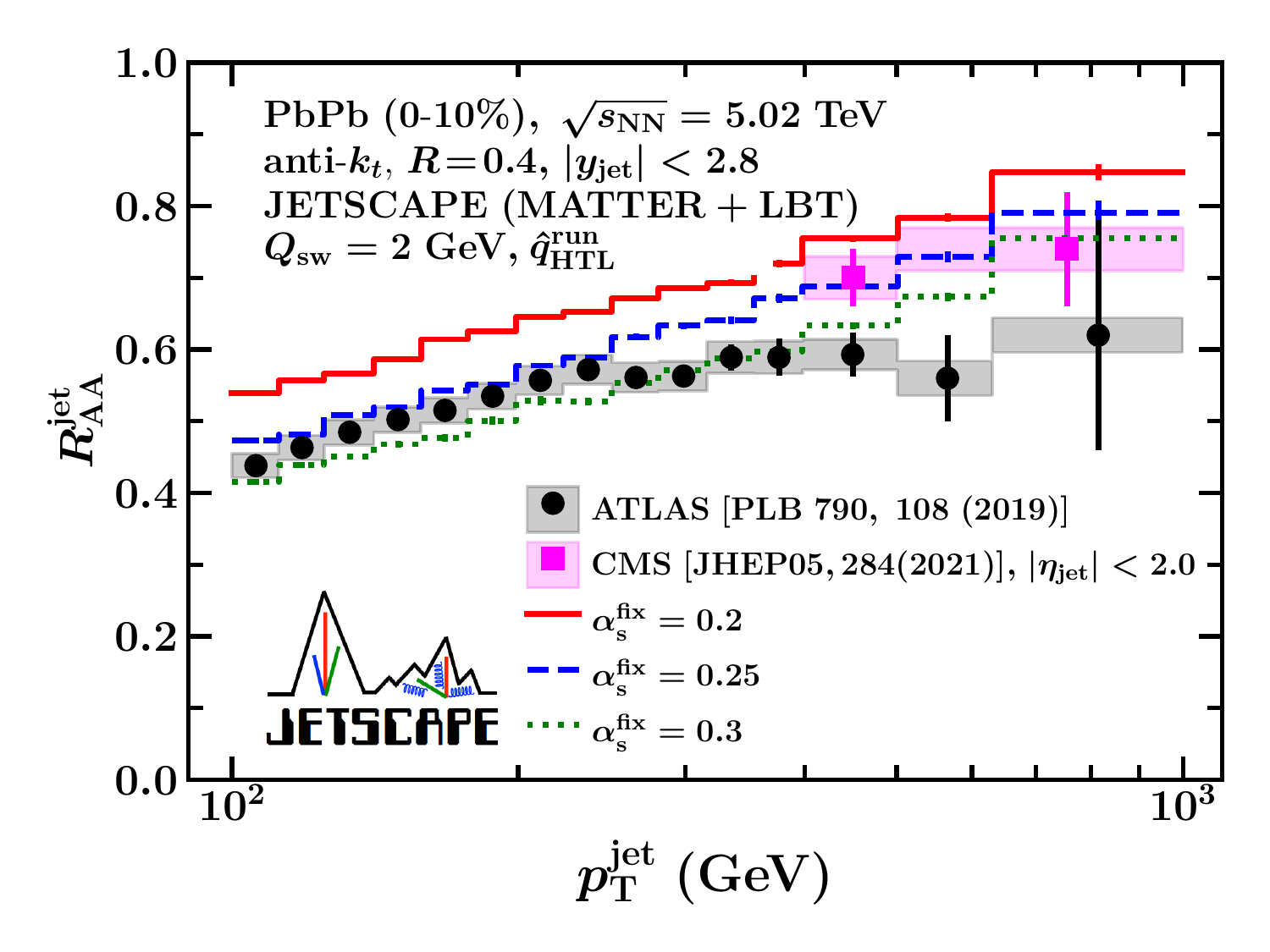}
\includegraphics[width=0.45\textwidth]{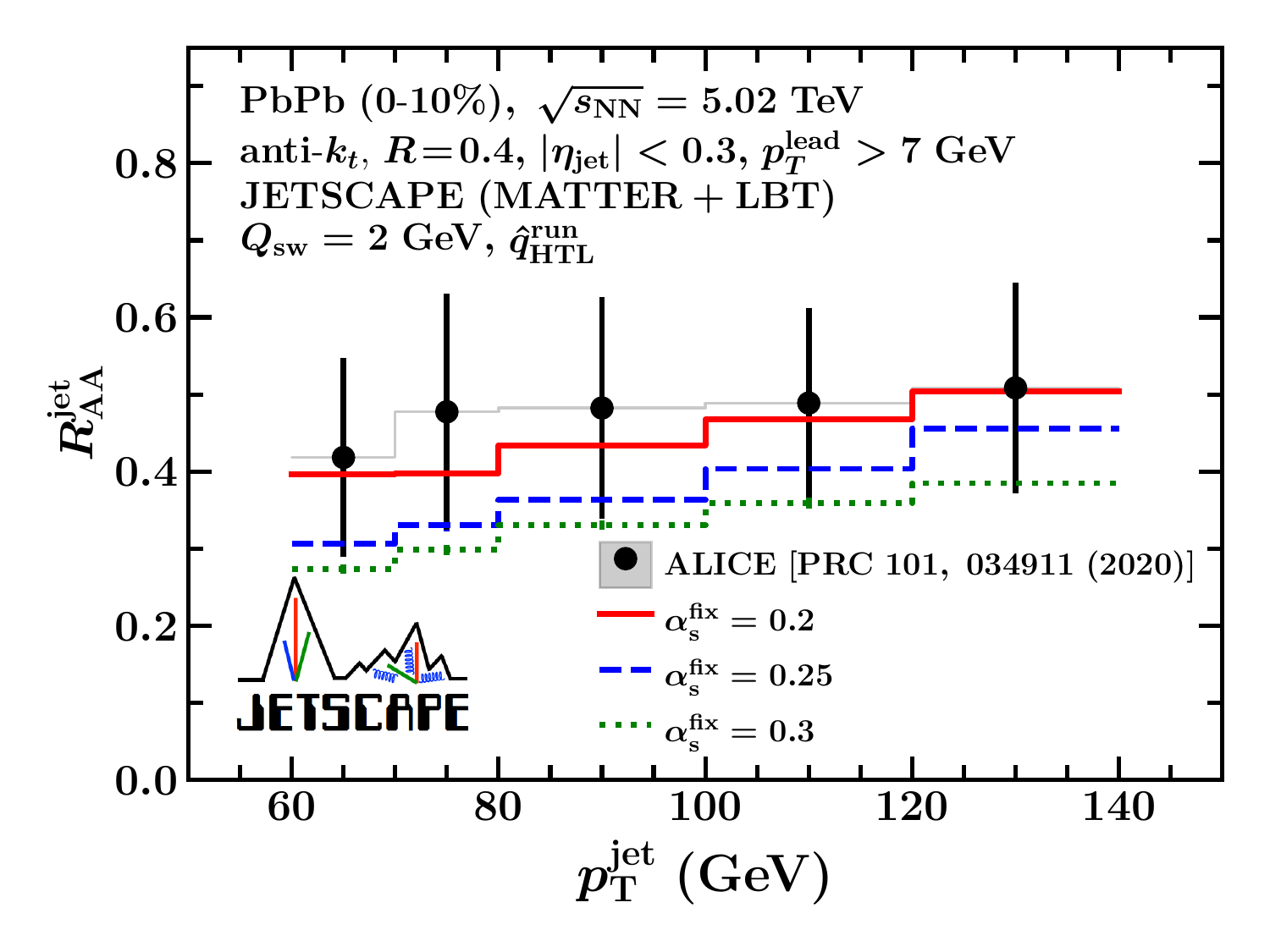}
\includegraphics[width=0.45\textwidth]{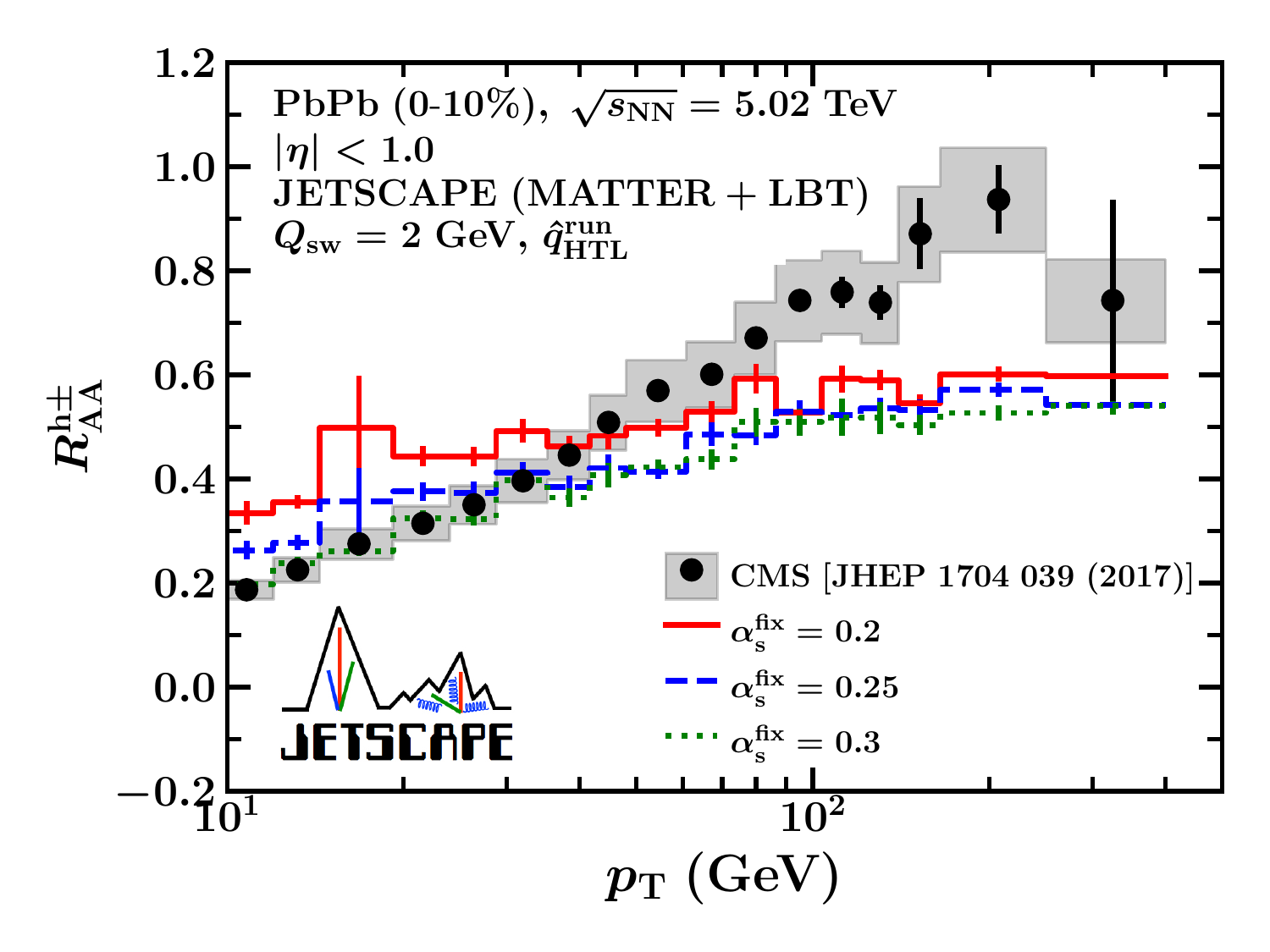}
\caption{Same as Fig.~\ref{fig:Effect_of_three_different_form_q-hat}. 
For the $\hat{q}$ formulation, the HTL with the running coupling (Type 2) is employed. 
The solid red, dashed blue, and dotted green lines show results 
with $\alpha^{\mathrm{fix}}_{\mathrm{s}}=0.2,\,0.25$ and $0.3$, respectively. 
Here we set $Q_{\mathrm{sw}}=2$~GeV.}
\label{fig:type2_alphas_dependence}
\end{figure}
Figure~\ref{fig:type2_alphas_dependence} presents the nuclear modification factor of inclusive jet with $R=0.4$ and charged particle in Pb+Pb collisions at $\sqrt{s_{\mathrm{NN}}} = 5.02$~TeV calculated using \textsc{jetscape} (\textsc{matter}+\textsc{lbt}) with the running-coupling HTL $\hat{q}$ (Type 2) 
for different coupling strengths $\alpha_\mathrm{s}^\mathrm{fix}=0.2,0.25$, and $0.3$. 

The same trend as that from the other formulations can be seen also here: 
stronger suppression with increasing $\alpha^\mathrm{fix}_\mathrm{s}$ in both inclusive jet and charged particle $R_\mathrm{AA}$. 
Here the strength of the dependence on $\alpha_\mathrm{s}^\mathrm{fix}$ in jet suppression is closer to 
that for the fixed coupling case (Type 1) than for the virtuality-dependent case (Type 3). 
This means that the effect of the running coupling does not affect the jet inner structure drastically.

\begin{figure}[!htb]
\centering
\includegraphics[width=0.45\textwidth]{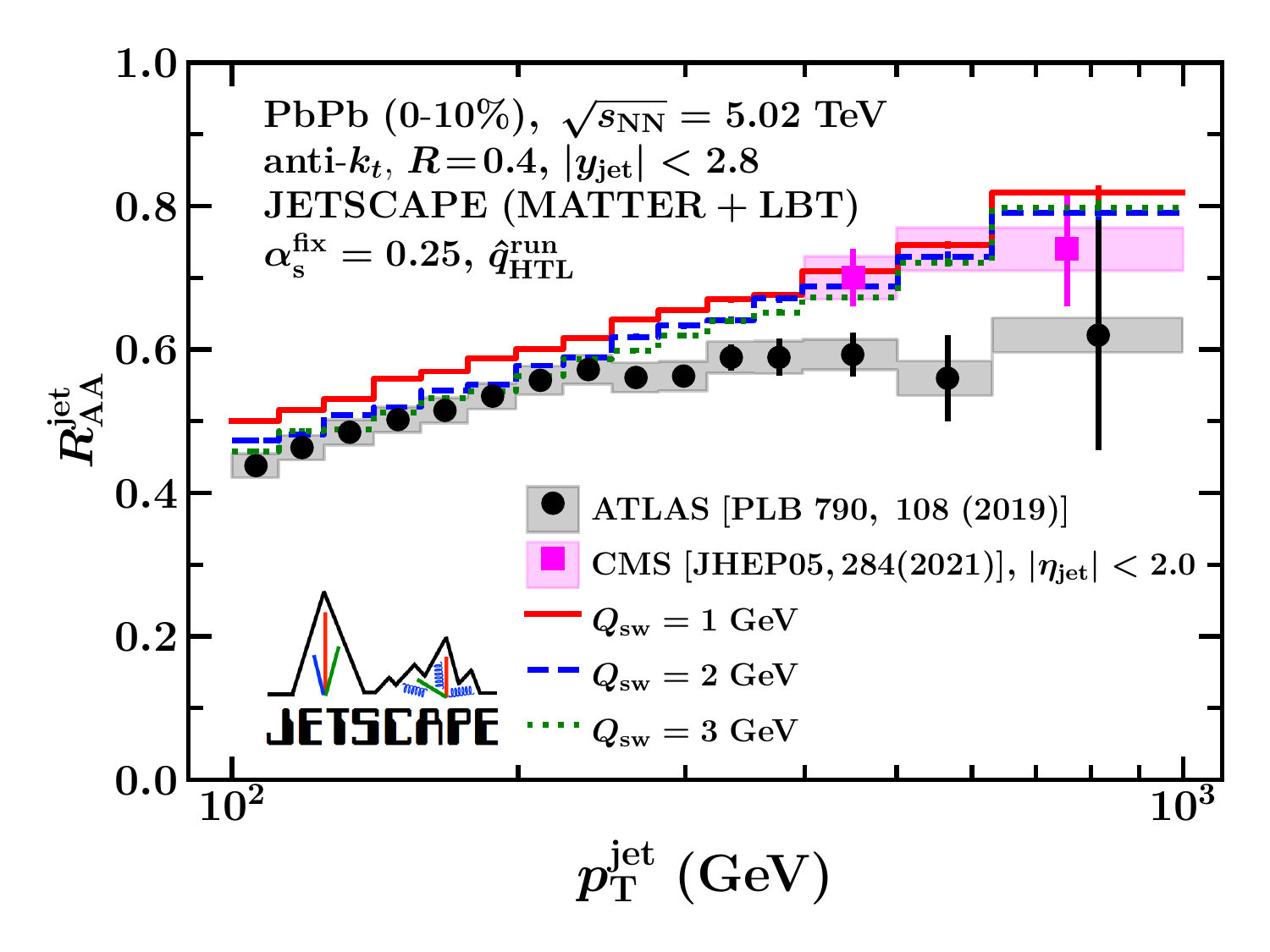}
\includegraphics[width=0.45\textwidth]{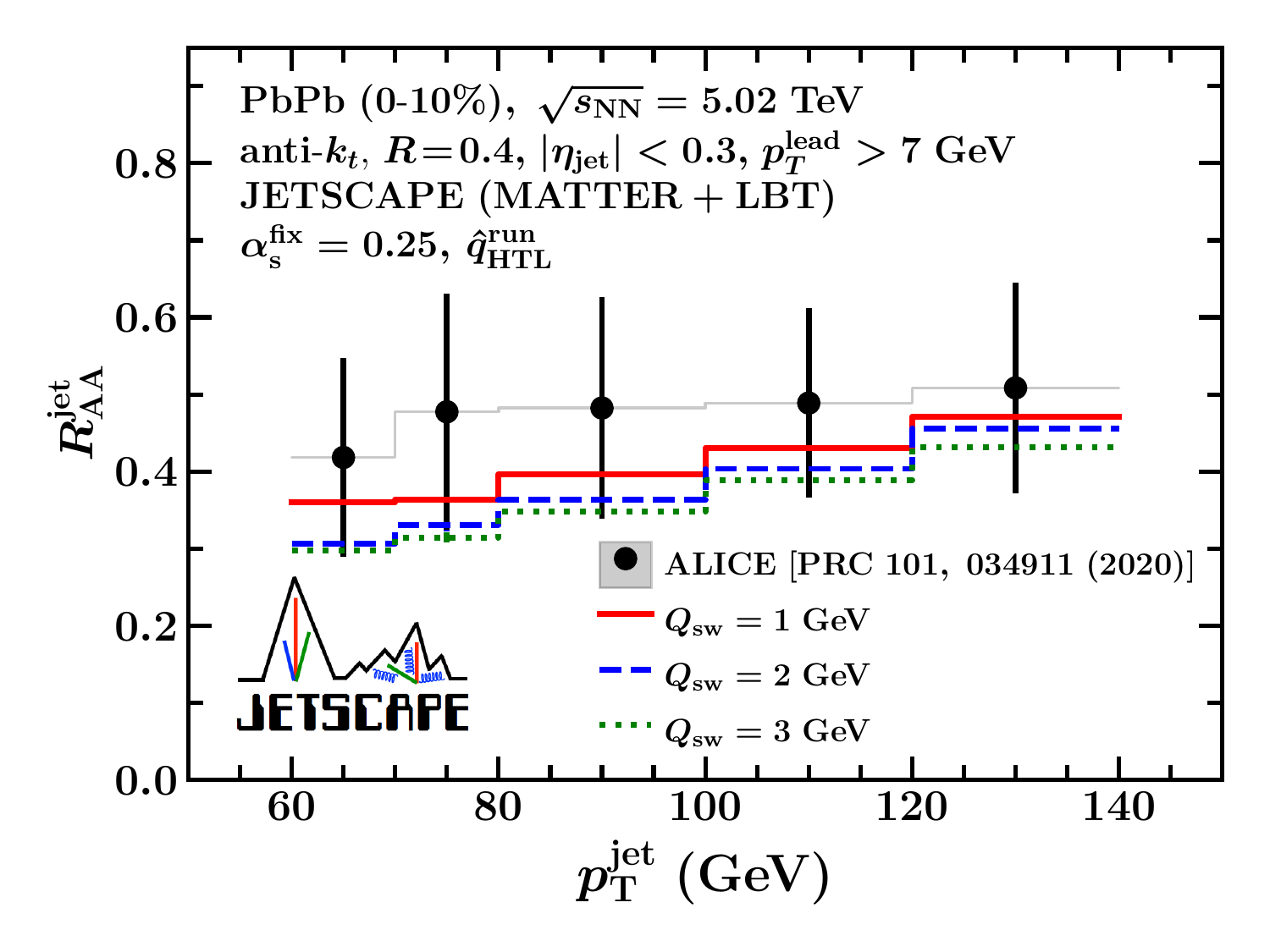}
\includegraphics[width=0.45\textwidth]{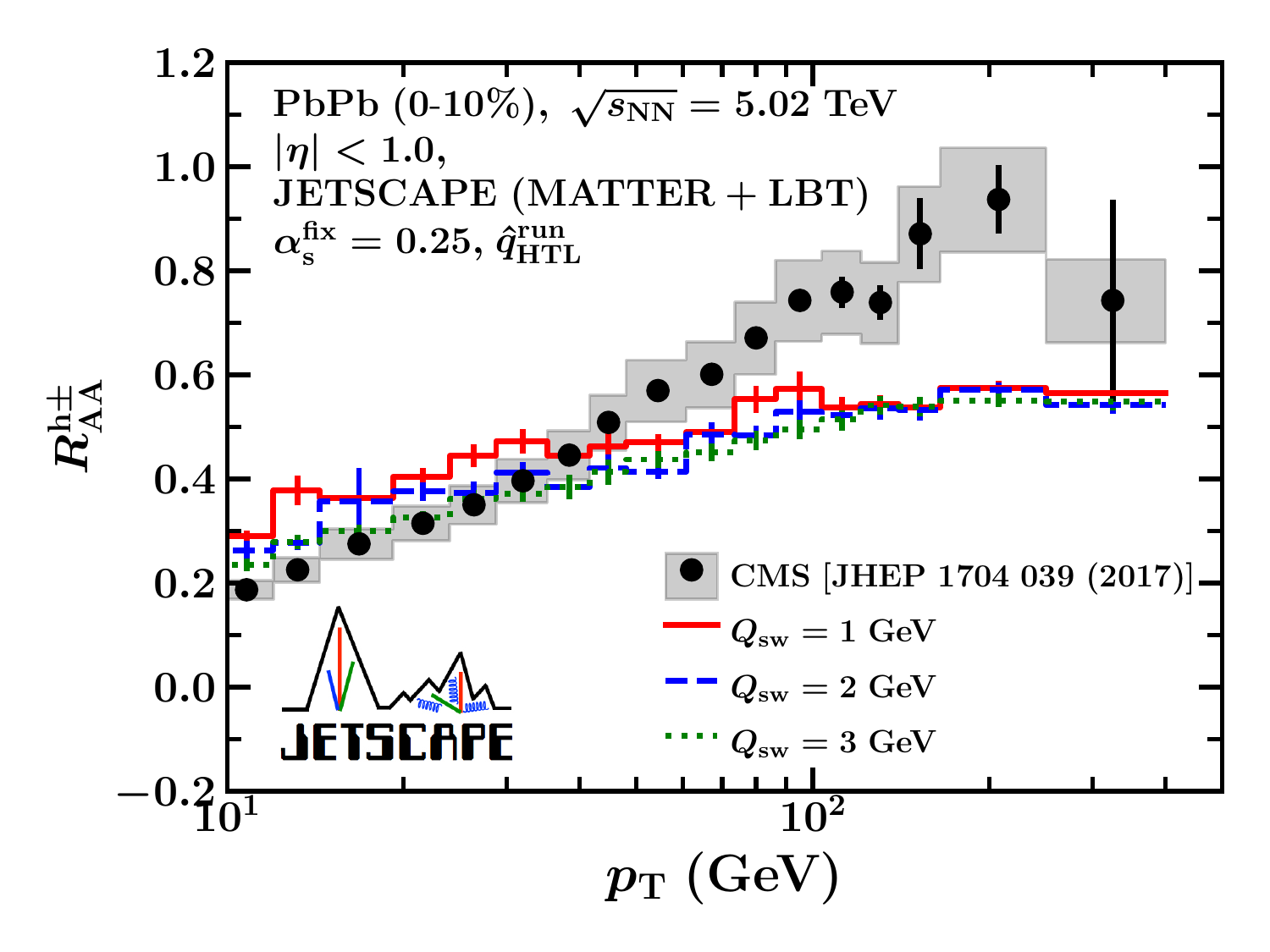}
\caption{Same as Fig.~\ref{fig:Effect_of_three_different_form_q-hat}. 
For the $\hat{q}$ formulation, the HTL with the running coupling (Type 2) is employed. The solid red, dashed blue, and dotted green lines show results 
with $Q_{\mathrm{sw}}=1,\,2$ and $3$~GeV, respectively. 
Here we set $\alpha^{\mathrm{fix}}_{\mathrm{s}}=0.25$.
}
\label{fig:type2_qsw_dependence}
\end{figure}
In Fig.~\ref{fig:type2_qsw_dependence}, 
we show 
the nuclear modification factor of inclusive jet with $R=0.4$ and 
charged particle in Pb+Pb collisions at $\sqrt{s_{\mathrm{NN}}} = 5.02$~TeV 
obtained from \textsc{jetscape} (\textsc{matter}+\textsc{lbt}) calculations with 
the running-coupling HTL $\hat{q}$ (Type 2) 
for different switching virtuality $Q_\mathrm{sw}=1,2$, and $3$~GeV. 
The other free parameters are set to their default values shown in Tab.~\ref{tab:aa_parameter_set}. 
The trend of the stronger suppression with increasing $Q_\mathrm{sw}$ is also shown here in both inclusive jet and charged particle $R_\mathrm{AA}$. 

Again we would like to note that 
the data for the $p_{\mathrm{T}}$ dependence of the charged particle $R_\mathrm{AA}$ 
cannot be described with any values of the main free parameters $\alpha_{\mathrm{s}}$ and $Q_{\mathrm{sw}}$ 
unless the virtuality-dependence via $f(Q^2)$ is introduced.

\section{Comparison of Partonic Jet $R_{\mathrm{AA}}$ and Hadronic Jet $R_{\mathrm{AA}}$}
\label{parton_jet}
In this appendix, we discuss how nonperturbative effects can modify the reconstructed jet spectra. An analysis of the $p_{\mathrm{T}}$ distribution of partons in jets, that have propagated through a dense medium, shows a large population of soft partons within quenched jets. This leads to noticeable shifts in the jet spectrum during hadronization carried out by string-based models. As outlined below, one can highlight these shifts by comparing distributions of jets constructed using final partons versus those constructed using final hadrons. In this section, we study these differences and discuss alternative approaches.  

\subsection{Comparison with Partonic Jet}
\begin{figure}[!htb]
\centering
\includegraphics[width=0.45\textwidth]{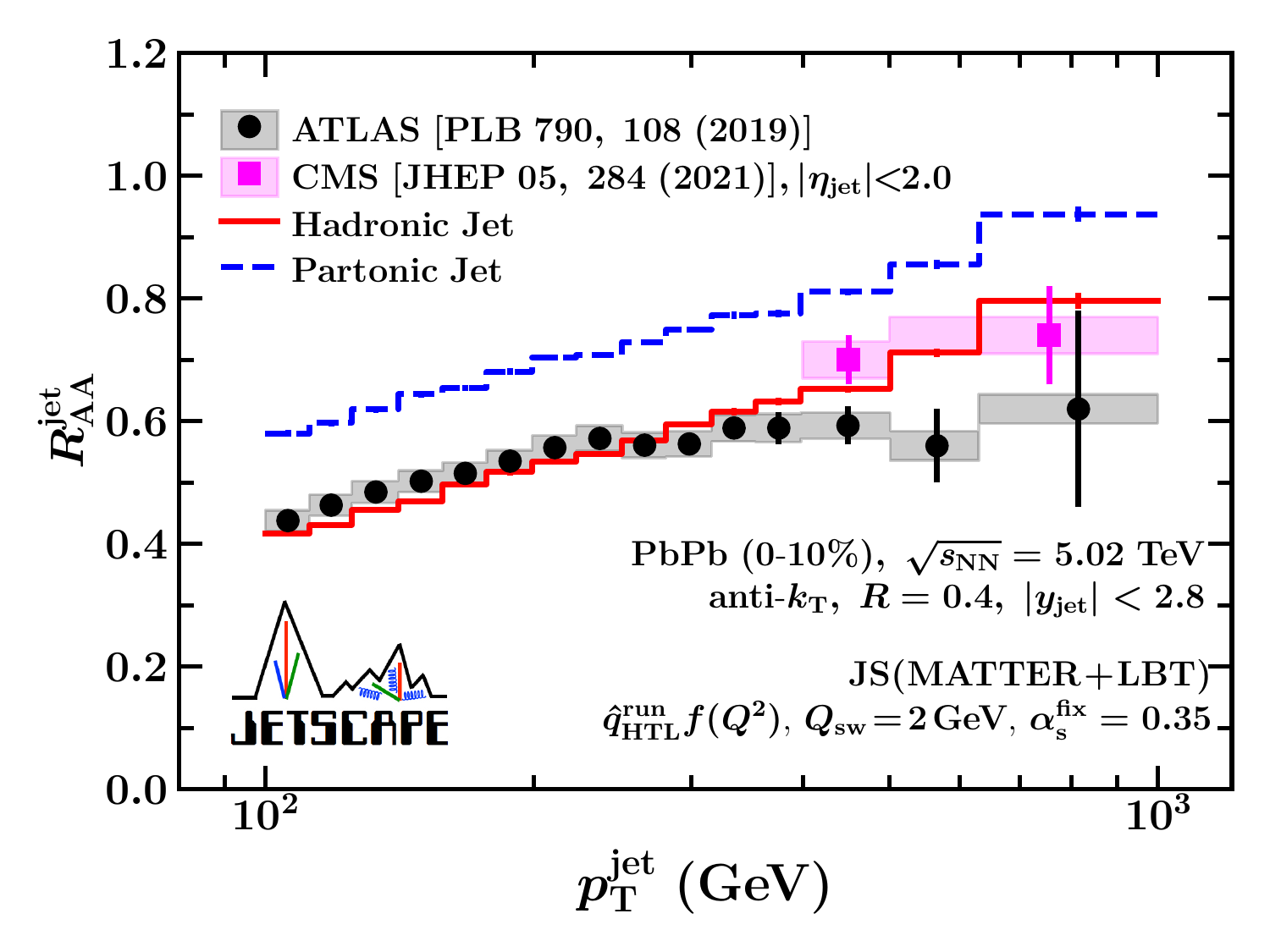}
\includegraphics[width=0.45\textwidth]{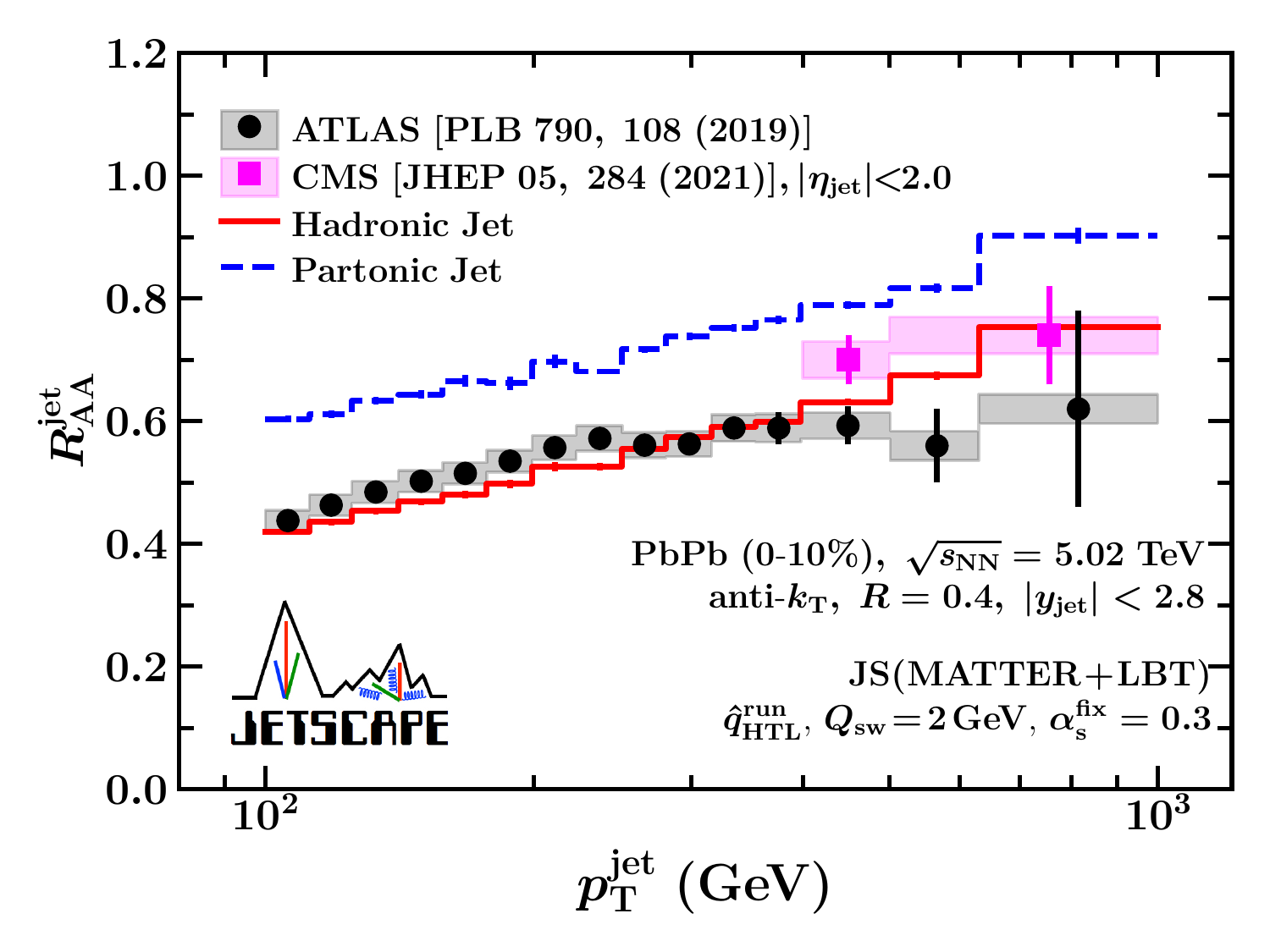}
\caption{Comparison of partonic jet $R_{\mathrm{AA}}$ with hadronic jet $R_{\mathrm{AA}}$ for two different parametrization of $\hat{q}$. 
Jets are reconstructed with cone $R=0.4$ at $y_{\mathrm{jet}} < 2.8$. 
The black circles are ATLAS data~\cite{ATLAS:2018gwx} 
and the dark red squares are 
CMS data for 
$\eta_{\mathrm{jet}} < 2.0$~\cite{CMS:2021vui}. 
Top panel:
Results for the formulation with  virtuality dependence (Type 3). 
Bottom panel:
Results for the $\hat{q}$ formulation of the HTL with the running coupling (Type 2).  
}
\label{fig:PbPb_parton_jet_hadron_jet}
\end{figure}

\begin{figure}[!htb]
\centering
\includegraphics[width=0.45\textwidth]{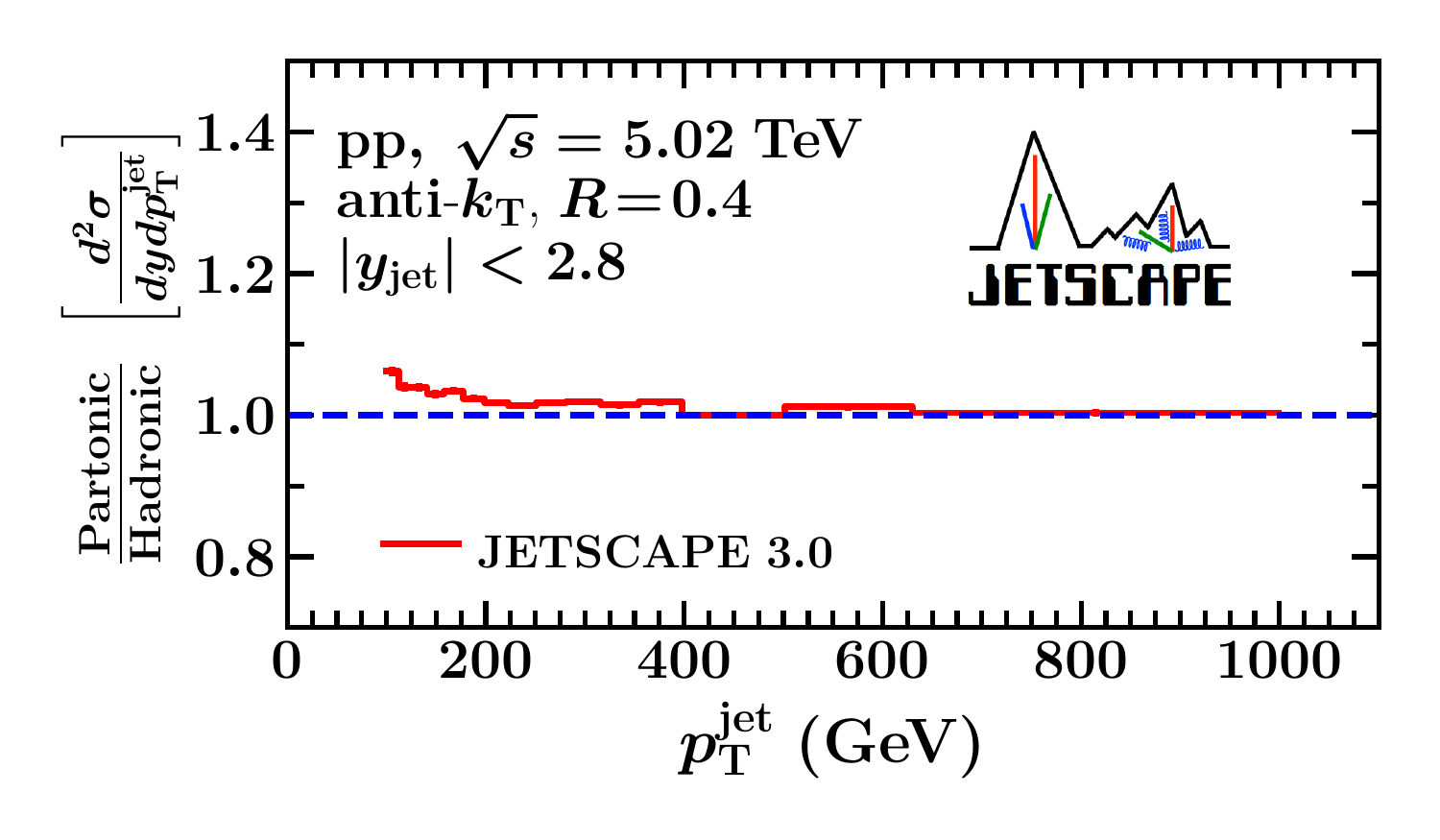}
\caption{Ratio of inclusive jet cross section at parton-level to hadron-level for $p+p$ collisions at $\sqrt{s}=5.02$~TeV. 
}
\label{fig:pp_parton_jet_hadron_jet}
\end{figure}

The contribution of nonperturbative effects 
in our model is limited mostly in the hadronization process. 
To quantify the effect, 
we conduct an analysis also for the jets reconstructed from the final state partons, just before being passed to the hadronization module. 
In Fig.~\ref{fig:PbPb_parton_jet_hadron_jet}, 
the $R_{\mathrm{AA}}$ of the inclusive partonic jet 
for most-central (0-10\%) Pb+Pb collisions at $\sqrt{s_{\mathrm{NN}}} = 5.02$~TeV is compared with the full calculation results with hadronization, 
for the case of running coupling, with (Type 3) and without (Type 2) the virtuality dependent modulation factor $f(Q^2)$. 
Note that the denominator of $R_{\mathrm{AA}}$, for partonic jets, is the jet spectrum in $p+p$ collisions calculated by turning off the hadronization module. 
The ratio between the spectra for parton jets and hadron jets in $p+p$ collisions is shown in Fig.~\ref{fig:pp_parton_jet_hadron_jet}. 
Thus, we find that hadronization effects are not significant in $p+p$ collisions and gives almost no modification for jets with $p^{\mathrm{jet}}_{\mathrm{T}}\gtrsim 200$~GeV. 

In both cases, with and without the virtuality dependence, 
additional suppression from the colorless string hadronization can be seen. 
This additional suppression comes from the modification of the soft parton spectrum prior to hadronization. 
As jets propagate through a dense medium, a considerable number of low-energy partons are branched off collinearly in the jet shower. 
In string hadronization, strings connecting such collinear soft partons do not have the minimum mass necessary to produce hadrons, and cannot be included in the hadronization process \emph{as is}.


In this work, we follow the methodology devised in our prior work on jets in $p+p$ collisions~\cite{JETSCAPE:2019udz}: Consider a pair of partons that possesses a $| \delta \vec{p} | < 4\Lambda_{\mathrm{QCD}}$. The parton with larger $|p_z|$ in this pair has its $p_z$ shifted to $p_z \pm 4\Lambda_{\mathrm{QCD}}$. The sign of the added momenta is set in a way to increase the relative momentum $|\delta \vec{p} |$. As a result, no net $z$-component of momentum is added to jets on average. The energy of the modified parton is changed accordingly to ensure that it remains on-shell. One continuously iterates through the entire parton list until no pairs meet the condition $| \delta \vec{p} | < 4\Lambda_{\mathrm{QCD}}$. While cast differently, in terms of the three-momentum, this condition is consistent with the minimum mass condition in Ref.~\cite{Andersson:1983ia}. 
The reader will note that the $p_{\mathrm{T}}$ of the jet is not affected by this procedure, though the energy and mass of the jet may be affected. 
The predominant effect of this procedure is that a large fraction of the soft partons are pushed to larger rapidities and outside the jet cone, leading to extra jet suppression (energy loss). This procedure does not affect the hard partons in a jet and thus has no effect on the single hadron spectrum or its nuclear modification.

In Fig.~\ref{fig:PbPb_parton_jet_hadron_jet}, 
the effect of this prescription shows up as a noticeable difference between the spectra of jets clustered with partons and those clustered with hadrons. Whereas this prescription shows almost no difference between partonic jets and hadronic jets in $p+p$ collisions, as shown in Fig.~\ref{fig:pp_parton_jet_hadron_jet}. These shifts for jets modified in heavy-ion environments are indicative of the limitations of applying Lund string hadronization to systems with several soft partons. 

A large fraction of these soft partons are included with the hard jet in the process of scattering and recoil. In this sense, they represent extra energy and momentum from the medium which has become correlated with the jet shower. As a result, these partons are subject to background subtraction. The exact method in which this background subtraction is carried out affects the difference between the partonic and hadronic spectra but has minimal effect on the final hadronic jet spectrum. While the conventional method of background subtraction was described in Sec.~\ref{Subsection:medium-response}, we describe alternate mechanisms and their effect on the parton-hadron offset in jet spectra in the subsequent subsection.

\subsection{Background Subtraction at Source with $E_{\mathrm{cut}}$}
In this sub-section, we study the effects of background subtraction of jet partons that are at the thermal scale. 
Throughout this paper, the method of background subtraction has been that of Eq.~\eqref{eq:neg_sub}. All partons generated by the jet shower, and those from the medium which scatter with a jet parton, and are included in the modified shower, are retained all the way down to $p_{\mathrm{T}} \to 0$. All these partons are assumed to be weakly coupled with the medium and can escape the medium. The four momentum of the  incoming partons from the medium, which scatter with the jet partons, are also retained (referred to as holes). These holes are assumed to be free from medium interactions. Once jets have been clustered, the four momenta of these incoming partons (holes) which fall within a jet's area are subtracted from the full four momentum of the jet. 

A more approximate algorithm, which allows for faster simulations is to remove all holes from the event record, as they appear in the simulation, along with the softest partons (with energy $E$ in the rest frame of the fluid cell) which range from $0< E <E_{\mathrm{cut}}$. The remaining partons with $E > E_{\mathrm{cut}}$ are clustered to form jets. The upper limit of $E_{\mathrm{cut}}$ is varied until the jet spectrum at the parton level is almost identical to that obtained by the subtraction of holes within clustered jets, as outlined in the paragraph above [i.e., using Eq.~\eqref{eq:neg_sub}]. 

The physical picture underlying this methodology is that holes, partons that arise from the medium and are scattered by the jet to form recoil partons, are removed from the medium, thus constituting ``holes". Also, one would expect the softest partons in the jet to thermalize within the strongly interacting medium. Thus, we are assuming that as the energy lost from the medium to the jet, thermalizes, it balances this negative contribution with the positive contribution of the softest partons correlated with the jet. The upper limit $E_{\mathrm{cut}}$, of the soft parton spectrum which is balanced by the hole contribution is varied to ensure that both methods for background subtraction yield identical results. Our simulations indicate that an $E_{\mathrm{cut}} = 3.2 T$, where $T$ is the ambient temperature in the rest frame of the unit cell where the scattering takes place, yields the same jet spectrum as that obtained from Eq.~\eqref{eq:neg_sub}, across all energies and centralities. 

While the presence of a single such value may be surprising, it should be pointed out that the mean momentum of the hole parsons emanating from the medium, for a given thermal distribution, is of the order of $\approx 2.5T$-$3.5T$. We also point out that since $E_{\mathrm{cut}}$ is determined by the comparison between two methods of background subtraction, it does not constitute a new parameter in jet modification.

\begin{figure}[!htb]
\centering
\includegraphics[width=0.45\textwidth]{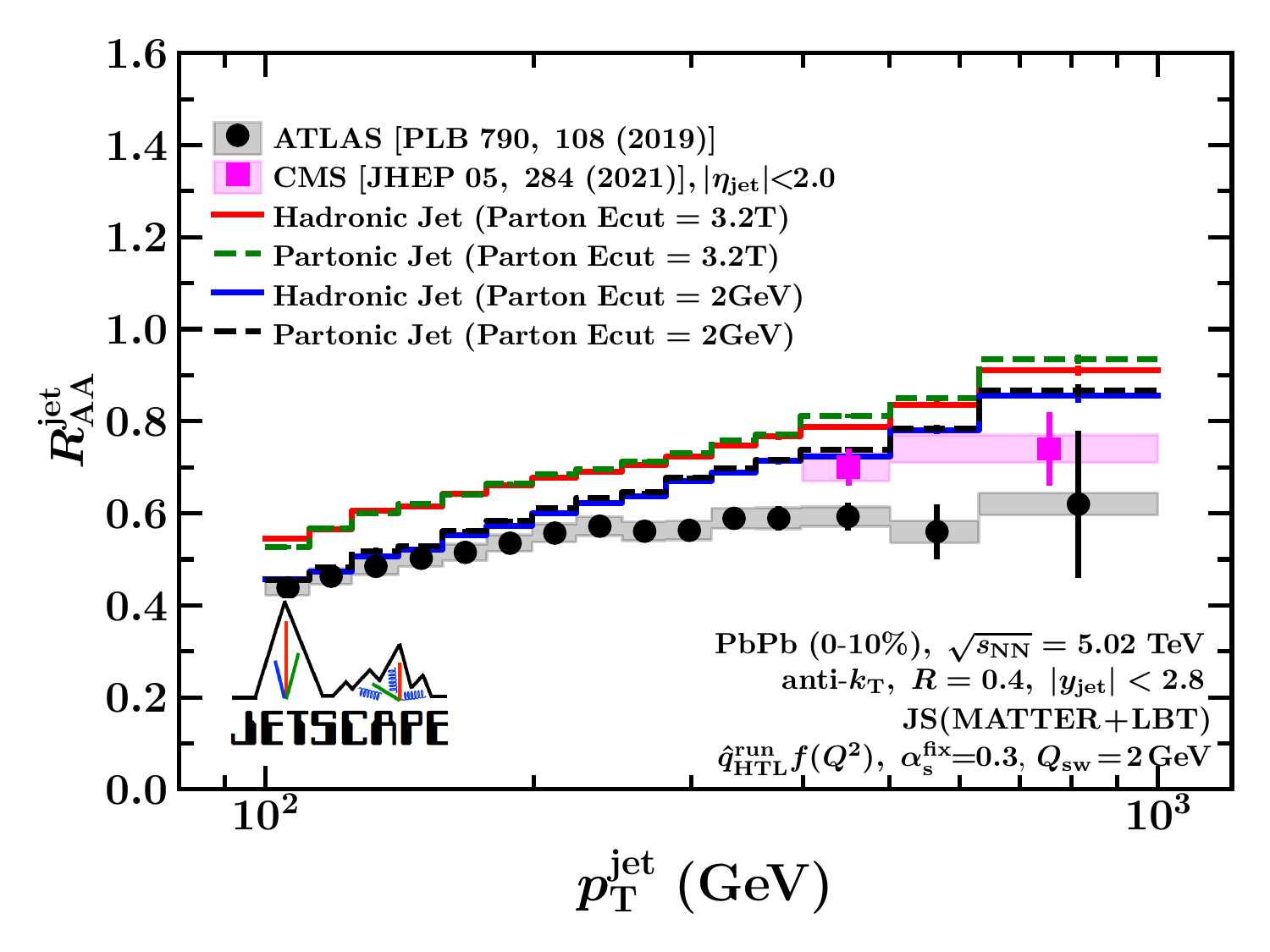}
\caption{Nuclear modification factor for inclusive jets before hadronization and after hadronization is shown. Here, the partons with energy $E\le E_{\mathrm{cut}}$ (in the rest frame of fluid cell) have been removed from the parton shower. 
For the functional form of $\hat{q}$, 
the virtuality dependent formulation (Type 3) is employed. 
Results for 
inclusive jets with $R=0.4$ and $y_{\mathrm{jet}} < 2.8$
are compared to 
ATLAS data~\cite{ATLAS:2018gwx} (black circles) 
and 
CMS data for 
$\eta_{\mathrm{jet}} < 2.0$~\cite{CMS:2021vui} (dark red squares). 
}
\label{fig:LiquefierAlone_parton_jet_hadron_jet}
\end{figure}

Along with the increased speed in the simulation, brought on by neglect of holes and soft partons with $E < E_{\mathrm{cut}}$, this second method of background subtraction has one further advantage. Since a large number of the soft partons have been removed from the parton showers, the offset between the partonic jet spectrum and the hadronic jet spectrum is greatly reduced. Note that as discussed in the preceding subsection, the major contribution to the offset is the presence of a large number of soft and collinear partons within the jet shower. 
In Fig. \ref{fig:LiquefierAlone_parton_jet_hadron_jet}, we present inclusive jet $R_{\mathrm{AA}}$ for two different choices of parton energy cuts: $E_{\mathrm{cut}}=3.2T$ and $E_{\mathrm{cut}}=2$~GeV. The partons with energy $E \le E_{\mathrm{cut}}$ (in the rest frame of fluid cell) have been removed from the parton shower. The results in Fig. \ref{fig:LiquefierAlone_parton_jet_hadron_jet} shows that the offset between the parton-level and hadron-level jets goes away  if one removes the thermal partons from the parton shower. Moreover, the results also demonstrate that the contribution of such nonperturbative effects is essential to describe the experimental data of $R_{\mathrm{AA}}$ for reconstructed jet. 

\subsection{Energy Loss due to Medium Response}
\label{subsect:approximate-medium-response}
As shown in Fig.~\ref{fig:LiquefierAlone_parton_jet_hadron_jet}, the removal of holes and partons with $E< E_{\mathrm{cut}} = 3.2T$, yields coinciding hadronic and partonic jet spectra. Also with this value of $E_{\mathrm{cut}}$ the two methods of background subtraction yield identical partonic jet spectra. However, with only an $E_{\mathrm{cut}} = 3.2T$ the suppression observed in the jet spectrum is not consistent with the data on jet $R_{\mathrm{AA}}$. Comparison with the hadronic jet spectrum in Fig.~\ref{fig:PbPb_parton_jet_hadron_jet}, indicates the presence of further jet modification by medium response. 

\begin{figure}[!htb]
\centering
\includegraphics[width=0.45\textwidth]{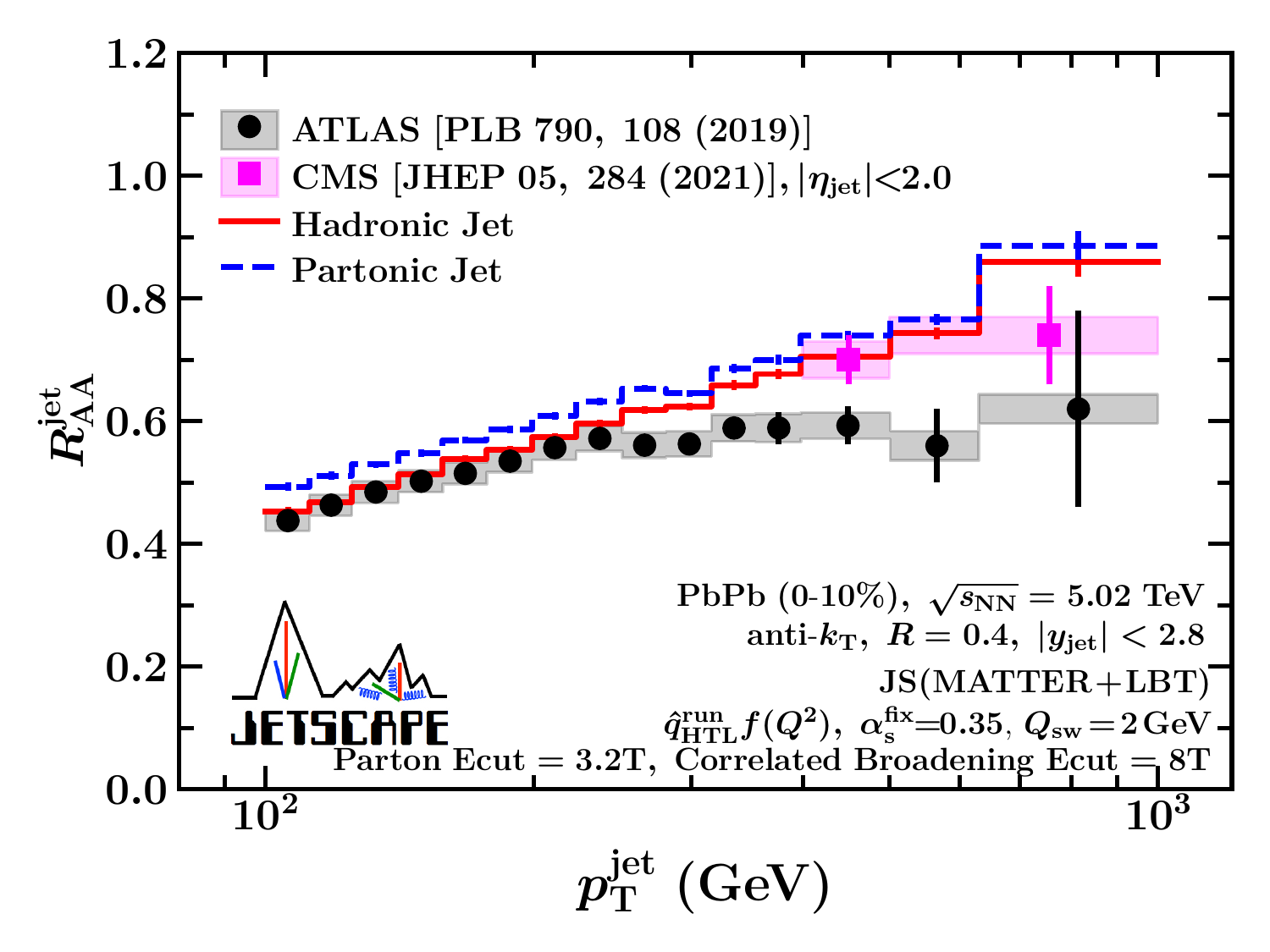}
\includegraphics[width=0.45\textwidth]{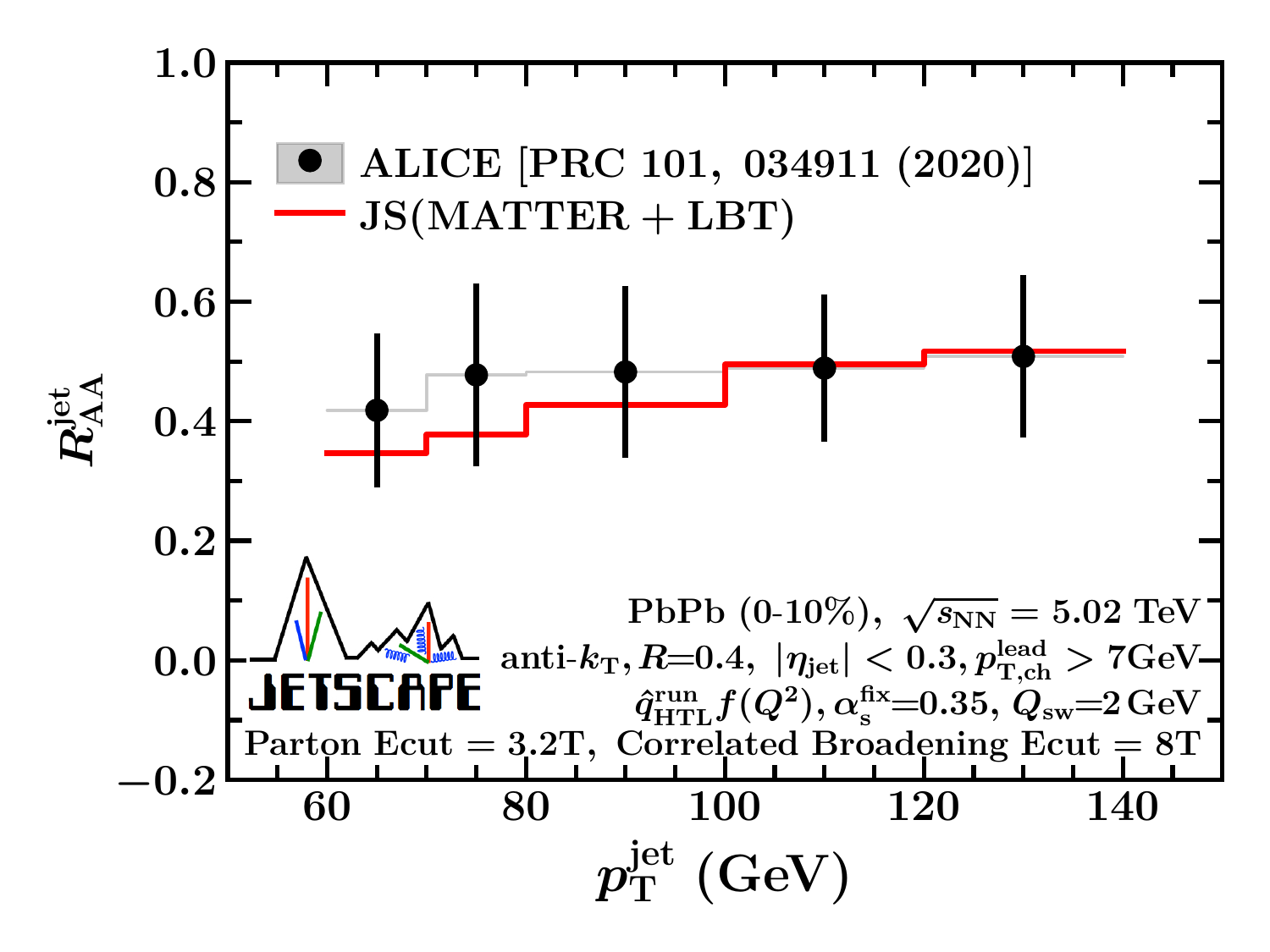}
\includegraphics[width=0.45\textwidth]{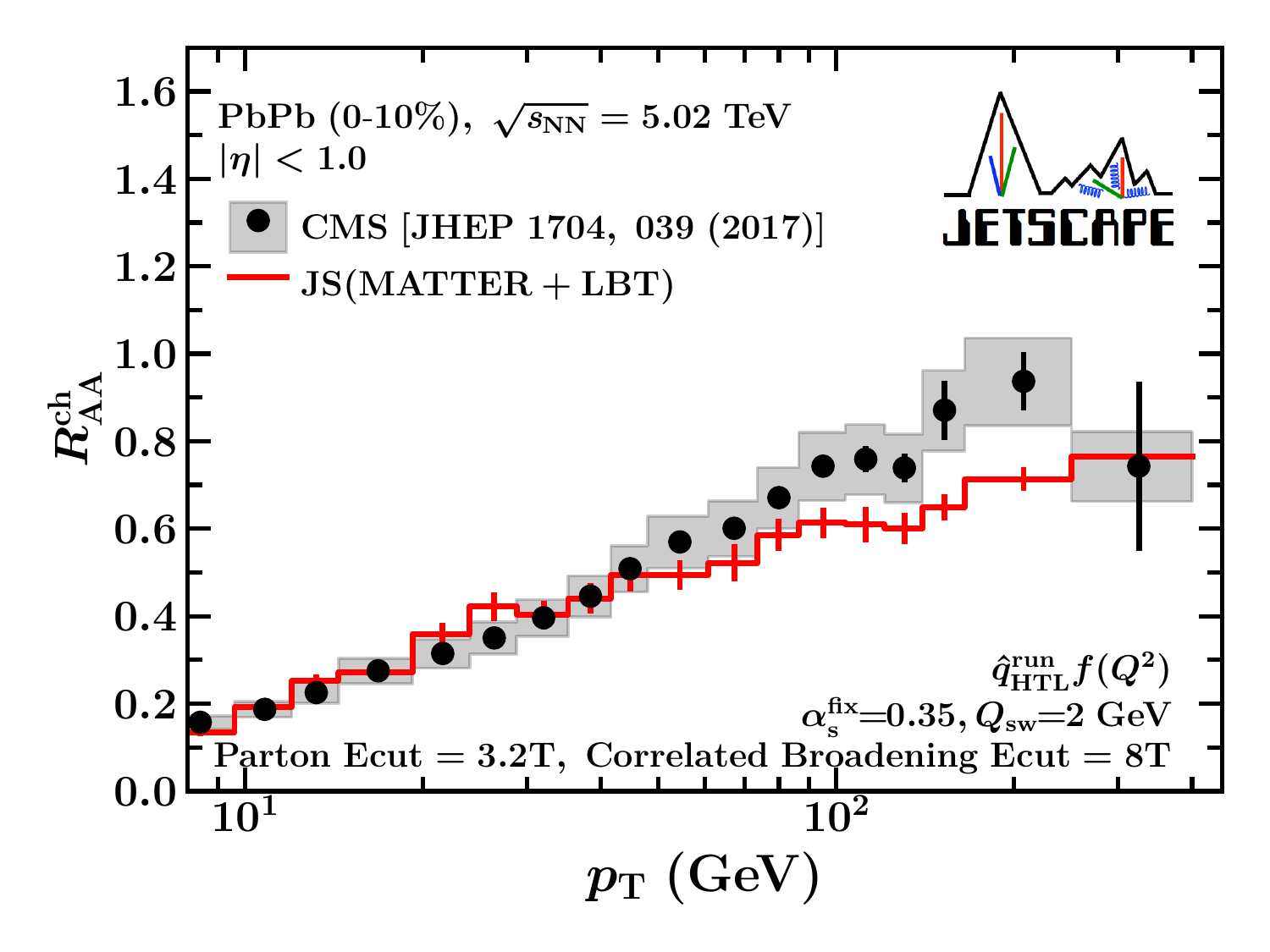}
\caption{
Same as Fig.~\ref{fig:Effect_of_three_different_form_q-hat}, but theoretical curves shows the results when partons with energy $E\le E_{\mathrm{cut}}=8T$ undergo nonperturbative energy loss modeled using correlated broadening. 
For the functional form of $\hat{q}$, 
the virtuality dependent formulation (Type 3) is employed. 
The solid red lines show the results with hadronization, and the dashed green line in the top panel shows the result for partonic jets. 
}
\label{fig:correlated_broadening_5TeV}
\end{figure}

In a complete simulation of jets in a heavy-ion collision (via the \textsc{jetscape} package), partons with $E \leq E_{\mathrm{med}} \approx 10 T$ would be considered soft enough to be thermalized within the medium~\cite{He:2018xjv,Tachibana:2020mtb,Zhao:2021vmu,Yang:2021qtl,Yang:2022nei}, 
the four-momentum of these soft partons would then become part of an energy-momentum source term for a subsequent bulk medium simulation, which would start with the exact initial state that generated both the primary bulk simulation and the distribution of hard scattering, that led to jet production. The space-time-dependent energy-momentum source term enacts a boundary between the portion of jet modification that can be carried out using perturbation theory and the part that should be carried out nonperturbatively. This source term takes the four momentum of the partons with $E < E_{\mathrm{med}}$ and diffuses these out in space-time using a causal diffusion equation~\cite{Tachibana:2020mtb}. The second hydro simulation, while starting out with the same initial state as the prior simulation, is modified by the presence of the source term. The hadrons produced in the freeze-out of this new bulk simulation are then combined with the hadrons from the fragmentation of the jet. Jet reconstruction algorithms will have to be run on the ``full'' event. Background subtraction and unfolding can be carried out by statistical subtraction of the jet distribution clustered from the hadrons produced using the primary bulk simulation (without jets).

What we outline above is a computationally challenging problem. A secondary bulk simulation would have to be run for every hard scattering event. One would not be able to avail of the standard methodology of running up to 1000 hard scattering events per a single bulk event. However, due to the diffusion of the source term and the hydrodynamical response of the bulk medium to supersonic energy deposition, one would expect the formation of a Mach cone at an 
$R \gtrsim 1$ from the jet axis~\cite{Tachibana:2015qxa}. The exact Mach angle depends on the equation of state and thus varies as the jet passes through the plasma. However, energy and momentum from partons with $E<E_{\mathrm{med}}$ would, over time, propagate away from the jet, at the Mach angle. The loss of these partons from the clustered jet would constitute an additional source of jet energy loss. 

In the last part of this appendix we model this process of energy loss via medium response by instituting an additional correlated broadening on partons with $E_{\mathrm{cut}} < E < E_{\mathrm{med}}$. Partons with energy in the local fluid rest frame that lie within this range receive a transverse momentum kick such that over a small length $\delta l$, $\langle k_\perp^2 \rangle = \hat{q} \delta l$, where $\hat{q}$ is the local jet transport coefficient. The difference from the regular process of transverse broadening is that subsequent kicks are always radially away from the jet cone and always in the same direction. These \emph{correlated} kicks continue until the parton is at or beyond the Mach angle from the jet axis, or it exits the medium, whichever occurs first. After crossing the Mach angle it continues to undergo transverse kicks from the medium, but with subsequent kick directions randomized (regular 2-dimensional diffusion).

This approximate method yields both a loss of partons from within the jet cone and also their reappearance at the Mach angle. The result of this shift on the partonic and eventual hadronic $R_{\mathrm{AA}}$ is shown in Fig.~\ref{fig:correlated_broadening_5TeV}. As one would notice, the effect of additional medium response with changing $\alpha^\mathrm{fix}_{\mathrm{s}} = 0.35$ allows us to obtain a simultaneous fit to both the  
inclusive jet and hadron suppression, and leads to consistent partonic and hadronic $R_{\mathrm{AA}}$.

As the reader will note, the effect of energy loss via medium response seems to have little effect on the hadronic jet $R_{\mathrm{AA}}$; comparing to the simulations in the earlier sections, its primary effect is to make the partonic $R_{\mathrm{AA}}$ consistent with the hadronic one which has been presented in the earlier sections.
As a result, for the current work, which only focuses on the inclusive jet and single hadron $R_{\mathrm{AA}}$, the new at-source method of background subtraction, followed by the jet modification due to medium response, has little import. Hence we discuss these in the appendix. The exact nature of how energy is transferred out of the jet cone is relevant to substructure observables such as the jet shape, which will be discussed in our future efforts.

\bibliography{main,misc}

\end{document}